\documentclass[aps,a4paper,superscriptaddress,showpacs,preprintnumbers,amsmath,amssymb]{revtex4}

\usepackage{ulem}

\usepackage{psfrag} \usepackage{graphicx} \usepackage{dcolumn}
\usepackage{color} \usepackage{latexsym,amsfonts} \usepackage{bm}
\usepackage{amssymb}
\begin{document}

\title{Gauge Properties of Hadronic Structure of Nucleon\\ in Neutron
  Radiative Beta Decay to Order $O(\alpha/\pi)$ in Standard $V - A$
  Effective Theory\\with QED and Linear Sigma Model of Strong
  Low--Energy Interactions}

\author{A. N. Ivanov}\email{ivanov@kph.tuwien.ac.at}
\affiliation{Atominstitut, Technische Universit\"at Wien, Stadionallee
  2, A-1020 Wien, Austria}
\author{R.~H\"ollwieser}\email{roman.hoellwieser@gmail.com}
\affiliation{Atominstitut, Technische Universit\"at Wien, Stadionallee
  2, A-1020 Wien, Austria}\affiliation{Department of Physics,
  Bergische Universit\"at Wuppertal, Gaussstr. 20, D-42119 Wuppertal,
  Germany} \author{N. I. Troitskaya}\email{natroitskaya@yandex.ru}
\affiliation{Atominstitut, Technische Universit\"at Wien, Stadionallee
  2, A-1020 Wien, Austria}
\author{M. Wellenzohn}\email{max.wellenzohn@gmail.com}
\affiliation{Atominstitut, Technische Universit\"at Wien, Stadionallee
  2, A-1020 Wien, Austria} \affiliation{FH Campus Wien, University of
  Applied Sciences, Favoritenstra\ss e 226, 1100 Wien, Austria}
\author{Ya. A. Berdnikov}\email{berdnikov@spbstu.ru}\affiliation{Peter
  the Great St. Petersburg Polytechnic University, Polytechnicheskaya
  29, 195251, Russian Federation}

\date{\today}

\begin{abstract}
Within the standard $V - A$ theory of weak interactions, Quantum
Electrodynamics (QED) and the linear $\sigma$--model (L$\sigma$M) of
strong low--energy hadronic interactions we analyse gauge properties
of hadronic structure of the neutron and proton in the neutron
radiative $\beta^-$--decay. We show that the Feynman diagrams,
describing contributions of hadronic structure to the amplitude of the
neutron radiative $\beta^-$--decay in the tree--approximation for
strong low--energy interactions in the L$\sigma$M, are gauge
invariant. In turn, the complete set of Feynman diagrams, describing
the contributions of hadron--photon interactions in the
one--hadron--loop approximation, is not gauge invariant. In the
infinite limit of the scalar $\sigma$--meson, reproducing the current
algebra results (Weinberg, Phys. Rev. Lett. {\bf 18}, 188 (1967)), and
to leading order in the large nucleon mass expansion the Feynman
diagrams, violating gauge invariance, do not contribute to the
amplitude of the neutron radiative $\beta^-$--decay in agreement with
Sirlin's analysis of strong low--energy interactions in neutron
$\beta^-$ decays.
We assert that the problem of appearance of gauge non--invariant
Feynman diagrams of hadronic structure of the neutron and proton is
related to the following. The vertex of the effective $V-A$ weak
interactions does not belong to the combined quantum field theory
including the L$\sigma$M and QED. We argue that gauge invariant set of
Feynman diagrams of hadrons, coupled to real and virtual photons in
neutron $\beta^-$ decays, can be obtained within the combined quantum
field theory including the Standard Electroweak Model (SEM) and the
L$\sigma$M, where the effective $V-A$ vertex of weak interactions is a
result of the $W^-$--electroweak boson exchange.
\end{abstract}
 \maketitle

\section{Introduction}
\label{sec:introduction}

It is well--known that radiative corrections of order $O(\alpha/\pi)$
\cite{Berman1958}-\cite{Ivanov2017a}, where $\alpha$ is the
fine--structure constant \cite{PDG2016}, play an important role for
correct description of the properties of neutron $\beta^-$ decays. In
turn, an important role of strong low--energy interactions in weak
decays has been pointed out by Weinberg \cite{Weinberg1957}.  Nowadays
the contributions of strong low--energy interactions to neutron
$\beta^-$ decays reduce to the axial coupling constant $g_A =
1.2750(9)$ \cite{Abele2008} (see also \cite{Nico2009}), agreeing well
with recent value $g^{(\rm favoured)}_A = 1.2755(11)$, which was
recommended by Czarnecki {\it et al.}  \cite{Sirlin2018} as a {\it
  favoured} one. We would like to remind that the axial coupling
constant $g_A$ appears in the standard $V - A$ theory of weak
interactions as a trace of strong low--energy interactions in the
matrix element of the hadronic $n \to p$ transition after
renormalization of the matrix element of the axial--vector hadronic
current \cite{DeAlfaro1973}. In turn, according to Weinberg
\cite{Weinberg1957}, contributions of strong low--energy interactions
beyond the axial coupling constant $g_A$ seem to be important for
gauge invariant description of radiative corrections of order
$O(\alpha/\pi)$ to neutron $\beta^-$ decays. However, as has been
shown by Sirlin \cite{Sirlin1967,Sirlin1968,Sirlin1978} the
contribution of strong low--energy interactions to the radiative
corrections of order $O(\alpha/\pi)$ to the neutron lifetime,
calculated to leading order in the large nucleon mass expansion, is a
constant independent of the electron energy. Because of such a
property of strong low--energy interactions their contributions to
neutron $\beta^-$ decays have been left at the level of the axial
coupling constant $g_A$ and screened in the radiative corrections
\cite{Sirlin1967}-\cite{Ivanov2017a}. The necessity to take into
account contributions of electroweak--boson exchanges
\cite{Weinberg1971} for the calculation of radiative corrections of
order $O(\alpha/\pi)$ has been pointed out by Sirlin
\cite{Sirlin1974,Sirlin1975b,Sirlin1978,Sirlin1982}. The analysis of
electroweak--boson exchanges and QCD corrections has been continued by
Marciano and Sirlin \cite{Sirlin1986,Sirlin2006}, Degrassi and Sirlin
\cite{Sirlin1992}, Czarnecki, Marciano and Sirlin \cite{Sirlin2004},
and Sirlin and Ferroglia \cite{Sirlin2013}. As has been shown by
Czarnecki {\it et al.} \cite{Sirlin2004} the contributions of
electroweak-boson exchanges change crucially the value of the
radiative corrections of order $O(\alpha/\pi)$. Indeed, the radiative
corrections to the neutron lifetime, averaged over the
electron--energy spectrum, are equal to $\langle
(\alpha/\pi)\,g_n(E_e)\rangle = 0.015056$ and $\langle
(\alpha/\pi)\,g_n(E_e)\rangle = 0.0390(8)$ without and with the
contributions of the electroweak-boson exchanges and QCD corrections,
respectively \cite{Sirlin2004}, where the function $g_n(E_e)$
describes the radiative corrections to the neutron lifetime in
notation \cite{Ivanov2013,Ivanov2017a}. For the correct gauge
invariant calculation of radiative corrections of order
$O(\alpha^2/\pi^2)$ to the rate of the neutron radiative
$\beta^-$--decay $n \to p + e^- + \bar{\nu}_e + \gamma$ an appearance
of non--trivial contributions of strong low--energy interactions
dependent on the energies of decay particles has been pointed out in
\cite{Ivanov2017b}. As has been found in \cite{Ivanov2017b} the
interactions of real and virtual photons with hadronic structure of
the neutron and proton should provide not only gauge invariance of
radiative corrections of order $O(\alpha^2/\pi^2)$ but also
non--trivial dependence of these corrections on the electron $E_e$ and
photon $\omega$ energies. The problem of gauge invariant non--trivial
contributions of strong low--energy interactions to neutron $\beta^-$
decays is closely related to the analysis of corrections of order
$10^{-5}$, calculated in the Standard Model
\cite{Ivanov2017a,Ivanov2017b}.

The experimental analysis of the Standard Model (SM) in neutron
$\beta^-$ decays at the level of $10^{-4}$ \cite{Abele2016} makes
urgent theoretical investigations of neutron $\beta^-$ decays at the
level of $10^{-5}$. For predictions at the level of $10^{-4}$, it is
apparent that the higher order corrections of order $10^{-5}$ should
be included, and for an experimental search for interactions beyond
the SM, a "discovery" experiment with the required 5$\sigma$
sensitivity will require experimental uncertainties of a few parts in
$10^{-5}$ \cite{Ivanov2017a}. The complete set of corrections of order
$10^{-5}$ contains 1) Wilkinson's corrections \cite{Wilkinson1982}
such as i) the proton recoil in the Coulomb electron--proton
final--state interaction, ii) the finite proton radius, iii) the
proton--lepton convolution and iv) the higher--order {\it outer}
radiative corrections, which are of order $10^{-5}$
\cite{Ivanov2017a}, 2) the radiative corrections of order $O(\alpha
E_e/m_N)$, calculated to next--to--leading order in the large nucleon
mass $m_N$ expansion, 3) the radiative corrections of order
$O(\alpha^2/\pi^2)$, calculated to leading order in the large nucleon
mass $m_N$ expansion, and 4) the weak magnetism and proton recoil
corrections $O(E^2_e/m^2_N)$, calculated to
next--to--next--to--leading order in the large nucleon mass $m_N$
expansion \cite{Ivanov2018a}. A derivation of such a complete set of
corrections of order $10^{-5}$ should give a new impetus for
experimental searches of interactions beyond the SM, induced by first
class ($G$--even) \cite{Lee1956}--\cite{Ivanov2018a} and second class
($G$--odd) \cite{Gardner2001,Gardner2013a} (see also
\cite{Ivanov2018a}) hadronic currents. We remind that the $G$--parity
transformation, i.e. $G = C\,e^{\,i \pi I_2}$, where $C$ and $I_2$ are
the charge conjugation and isospin operators, was introduced by Lee
and Yang \cite{Lee1956a} as a symmetry of strong
interactions. According to the properties of hadronic currents with
respect to the $G$--parity transformation, Weinberg divided hadronic
currents into two classes \cite{Weinberg1958}, where the first class
and second class hadronic currents are $G$--even and $G$--odd,
respectively.

This paper addresses to the quantum field theoretic analysis of gauge
invariance of contributions of strong low--energy interactions or
hadronic structure of the neutron and proton to the amplitude of the
neutron radiative $\beta^-$--decay. We follow the standard $V - A$
effective theory of weak interactions with electromagnetic and strong
low--energy interactions described by Quantum Electrodynamics (QED)
and the linear $\sigma$--model (L$\sigma$M) \cite{GellMann1960} (see
also \cite{DeAlfaro1973}), respectively. The L$\sigma$M with chiral
$SU(2) \times SU(2)$ symmetry describing strong low--energy
meson--nucleon interactions possesses the following important
properties: i) it is unstable under spontaneous breaking of chiral
$SU(2) \times SU(2)$ symmetry leading to appearance of heavy nucleon,
light massive pions (the $\pi$--mesons) and heavy scalar
$\sigma$--meson, ii) in the chiral symmetry broken phase it provides a
proportionality of the divergence of the axial--vector hadronic
current to the $\pi$--meson field realizing at the quantum field
theoretic level the hypothesis of partial conservation of the
axial--vector hadronic current or the PCAC hypothesis, iii) it is
renormalizable \cite{Bernstein1960}--\cite{Strubbe1972}, and iv) it
reproduces the results of the current algebra approach in the limit of
the infinite $\sigma$--meson mass
\cite{Weinberg1967,Gasiorowicz1969}. These properties of the
L$\sigma$M should give a possibility to describe the contributions of
strong low--energy interactions to the neutron radiative
$\beta^-$--decay at the quantum field theoretic level and at the
confidence level of Sirlin's analysis of strong low--energy
interactions in neutron $\beta^-$ decays \cite{Sirlin1967,Sirlin1978}.

The paper is organized as follows. In section \ref{sec:sigma} we
follow \cite{GellMann1960,DeAlfaro1973} and formulate the L$\sigma$M
with chiral $SU(2)\times SU(2)$ symmetry, describing strong
low--energy interactions of the hadronic system including the scalar
$\sigma$--meson, $\pi$--mesons and nucleon (neutron and proton). We
outline the L$\sigma$M in the chirally symmetric and chiral symmetry
broken phases, and describe renormalization procedure in the
L$\sigma$M \cite{Bernstein1960}--\cite{Strubbe1972}. In section
\ref{sec:qed} we describe renormalization procedure of the quantum
field theory of the proton, $\pi^{\pm}$--mesons and electron in the
framework of Quantum Electrodynamics (QED)
\cite{Itzykson1980}--\cite{Ward1950}. In section \ref{sec:qedlsm} we
consider the quantum field theory of the scalar $\sigma$--meson,
$\pi$--mesons, nucleon and electron coupled within the L$\sigma$M and
QED, and describe renormalization procedure of such a combined quantum
field theory. In section \ref{sec:weak} we derive the general
expressions for the matrix elements of the neutron $\beta^-$--decay
and neutron radiative $\beta^-$--decay in the standard $V - A$
effective theory of weak interactions, where the vector $V$ and
axial--vector $A$ charged hadronic currents are defined within the
L$\sigma$M in the chirally broken phase, with strong low--energy and
electromagnetic interactions described by the L$\sigma$M and QED. In
section \ref{sec:strong} we calculate the matrix element of the
hadronic $n \to p$ transition of the neutron $\beta^-$--decay in the
tree-- and one--hadron--loop approximation for strong low--energy
interactions in the L$\sigma$M. We reproduce the standard Lorentz
structure of this matrix element, which has been earlier reproduced
within Yukawa's theory of strong pion--nucleon interactions
\cite{Ivanov2018}. In section \ref{sec:strahlung} we calculate the
amplitude of the neutron radiative $\beta^-$--decay in the tree-- and
one--hadron--loop approximation for strong low--energy interactions in
the L$\sigma$M and QED. In section \ref{sec:schluss} we discuss the
obtained results and perspectives of gauge invariant description of
contributions of hadronic structure of the neutron and proton in
neutron $\beta^-$ decays.

\section{Linear $\sigma$--model (L$\sigma$M) with chiral $SU(2) \times
 SU(2)$ symmetry}
\label{sec:sigma}

\subsection{Chirally symmetric phase}

The L$\sigma$M with chiral $SU(2)\times SU(2)$ symmetry describes
strong low--energy pion--nucleon interactions with a mediation of the
scalar $\sigma$--meson. As has been shown in \cite{Ivanov1992} by
example of the low--energy $\gamma + \gamma \to \pi + \pi$ processes
the contributions of the $\sigma$--meson can be screened in observable
processes. The Lagrangian of the L$\sigma$M in the chirally symmetric
phase is given by \cite{DeAlfaro1973}
\begin{eqnarray}\label{eq:1}
\hspace{-0.15in}{\cal L}_{\rm L\sigma M}(x) =
\bar{\psi}_N\big(i\gamma^{\mu}\partial_{\mu} + g_{\pi N}(\sigma + i
\vec{\tau}\cdot \vec{\pi}\,)\big)\psi_N +
\frac{1}{2}\,\big(\partial_{\mu}\sigma\partial^{\mu}\sigma +
\partial_{\mu}\vec{\pi}\cdot \partial^{\mu}\vec{\pi}\,\big) +
\frac{1}{2}\,\mu^2\,\big(\sigma^2 + \vec{\pi}^{\,2}\big) -
\frac{\gamma}{4}\,\big(\sigma^2 + \vec{\pi}^{\,2}\big)^2 ,
\end{eqnarray}
where $\psi_N$ is the isospin doublet of the nucleon field operator
with components $(\psi_p, \psi_n)$, where $\psi_p$ and $\psi_n$ are
the proton and neutron field operators, respectively, $\sigma$ and
$\vec{\pi} = (\pi^+, \pi^0, \pi^-)$ are the scalar $\sigma$-- and
pseudoscalar pion--meson field operators, $\mu^2$, $\gamma$ and
$g_{\pi N}$ are input parameters of the L$\sigma$M, and
$\vec{\tau} = (\tau_1, \tau_2,\tau_3)$ are the isospin $2\times 2$
Pauli matrices. The scalar $\sigma$ and pseudoscalar $\vec{\pi}$
fields describe the isoscalar and isovector meson states,
respectively.

Under isovector and isoaxial--vector (or chiral) infinitesimal
transformations with parameters $\vec{\alpha}_V$ and $\vec{\alpha}_A$,
respectively, the nucleon and meson fields transform as follows
\begin{eqnarray}\label{eq:2}
\hspace{-0.3in} \psi_N \stackrel{\vec{\alpha}_V}\longrightarrow
\psi'_N &=& \Big(1 + i\,\frac{1}{2}\,\vec{\alpha}_V\cdot
\vec{\tau}\,\Big) \psi_N \quad, \quad \bar{\psi}_N \stackrel{\vec{\alpha}_V}\longrightarrow \bar{\psi}'_N =
\bar{\psi}_N\Big(1 - i\,\frac{1}{2}\,\vec{\alpha}_V\cdot
\vec{\tau}\,\Big),\nonumber\\
\hspace{-0.3in}\sigma \stackrel{\vec{\alpha}_V}\longrightarrow \sigma'
&=& \sigma \quad,\quad \vec{\pi}
\stackrel{\vec{\alpha}_V}\longrightarrow \vec{\pi}^{\,'} =  \vec{\pi}
- \vec{\alpha}_V \times \vec{\pi},\nonumber\\
\hspace{-0.3in} N \stackrel{\vec{\alpha}_A}\longrightarrow N' &=&
\Big(1 + i\,\frac{1}{2}\,\gamma^5 \vec{\alpha}_A\cdot
\vec{\tau}\,\Big) N\quad, \quad \bar{\psi}_N
\stackrel{\vec{\alpha}_A}\longrightarrow \bar{\psi}'_N = \bar{\psi}_N
\Big(1 + i\,\frac{1}{2}\,\gamma^5 \vec{\alpha}_A\cdot
\vec{\tau}\,\Big),\nonumber\\
\hspace{-0.3in}\sigma \stackrel{\vec{\alpha}_A}\longrightarrow \sigma'
&=& \sigma + \vec{\alpha}_A \cdot \vec{\pi} \quad , \quad \vec{\pi}
\stackrel{\vec{\alpha}_A}\longrightarrow \vec{\pi}^{\,'} = \vec{\pi} -
\vec{\alpha}_A \sigma.
\end{eqnarray}
The Lagrangian Eq.(\ref{eq:1}) is invariant under global
transformations Eq.(\ref{eq:2}). Under local transformations
Eq.(\ref{eq:2}) the Lagrangian Eq.(\ref{eq:1}) acquires the following
corrections
\begin{eqnarray}\label{eq:3}
\delta {\cal L}_{\rm L\sigma M}(x) = - \partial^{\mu}\vec{\alpha}_V\cdot
\Big(\bar{\psi}_N \gamma_{\mu}\,\frac{1}{2}\,\vec{\tau}\,\psi_N +
\vec{\pi} \times \partial_{\mu}\vec{\pi}\,\Big) -
\partial^{\mu}\vec{\alpha}_A \cdot \Big(\bar{\psi}_N
\gamma_{\mu}\,\gamma^5 \frac{1}{2}\,\vec{\tau}\,\psi_N + \big(\sigma
\,\partial_{\mu}\vec{\pi} -
\vec{\pi}\,\partial_{\mu}\sigma\,\big)\Big),
\end{eqnarray}
which allow to define the vector and axial--vector hadronic currents
\cite{DeAlfaro1973}
\begin{eqnarray}\label{eq:4}
\vec{V}_{\mu} &=& - \frac{\delta {\cal L}_{\rm L\sigma M}}{\delta
  \partial^{\mu}\vec{\alpha}_V} = \bar{\psi}_N
\gamma_{\mu}\,\frac{1}{2}\,\vec{\tau}\,\psi_N + \vec{\pi} \times
\partial_{\mu}\vec{\pi},\nonumber\\ \vec{A}_{\mu} &=& - \frac{\delta
  {\cal L}_{\rm L\sigma M}}{\delta \partial^{\mu}\vec{\alpha}_A} =
\bar{\psi}_N \gamma_{\mu}\,\gamma^5 \frac{1}{2}\,\vec{\tau}\,\psi_N +
\big( \sigma \,\partial_{\mu}\vec{\pi} -
\vec{\pi}\,\partial_{\mu}\sigma\,\big).
\end{eqnarray}
Using the equations of motion for the nucleon, scalar and pseudoscalar
fields one may show that in the chirally symmetric phase the
divergences of the vector and axial--vector hadronic currents vanish 
\begin{eqnarray}\label{eq:5}
\partial^{\mu}\vec{V}_{\mu} = \partial^{\mu}\vec{A}_{\mu} = 0.
\end{eqnarray}
This means that in the chirally symmetric phase the vector and
axial--vector hadronic current are locally conserved.

\subsection{Phase of spontaneously broken chiral symmetry}

We would like to accentuate that the nucleon, scalar and pseudoscalar
fields in Eq.(\ref{eq:1}) are unphysical. Indeed, the nucleon is
massless and the mass term of the scalar and pseudoscalar fields
enters with incorrect sign. Hence, physical hadronic states can appear
in the L$\sigma$M only in the phase of spontaneously broken chiral
symmetry \cite{GellMann1960}. In the L$\sigma$M the phase of
spontaneously broken chiral $SU(2) \times SU(2)$ symmetry can be
described by the Lagrangian \cite{DeAlfaro1973}
\begin{eqnarray}\label{eq:6}
&&{\cal L}_{\rm L\sigma M}(x) =
  \bar{\psi}_N\big(i\gamma^{\mu}\partial_{\mu} + g_{\pi N}(\sigma +
  i\gamma^5 \vec{\tau}\cdot \vec{\pi}\,)\big) \psi_N +
  \frac{1}{2}\,\big(\partial_{\mu}\sigma\partial^{\mu}\sigma +
  \partial_{\mu}\vec{\pi}\cdot \partial^{\mu}\vec{\pi}\,\big) +
  \frac{1}{2}\,\mu^2\,\big(\sigma^2 + \vec{\pi}^{\,2}\big) -
  \frac{\gamma}{4}\,\big(\sigma^2 + \vec{\pi}^{\,2}\big)^2 + a
  \sigma,\nonumber\\
\end{eqnarray}
where the last term $a \sigma$ is non--invariant under chiral
transformations Eq.(\ref{eq:2}).

The phase of spontaneously broken chiral symmetry characterizes by a
non--vanishing vacuum expectation value of the $\sigma$--field
$\langle \sigma \rangle = b \neq 0$. The transition to the fields of
physical hadronic states goes through the change of the
$\sigma$--field $\sigma \to \sigma + b$, where in the
right--hand--side (r.h.s.) the $\sigma$--field possesses a vanishing
vacuum expectation value. After such a change of the $\sigma$--field
the dynamics of physical hadronic states is described by the
Lagrangian
\begin{eqnarray}\label{eq:7}
\hspace{-0.3in}{\cal L}_{\rm L\sigma M}(x) &=&
\bar{\psi}_N\big(i\gamma^{\mu}\partial_{\mu} - m_N + g_{\pi N}(\sigma
+ i\gamma^5 \vec{\tau}\cdot \vec{\pi}\,)\big)\, \psi_N +
\frac{1}{2}\,\big(\partial_{\mu}\sigma \partial^{\mu}\sigma -
m^2_{\sigma} \sigma^2\big) +
\frac{1}{2}\,\big(\partial_{\mu}\vec{\pi}\cdot \partial^{\mu}\vec{\pi}
- m^2_{\pi}\vec{\pi}^{\,2}\big)\nonumber\\
\hspace{-0.3in}&-&\gamma\,b\,\sigma\big(\sigma^2 +
\vec{\pi}^{\,2}\big) - \frac{\gamma}{4}\,\big(\sigma^2 +
\vec{\pi}^{\,2}\big)^2,
\end{eqnarray}
where the masses of physical hadrons and coupling constants are
determined by
\begin{eqnarray}\label{eq:8}
\hspace{-0.3in}m_N = - g_{\pi N} b\;,\; m^2_{\sigma} = 3\gamma b^2 - \mu^2\;,\; m^2_{\pi} = \gamma b^2 - \mu^2\;,\,a = m^2_{\pi}b.
\end{eqnarray}
In the phase of spontaneously broken chiral symmetry the vector
and axial--vector hadronic currents are equal to
\begin{eqnarray}\label{eq:9}
\vec{V}_{\mu} &=&\bar{\psi}_N \gamma_{\mu}\,\frac{1}{2}\,\vec{\tau}\,\psi_N +
\vec{\pi} \times \partial_{\mu}\vec{\pi},\nonumber\\ \vec{A}_{\mu} &=&
\bar{\psi}_N \gamma_{\mu}\,\gamma^5 \frac{1}{2}\,\vec{\tau}\,\psi_N + \big(
\sigma \,\partial_{\mu}\vec{\pi} -
\vec{\pi}\,\partial_{\mu}\sigma\,\big) + b\,\partial_{\mu}\vec{\pi}.
\end{eqnarray}
Using the equations of motion for the nucleon, scalar and pseudoscalar
fields one may show that the divergences of the vector and axial
vector hadronic currents are given by
\begin{eqnarray}\label{eq:10}
\partial^{\mu}\vec{V}_{\mu} &=&
0,\nonumber\\ \partial^{\mu}\vec{A}_{\mu} &=& - m^2_{\pi}
b\,\vec{\pi}.
\end{eqnarray}
According to the hypothesis of partial conservation of the
axial--vector hadronic current (the PCAC hypothesis)
\cite{GellMann1960} the divergence of the axial--vector hadronic
current is proportional to the pion--field operator
$\partial^{\mu}\vec{A}_{\mu} = m^2_{\pi}f_{\pi}\vec{\pi}$, where
$f_{\pi}$ is the PCAC constant or the pion decay constant
\cite{GellMann1960}. From the comparison of the divergence in
Eq.(\ref{eq:10}) and $\partial^{\mu}\vec{A}_{\mu} =
m^2_{\pi}f_{\pi}\vec{\pi}$ we get $b = - f_{\pi}$
\cite{GellMann1960,DeAlfaro1973}. Unlike the axial--vector hadronic
current the vector hadronic current is locally conserved even in the
phase of spontaneously broken chiral symmetry. Conservation of the
vector hadronic current in the L$\sigma$M can be violated only by
isospin symmetry breaking.

\subsection{Renormalization in the L$\sigma$M}

For the discussion of renormalization procedure in the L$\sigma$M we
rewrite the Lagrangian Eq.(\ref{eq:7}) as follows
\cite{Lee1969,Gervais1969,Mignaco1971}
\begin{eqnarray}\label{eq:11}
\hspace{-0.3in}{\cal L}^{(0)}_{\rm L\sigma M}(x) &=&
\bar{\psi}^{(0)}_N\big(i\gamma^{\mu}\partial_{\mu} - m^{(0)}_N +
g^{(0)}_{\pi N}(\sigma^{(0)} + i \gamma^5 \vec{\tau}\cdot
\vec{\pi}^{\,(0)})\big)\,\psi^{(0)}_N\nonumber\\
\hspace{-0.3in} &+& \frac{1}{2}\,\big(\partial_{\mu}\sigma^{(0)}
\partial^{\mu}\sigma^{(0)} - m^{(0)2}_{\sigma} \sigma^{(0)2}\big) +
\frac{1}{2}\,\big(\partial_{\mu}\vec{\pi}^{\,(0)}\cdot
\partial^{\mu}\vec{\pi}^{\,(0)} -
m^{(0)2}_{\pi}\vec{\pi}^{\,(0)2}\big)\nonumber\\
\hspace{-0.3in}&+&\gamma^{(0)}\,f^{(0)}_{\pi}\,\sigma^{(0)}\big(\sigma^{(0)2}
+ \vec{\pi}^{\,(0)2}\big) - \frac{\gamma^{(0)}}{4}\,\big(\sigma^{(0)2}
+ \vec{\pi}^{\,(0)2}\big)^2,
\end{eqnarray}
where $\psi^{(0)}_N$, $\sigma^{(0)}$ and $\vec{\pi}^{\,(0)}$ are {\it
  bare} hadronic fields, $m^{(0)}_N$, $m^{(0)}_{\sigma}$,
$m^{(0)}_{\pi}$ and $\gamma^{(0)}$, $f^{(0)}_{\pi}$ are {\it bare}
hadronic masses and coupling constants, respectively.  After the
calculation of hadron--loop contributions the dynamics of physical
fields is described by the Lagrangian
\begin{eqnarray}\label{eq:12}
\hspace{-0.3in}{\cal L}^{(r)}_{\rm L\sigma M}(x) &=&
\bar{\psi}^{(r)}_N\big(i\gamma^{\mu}\partial_{\mu} - m^{(r)}_N +
g^{(r)}_{\pi N}(\sigma^{(r)} + i \gamma^5 \vec{\tau}\cdot
\vec{\pi}^{\,(r)}\,)\big)\,\psi^{(r)}_N +
\frac{1}{2}\,\big(\partial_{\mu}\sigma^{(r)}
\partial^{\mu}\sigma^{(r)} - m^{(r)2}_{\sigma}
(\sigma^{(r)})^2\big)\nonumber\\
\hspace{-0.3in}&+&
\frac{1}{2}\,\big(\partial_{\mu}\vec{\pi}^{\,(r)}\cdot
\partial^{\mu}\vec{\pi}^{\,(r)} -
m^{(r)2}_{\pi}(\vec{\pi}^{\,(r)})^2\big) +
\gamma^{(r)}\,f^{(r)}_{\pi}\,\sigma^{(r)}\big((\sigma^{(r)})^2 +
(\vec{\pi}^{\,(r)})^2\big) -
\frac{\gamma^{(r)}}{4}\,\big((\sigma^{(r)})^2 +
(\vec{\pi}^{\,(r)})^2\big)^2\nonumber\\
\hspace{-0.3in}&+& {\cal L}^{(\rm CT)}_{\rm L\sigma M}(x),
\end{eqnarray}
where the Lagrangian ${\cal L}^{(\rm CT)}_{\rm L\sigma M}(x)$
is given by
\begin{eqnarray}\label{eq:13}
\hspace{-0.3in}&&{\cal L}^{(\rm CT)}_{\rm L\sigma M}(x) = (Z_N
- 1)\bar{\psi}^{(r)}_N\big(i\gamma^{\mu}\partial_{\mu} -
m^{(r)}_N\big)\,\psi^{(r)}_N - Z_N \delta
m^{(r)}_N\,\bar{\psi}^{(r)}_N \psi^{(r)}_N + \big(Z_{M N} -
1\big)\,g^{(r)}_{\pi N}\bar{\psi}^{(r)}_N \big(\sigma^{(r)} +
i\gamma^5 \vec{\tau}\cdot \vec{\pi}^{\,(r)}\big)\,
\psi^{(r)}_N\nonumber\\
\hspace{-0.3in}&& +\big(Z_M -
1\big)\,\frac{1}{2}\,\big(\partial_{\mu}\sigma^{(r)}
\partial^{\mu}\sigma^{(r)} - m^{(r)2}_{\sigma} (\sigma^{(r)})^2\big) -
Z_M \delta m^{(r)2}_{\sigma}(\sigma^{(r)})^2 + \big(Z_M - 1\big)\,
\frac{1}{2}\,\big(\partial_{\mu}\vec{\pi}^{\,(r)}\cdot
\partial^{\mu}\vec{\pi}^{\,(r)} -
m^{(r)2}_{\pi}(\vec{\pi}^{\,(r)})^2\big)\nonumber\\
\hspace{-0.3in}&& - Z_M\delta m^{(r)2}_{\pi}(\vec{\pi}^{\,(r)})^2 +
\big(Z_{3M} -
1\big)\,\gamma^{(r)}\,f^{(r)}_{\pi}\,\sigma^{(r)}\big((\sigma^{(r)})^2
+ (\vec{\pi}^{\,(r)})^2\big) - \big(Z_{4M} -
1\big)\,\frac{\gamma^{(r)}}{4}\,\big((\sigma^{(r)})^2 +
(\vec{\pi}^{\,(r)})^2\big)^2.
\end{eqnarray}
Here $Z_N$, $Z_M$ and $\delta m^{(r)}_N$, $\delta m^{(r)2}_{\sigma}$,
$\delta m^{(r)2}_{\pi}$ are renormalization constants of wave
functions and masses of the nucleon, scalar and pseudoscalar fields,
respectively. Then, $Z_{M N}$, $Z_{3 M}$ and $Z_{4 M}$ are
renormalization constants of the corresponding vertices of
meson--nucleon and meson--meson field interactions. The abbreviation
``CT'' means ``Counter--Terms''. If the fields, masses, coupling
constants and renormalization constants satisfy the relations
\begin{eqnarray}\label{eq:14}
\hspace{-0.3in}\psi^{(0)}_N &=& \sqrt{Z_N}\,\psi^{(r)}_N\;,\; \sigma^{(0)} =
\sqrt{Z_M}\,\sigma^{(r)}\;,\; \vec{\pi}^{\,(0)} =
\sqrt{Z_M}\,\vec{\pi}^{\,(r)},\nonumber\\
\hspace{-0.3in}m^{(0)}_N &=& m^{(r)}_N + \delta m^{(r)}_N\;,\;
m^{(0)2}_{\sigma} = m^{(r)2}_{\sigma} + \delta m^{(r)2}_{\sigma}\;,\;
m^{(0)2}_{\pi} = m^{(r)2}_{\pi} + \delta m^{(r)2}_{\pi},\nonumber\\
\hspace{-0.3in} g^{(0)}_{\pi N} &=& Z_{M N } Z^{-1}_N Z^{-1/2}_M
g^{(r)}_{\pi N}\;,\; f^{(0)}_{\pi} = Z_{3 M} Z^{-1}_{4 M}Z^{1/2}_M
f^{(r)}_{\pi} \;,\;\gamma^{(0)} = Z_{4 M}Z^{-2}_M \gamma^{(r)},\nonumber\\
Z_{3 M} &=& Z_{4 M}.
\end{eqnarray}
the Lagrangian Eq.(\ref{eq:12}) reduces to the Lagrangian
Eq.(\ref{eq:11}). The relation $Z_{3 M} = Z_{4 M}$ implies that the
pion decay constant $f^{(r)}_{\pi}$ is renormalized only
renormalization of the wave function of the $\vec{\pi}$--meson,
i.e. $f^{(0)}_{\pi} = Z^{1/2}_M f^{(r)}_{\pi}$.

\section{Quantum Electrodynamics (QED) of electron, proton and charged 
pions}
\label{sec:qed}

The Lagrangian of the electron--photon and charged hadron--photon
interactions is given by
\begin{eqnarray}\label{eq:15}
\hspace{-0.3in}&&{\cal L}^{(0)}_{\rm QED}(x) = -
\frac{1}{4}\,F^{(0)}_{\mu\nu} F^{(0)\mu\nu} -
\frac{1}{2\xi_0}\,\big(\partial_{\mu}A^{(0)\mu} \big)^2 +
\bar{\psi}^{(0)}_e \big(i\gamma^{\mu}\partial_{\mu} -
m^{(0)}_e\big)\psi^{(0)}_e - (- e_0)\, \bar{\psi}^{(0)}_e
\gamma^{\mu}\psi^{(0)}_e\, A^{(0)}_{\mu} \nonumber\\
\hspace{-0.3in}&&+ \bar{\psi}^{(0)}_p \big(i\gamma^{\mu}\partial_{\mu}
- m^{(0)}_p\big)\psi^{(0)}_p - (+ e_0)\, \bar{\psi}^{(0)}_p
\gamma^{\mu}\psi^{(0)}_p\, A^{(0)}_{\mu} + \big(\partial_{\mu} + i (+
e_0)\, A^{(0)}_{\mu}\big) \pi^{(0)+}\big(\partial^{\mu} + i (- e_0)\,
A^{(0)\mu}\big) \pi^{(0)-}\nonumber\\
\hspace{-0.3in}&& - m^{(0)2}_{\pi}\pi^{(0)+}\pi^{(0)-},
\end{eqnarray}
where $F^{(0)}_{\mu\nu}(x) = \partial_{\mu}A^{(0)}_{\nu}(x) -
\partial_{\nu}A^{(0)}_{\mu}(x)$ is the electromagnetic field strength
tensor of the {\it bare} electromagnetic field operator
$A^{(0)}_{\mu}(x)$, $\xi_0$ is a {\it bare} gauge parameter;
$\psi^{(0)}_e$, $\psi^{(0)}_p$ and $\pi^{(0)\pm}$ are {\it bare}
electron, proton and charged pion fields with {\it bare} masses
$m^{(0)}_e$, $m^{(0)}_p$ and $m^{(0)}_{\pi}$ and {\it bare} electric
charges $\mp e_0$ in units of the proton charge $e_0$.

After the calculation of loop corrections a transition to the
renormalized field operators, masses and electric charges is defined
by the Lagrangian
\begin{eqnarray}\label{eq:16}
\hspace{-0.3in}{\cal L}^{(r)}_{\rm QED}(x) &=& -
\frac{1}{4}\,F^{(r)}_{\mu\nu}(x) F^{(r)\mu\nu} -
\frac{1}{2\xi}\,\big(\partial_{\mu}A^{(r)\mu}\big)^2\nonumber\\
\hspace{-0.3in}&+& \bar{\psi}^{(r)}_e(i\gamma^{\mu}\partial_{\mu} -
m^{(r)}_{e})\psi^{(r)}_e - (- e_r)\, \bar{\psi}^{(r)}_e
\gamma^{\mu} \psi^{(r)}_e A^{(r)}_{\mu}\nonumber\\ &+&
\bar{\psi}^{(r)}_p(i\gamma^{\mu}\partial_{\mu} - m^{(r)}_p
)\psi^{(r)}_p - (+ e_r)\, \bar{\psi}^{(r)}_p
\gamma^{\mu}\psi^{(r)}_p A^{(r)}_{\mu}\nonumber\\
\hspace{-0.3in}&+& \big(\partial_{\mu} + i (+ e_r)\,
A^{(r)}_{\mu}\big) \pi^{(r)+}\big(\partial^{\mu} + i (- e_r)\,
A^{(r)\mu}\big) \pi^{(r)-} - m^{(r)2}_{\pi}\pi^{(r)+}\pi^{(r)-} +
{\cal L}^{(\rm CT)}_{\rm QED}(x),
\end{eqnarray}
where $A^{(r)}_{\mu}$, $\psi^{(r)}_e$, $\psi^{(r)}_p$ and
$\pi^{(r)\mp}$ are renormalized operators of the electromagnetic,
electron, proton and charged pion fields, respectively; $m^{(r)}_e$,
$m^{(r)}_p$ and $m^{(r)}_{\pi}$ are renormalized masses of the
electron, proton and charged pions; $e_r$ is the renormalized electric
charge; and $\xi_r$ is the renormalized gauge parameter. The
Lagrangian ${\cal L}^{(\rm CT)}_{\rm QED}(x)$ contains a complete set
of counter--terms
\cite{Itzykson1980,Weinberg1995,Bogoliubov1959,Salam1952,Matthews1954,Fry1973},
\begin{eqnarray}\label{eq:17}
\hspace{-0.3in}&&{\cal L}^{(\rm CT}_{\rm QED}(x) = -
\frac{1}{4}\,(Z_3 - 1)\,F^{(r)}_{\mu\nu}F^{(r)\mu\nu} - \frac{Z_3 -
  1}{Z_{\xi}}\,\frac{1}{2\xi^{(r)}}\,\big(\partial_{\mu}A^{(r)\mu}\big)^2
\nonumber\\\hspace{-0.3in} && + (Z^{(e)}_2 -
1)\,\bar{\psi}^{(r)}_e(i\gamma^{\mu}\partial_{\mu} -
m^{(r)}_{e})\psi^{(r)}_e - (Z^{(e)}_1 - 1)\,(- e_r)\,\bar{\psi}^{(r)}_e
\gamma^{\mu} \psi_e A_{\mu} - Z^{(e)}_2 \delta m^{(r)}_e
\bar{\psi}^{(r)}_e\psi^{(r)}_e \nonumber\\ \hspace{-0.3in}&& +
(Z^{(p)}_2 - 1)\,\bar{\psi}^{(r)}_p(i\gamma^{\mu}\partial_{\mu} -
m^{(r)}_p )\psi^{(r)}_p - (Z^{(p)}_1 - 1)\,( + e_r)
\,\bar{\psi}^{(r)}_p \gamma^{\mu}\psi^{(r)}_p A^{(r)}_{\mu} -
Z^{(p)}_2 \delta m^{(r)}_p \bar{\psi}_p \psi^{(r)}_p\nonumber\\
\hspace{-0.3in}&&+ (Z^{(\pi)}_2 - 1) \big(\partial_{\mu}
\pi^{(r)+}\partial^{\mu} \pi^{(r)-} -
m^{(r)2}_{\pi}\pi^{(r)+}\pi^{(r)-}\big) + (Z^{(\pi)}_1 -
1)\,i\,e_r\,\big(\pi^{(r)+}\partial^{\mu}\pi^{(r)-} -
\partial^{\mu}\pi^{(r)+}\,\pi^{(r)-}\big)\,A^{(r)}_{\mu}\nonumber\\
\hspace{-0.3in}&& + (Z^{(\pi)}_4 -
1)\,e^2_r\,\pi^{(r)+}\,\pi^{(r)-}\,A^{(r)}_{\mu}A^{(r)\mu} -
Z^{(\pi)}_2\,\delta m^{(r)2}_{\pi} \pi^{(r)+}\,\pi^{(r)-},
\end{eqnarray}
where $Z_3$, $Z^{(j)}_2$, $Z^{(j)}_1$ and $Z^{(\pi)}_4$ for $j =
e,p,\pi^{\mp}$, $\delta m^{(r)}_e$, $\delta m^{(r)}_p$ and $\delta
m^{(r)2}_{\pi}$ are the counter--terms. Here $Z_3$ is the
renormalization constant of the electromagnetic field  operator
$A_{\mu}$, $Z^{(j)}_2$ and $Z^{(j)}_1$ and $Z^{(\pi)}_4$ for $j =
e,p,\pi^{\mp}$ are renormalization constants of the electron, proton
and charged pion field operators and photon--electron and
photon--proton and photon--pion vertices, respectively, where
$Z^{(\pi)}_4 = Z^{(\pi)2}_1Z^{(\pi)- 1}_2$ \cite{Salam1952}. Then,
$(\mp e_r)$, $m^{(r)}_e$, $m^{(r)}_p$, $m^{(r)2}_{\pi}$ and $\delta
m^{(r)}_e$, $\delta m^{(r)}_p$ and $\delta m^{(r)2}_{\pi}$ are
renormalized electric charges and masses and the mass--counter--terms of
the electron, proton and charged pions, respectively. Rescaling the
field operators
\cite{Itzykson1980,Weinberg1995,Bogoliubov1959,Salam1952,Matthews1954,Fry1973}
\begin{eqnarray}\label{eq:18}
\sqrt{Z_3}\, A^{(r)}_{\mu} = A^{(0)}_{\mu}\;,\;
\sqrt{Z^{(e)}_2}\,\psi^{(r)}_e = \psi^{(0)}_e \;,\;
\sqrt{Z^{(p)}_2}\,\psi^{(r)}_p = \psi^{(0)}_p \;,\;
\sqrt{Z^{(\pi)}_2}\,\pi^{(r)\mp}_e = \pi^{(0)\mp},
\end{eqnarray}
and denoting $m^{(r)}_e + \delta m^{(r)}_e = m^{(0)}_e$, $m^{(r)}_p +
\delta m^{(r)}_p = m^{(0)}_p $, $m^{(r)2}_{\pi} + \delta
m^{(r)2}_{\pi} = m^{(0)2}_{\pi}$ and $Z_{\xi} \xi_r = \xi_0$ we arrive
at the Lagrangian
\begin{eqnarray}\label{eq:19}
  \hspace{-0.3in}{\cal L}_{\rm QED}(x) &=& -
  \frac{1}{4}\,F^{(0)}_{\mu\nu}F^{(0)\mu\nu} -
  \frac{1}{2\xi_0}\,\big(\partial_{\mu} A^{(0)\mu}\big)^2\nonumber\\
\hspace{-0.3in}&&+ \bar{\psi}^{(0)}_e(i\gamma^{\mu}\partial_{\mu} -
m^{(r)}_e)\psi^{(0)}_e - ( - e_r)\,Z^{(e)}_1 (Z^{(e)}_2)^{-1} Z^{-1/2}_3
\bar{\psi}^{(0)}_e \gamma^{\mu}\psi^{(0)}_e A^{(0)}_{\mu}\nonumber\\
\hspace{-0.3in}&& + \bar{\psi}^{(0)}_p (i\gamma^{\mu}\partial_{\mu} -
m^{(0)}_p)\psi^{(0)}_p - (+ e_r)\, Z^{(p)}_1(Z^{(p)}_2)^{-1}
Z^{-1/2}_3 \bar{\psi}^{(0)}_p \gamma^{\mu}\psi^{(0)}_p
A^{(0)}_{\mu}\nonumber\\
\hspace{-0.3in}&& + \big(\partial_{\mu} + i (+ e_r)\,Z^{(\pi)}_1
(Z^{(\pi)}_2)^{-1} Z^{-1/2}_3\, A^{(0)}_{\mu}\big)
\pi^{(0)+}\big(\partial^{\mu} + i (- e_r)\,Z^{(\pi)}_1
(Z^{(\pi)}_2)^{-1} Z^{-1/2}_3\, A^{(0)\mu}\big) \pi^{(0)-}\nonumber\\
\hspace{-0.3in}&& -
m^{(0)2}_{\pi}\pi^{(0)+}\pi^{(0)-}.
\end{eqnarray}
Because of the Ward identities $Z^{(e)}_1 = Z^{(e)}_2$ and $Z^{(p)}_1
= Z^{(p)}_2$ \cite{Itzykson1980,Weinberg1995,Bogoliubov1959} and
$Z^{(\pi)}_1 = Z^{(\pi)}_2$ \cite{Salam1952}, we may replace $(\mp
e_r)\,Z^{-1/2}_3 = \mp e_0$. This brings Eq.(\ref{eq:19}) to the form
of Eq.(\ref{eq:15}). 

\section{Quantum field theory of photons, electrons, nucleons, scalar 
and pseudoscalar mesons in QED and L$\sigma$M}
\label{sec:qedlsm}

The dynamics of the system of photons, electrons, nucleons, scalar and
pseudoscalar mesons within QED and L$\sigma$M, taken in the chirally
broken phase, we describe by the Lagrangian
\begin{eqnarray}\label{eq:20}
\hspace{-0.3in}&&{\cal L}^{(0)}_{\rm QED + L\sigma M}(x) = -
\frac{1}{4}\,F^{(0)}_{\mu\nu} F^{(0)\mu\nu} -
\frac{1}{2\xi_0}\,\big(\partial_{\mu}A^{(0)\mu} \big)^2 +
\bar{\psi}^{(0)}_e \big(i\gamma^{\mu}\partial_{\mu} -
m^{(0)}_e\big)\psi^{(0)}_e - (- e_0)\, \bar{\psi}^{(0)}_e
\gamma^{\mu}\psi^{(0)}_e\, A^{(0)}_{\mu} \nonumber\\
\hspace{-0.3in}&& + \bar{\psi}^{(0)}_p \big(i\gamma^{\mu}\partial_{\mu}
- m^{(0)}_N\big)\psi^{(0)}_p - (+ e_0)\, \bar{\psi}^{(0)}_p
\gamma^{\mu}\psi^{(0)}_p\, A^{(0)}_{\mu} + \big(\partial_{\mu} + i (+
e_0)\, A^{(0)}_{\mu}\big) \pi^{(0)+}\big(\partial^{\mu} + i (- e_0)\,
A^{(0)\mu}\big) \pi^{(0)-}\nonumber\\
\hspace{-0.3in}&& - m^{(0)2}_{\pi}\pi^{(0)+}\pi^{(0)-} +
\bar{\psi}^{(0)}_n\big(i\gamma^{\mu}\partial_{\mu} -
m^{(0)}_N\big)\psi^{(0)}_n +
\frac{1}{2}\,\big(\partial_{\mu}\sigma^{(0)}
\partial^{\mu}\sigma^{(0)} - m^{(0)2}_{\sigma} \sigma^{(0)2}\big)\nonumber\\
\hspace{-0.3in}&& + \frac{1}{2}\,\big(\partial_{\mu}\pi^{(0)0}
\partial^{\mu}\pi^{(0)0} - m^{(0)2}_{\pi}(\pi^{(0)
  0})^2\big) + g^{(0)}_{\pi
  N}\,\big(\bar{\psi}^{(0)}_p\psi^{(0)}_p +
\bar{\psi}^{(0)}_n\psi^{(0)}_n\big)\,\sigma^{(0)} + g^{(0)}_{\pi
  N}\,\big(\bar{\psi}^{(0)}_p i\gamma^5\psi^{(0)}_p -
\bar{\psi}^{(0)}_n i\gamma^5\psi^{(0)}_n\big)\,\pi^{(0)0}\nonumber\\
\hspace{-0.3in}&&+ \sqrt{2}\, g^{(0)}_{\pi N}\,\bar{\psi}^{(0)}_p i
\gamma^5\psi^{(0)}_n\, \pi^{(0)+} + \sqrt{2}\,g^{(0)}_{\pi
  N}\,\bar{\psi}^{(0)}_n i \gamma^5\psi^{(0)}_p\,\pi^{(0)-} +
\gamma^{(0)} f^{(0)}_{\pi}\,\sigma^{(0)} \big(\sigma^{(0) 2} +
2\,\pi^{(0)+}\pi^{(0)-} + (\pi^{(0) 0})^2\big)\nonumber\\
\hspace{-0.3in}&& - \frac{\gamma^{(0)}}{4}\,\big(\sigma^{(0) 2} +
2\,\pi^{(0)+}\pi^{(0)-} + (\pi^{(0) 0})^2\big)^2,
\end{eqnarray}
where all fields, their masses and coupling constants are {\it
  bare}. The vector and axial--vector hadronic currents are equal to
\begin{eqnarray}\label{eq:21}
V^{(0)+}_{\mu} &=& \bar{\psi}^{(0)}_p \gamma_{\mu} \psi^{(0)}_n +
i\,\sqrt{2}\,\Big(\pi^{(0)0}\big(\partial_{\mu} + i (- e_0)\,
A^{(0)}_{\mu}\big) \pi^{(0)-} -
\pi^{(0)-}\partial_{\mu}\pi^{(0)0}\Big),\nonumber\\ V^{(0)-}_{\mu} &=&
\bar{\psi}^{(0)}_n \gamma_{\mu} \psi^{(0)}_p -
i\,\sqrt{2}\,\Big(\pi^{(0)0}\big(\partial_{\mu} + i (+ e_0)\,
A^{(0)}_{\mu}\big) \pi^{(0)+} -
\pi^{(0)+}\partial_{\mu}\pi^{(0)0}\Big),\nonumber\\
\hspace{-0.3in}V^{(0)0}_{\mu} &=& \frac{1}{2}\,\big(\bar{\psi}^{(0)}_p
\gamma_{\mu}\psi^{(0)}_p - \bar{\psi}^{(0)}_n \gamma_{\mu}\psi^{(0)}_n
\big) + i\Big(\pi^{(0)-}\big(\partial_{\mu} + i (+ e_0)\,
A^{(0)}_{\mu}\big) \pi^{(0)+} - \pi^{(0)+}\big(\partial_{\mu} + i (-
e_0)\, A^{(0)}_{\mu}\big) \pi^{(0)-}\Big),\nonumber\\ A^{(0)+}_{\mu}
&=& \bar{\psi}^{(0)}_p \gamma_{\mu} \gamma^5\psi^{(0)}_n +
\sqrt{2}\,\Big(\sigma^{(0)}\big(\partial_{\mu} + i (- e_0)\,
A^{(0)}_{\mu}\big) \pi^{(0)-} -
\pi^{(0)-}\partial_{\mu}\sigma^{(0)}\Big) -
\sqrt{2}\,f^{(0)}_{\pi}\big(\partial_{\mu} + i (- e_0)\,
A^{(0)}_{\mu}\big) \pi^{(0)-} ,\nonumber\\ A^{(0)-}_{\mu}
&=&\bar{\psi}^{(0)}_n \gamma_{\mu} \gamma^5\psi^{(0)}_p +
\sqrt{2}\,\Big(\sigma^{(0)}\big(\partial_{\mu} + i (+ e_0)\,
A^{(0)}_{\mu}\big) \pi^{(0)+} -
\pi^{(0)+}\partial_{\mu}\sigma^{(0)}\Big) -
\sqrt{2}\,f^{(0)}_{\pi}\big(\partial_{\mu} + i (+ e_0)\,
A^{(0)}_{\mu}\big) \pi^{(0)+} ,\nonumber\\
\hspace{-0.3in}A^{(0)0}_{\mu} &=& \frac{1}{2}\,\big(\bar{\psi}^{(0)}_p
\gamma_{\mu}\psi^{(0)}_p - \bar{\psi}^{(0)}_n \gamma_{\mu}\psi^{(0)}_n
\big) + \Big(\sigma^{(0)} \partial_{\mu} \pi^{(0)0} - \pi^{(0)0}
\partial_{\mu}\sigma^{(0)}\Big) - f^{(0)}_{\pi}\partial_{\mu}
\pi^{(0)0}.
\end{eqnarray}
After the calculation of loop corrections the dynamics of the fields
is described by the Lagrangian
\begin{eqnarray}\label{eq:22}
\hspace{-0.3in}&&{\cal L}_{\rm QED + L\sigma M}(x) = -
\frac{1}{4}\,F_{\mu\nu} F^{\mu\nu} -
\frac{1}{2\xi}\,\big(\partial_{\mu}A^{\mu} \big)^2 + \bar{\psi}_e
\big(i\gamma^{\mu}\partial_{\mu} - m_e\big)\psi_e - (- e)\,
\bar{\psi}_e \gamma^{\mu}\psi_e\, A_{\mu} \nonumber\\
\hspace{-0.3in}&& + \bar{\psi}_p \big(i\gamma^{\mu}\partial_{\mu}
- m_p\big)\psi_p - (+ e)\, \bar{\psi}_p
\gamma^{\mu}\psi_p\, A_{\mu} + \big(\partial_{\mu} + i (+
e)\, A_{\mu}\big) \pi^+\big(\partial^{\mu} + i (- e)\,
A^{\mu}\big) \pi^- - m^2_{\pi}\pi^+ \pi^- \nonumber\\
\hspace{-0.3in}&& + \bar{\psi}_n\big(i\gamma^{\mu}\partial_{\mu} -
m_n\big)\psi_n + \frac{1}{2}\,\big(\partial_{\mu}\sigma
\partial^{\mu}\sigma - m^2_{\sigma} \sigma^2\big) +
\frac{1}{2}\,\big(\partial_{\mu}\pi^0 \partial^{\mu}\pi^0 -
m^2_{\pi^0}(\pi^0)^2\big)\nonumber\\
\hspace{-0.3in}&& + g_{\pi N}\,\big( \bar{\psi}_p\psi_p +
\bar{\psi}_n\psi_n\big)\,\sigma + g_{\pi N}\,\big(\bar{\psi}_p
i\gamma^5\psi_p - \bar{\psi}_n i\gamma^5\psi_n\big)\,\pi^0
+ \sqrt{2}\, g_{\pi N}\,\bar{\psi}_p i \gamma^5\psi_n\, \pi^+
\nonumber\\
\hspace{-0.3in}&& + \sqrt{2}\,g_{\pi N}\,\bar{\psi}_n i
\gamma^5\psi_p\,\pi^- + \gamma f_{\pi}\,\sigma \big(\sigma^2 +
(\pi^0)^2\big) + 2\, \gamma f_{\pi} \sigma \,\pi^+ \pi^- -
\frac{\gamma}{4}\,\big(\sigma^2 + (\pi^0)^2\big)^2 \nonumber\\
\hspace{-0.3in}&&- \gamma  \big(\sigma^2 + (\pi^0)^2\big)\,\pi^+
\pi^- - \gamma \,(\pi^+ \pi^-)^2 + {\cal L}^{(\rm CT)}_{\rm QED +
  L\sigma M}(x),
\end{eqnarray}
where all fields, their masses and coupling constants are renormalized
or {\it physical}. The Lagrangian ${\cal L}^{(\rm CT)}_{\rm QED +
  L\sigma M}(x)$ contains the complete set of counter--terms
\begin{eqnarray}\label{eq:23}
\hspace{-0.3in}&&{\cal L}^{(\rm CT)}_{\rm QED + L\sigma M} = - (Z_3 -
1)\,\frac{1}{4}\,F_{\mu\nu} F^{\mu\nu} - \frac{Z_3 -
  1}{Z_{\xi}}\,\frac{1}{2\xi}\,\big(\partial_{\mu} A^{\mu} \big)^2\nonumber\\
\hspace{-0.3in}&& + (Z^{(e)}_2 - 1)\bar{\psi}_e
\big(i\gamma^{\mu}\partial_{\mu} - m_e\big)\psi_e - (Z^{(e)}_1 -
1)\,(- e)\, \bar{\psi}_e \gamma^{\mu}\psi_e\, A_{\mu} - Z^{(e)}_2
\delta m_e \bar{\psi}_e\psi_e\nonumber\\
\hspace{-0.3in}&& + (Z_N Z^{(p)}_2 - 1)\,\bar{\psi}_p
\big(i\gamma^{\mu}\partial_{\mu} - m_p\big)\psi_p - (Z_N Z^{(p)}_1 -
1)\,(+ e)\, \bar{\psi}_p \gamma^{\mu}\psi_p\, A_{\mu} - Z_N Z^{(p)}_2
\delta m_p \bar{\psi}_p\psi_p\nonumber\\
\hspace{-0.3in}&& + (Z_M Z^{(\pi)}_2 - 1)\,\big(\partial_{\mu}
\pi^+ \partial^{\mu}\pi^- - m^2_{\pi}\pi^+ \pi^-\big) + (Z_M
Z^{(\pi)}_1 - 1)\,i\,e\,\big(\pi^+\partial^{\mu}\pi^- -
\partial^{\mu}\pi^+\,\pi^-\big)\,A_{\mu}\nonumber\\
\hspace{-0.3in}&& + (Z_M Z^{(\pi)}_4 -
1)\,e^2\,\pi^+\,\pi^-\,A_{\mu}A^{\mu} - Z_M Z^{(\pi)}_2\,\delta
m^2_{\pi} \pi^+\,\pi^- + (Z_N - 1)\,\bar{\psi}_n
\big(i\gamma^{\mu}\partial_{\mu} - m_n\big)\psi_n - Z_N \delta m_n
\bar{\psi}_n\psi_n\nonumber\\
\hspace{-0.3in}&& + \big(Z_M -
1\big)\,\frac{1}{2}\,\big(\partial_{\mu}\sigma \partial^{\mu}\sigma -
m^2_{\sigma} \sigma^2\big) - Z_M\delta m^2_{\sigma}\sigma^2 + \big(Z_M
- 1\big)\, \frac{1}{2}\,\big(\partial_{\mu}\pi^0 \partial^{\mu}\pi^0 -
m^2_{\pi^0}(\pi^0)^2\big) - Z_M\delta
m^2_{\pi^0}(\pi^0)^2\nonumber\\\hspace{-0.3in}&&+ (Z_{M N} Z^{(p)}_2 -
1)\,g_{\pi N}\,\bar{\psi}_p \psi_p\,\sigma + (Z_{M N} - 1)\,g_{\pi
  N}\,\bar{\psi}_n \psi_n\,\sigma\nonumber\\\hspace{-0.3in}&&+ (Z_{M
  N}Z^{(p)}_2 - 1)\,\frac{1}{2}\,g_{\pi N}\,\bar{\psi}_pi\gamma^5
\psi_p\,\pi^0 - (Z_{M N} - 1)\,\frac{1}{2}\,g_{\pi N}\,\bar{\psi}_n
i\gamma^5 \psi_n\,\pi^0\nonumber\\
\hspace{-0.3in}&&+ (Z_{M N} Z^{(p)1/2}_2 Z^{(\pi)1/2}_2 -
1)\,g_{\pi N}\,\bar{\psi}_pi\gamma^5 \psi_n\,\pi^+ + (Z_{M N}
Z^{(p)1/2}_2 Z^{(\pi)1/2}_2 - 1)\,g_{\pi N}\,\bar{\psi}_n i \gamma^5
\psi_p\,\pi^-\nonumber\\
\hspace{-0.3in}&&+ \big(Z_{3 M} -
1\big)\,\gamma\,f_{\pi}\,\sigma\, (\sigma^2 + (\pi^0)^2) +
\big(Z_{3 M} Z^{(\pi)}_2 - 1\big)\,2\,
\gamma\,f_{\pi}\,\sigma\, \pi^+\,\pi^- - \big(Z_{4 M} -
1\big)\,\frac{\gamma}{4}\,\big(\sigma^2 + (\pi^0)^2\big)^2\nonumber\\
\hspace{-0.3in}&& - \big(Z_{4 M} Z^{(\pi)}_2 -
1\big)\,\gamma\,\sigma^2\,\pi^+ \pi^- - \big(Z_{4 M}Z^{(\pi)2}_2 -
1\big)\,\gamma\,(\pi^+ \pi^-)^2.
\end{eqnarray}
Rescaling the field operators 
\begin{eqnarray}\label{eq:24}
\sqrt{Z_3}\, A_{\mu} &=& A^{(0)}_{\mu}\;,\; \sqrt{Z^{(e)}_2}\,\psi_e =
\psi^{(0)}_e \;,\; \sqrt{Z_N Z^{(p)}_2}\,\psi_p = \psi^{(0)}_p
\;,\;\sqrt{Z_N}\,\psi_n = \psi^{(0)}_n,\nonumber\\
\hspace{-0.3in}\sqrt{Z_M Z^{(\pi)}_2}\,\pi^{\mp} &=&
\pi^{(0)\mp}\;,\;\sqrt{Z_M}\,\sigma = \sigma^{(0)}\;,\;
\sqrt{Z_M}\,\pi^0 = \pi^{(0)0}
\end{eqnarray}
and plugging Eq.(\ref{eq:24}) into Eq.(\ref{eq:22}) we arrive at the
Lagrangian
\begin{eqnarray}\label{eq:25}
\hspace{-0.3in}&&{\cal L}_{\rm QED + L\sigma M} = -
\frac{1}{4}\,F^{(0)}_{\mu\nu} F^{(0)\mu\nu} - \frac{1}{2
  Z_{\xi}\xi}\,\big(\partial_{\mu}A^{(0)\mu} \big)^2 +
\bar{\psi}^{(0)}_e \big(i\gamma^{\mu}\partial_{\mu} - m^{(0)}_e - (-
e)\,Z^{(e)}_1 Z^{(e)-1}_2 Z^{1/2}_3\gamma^{\mu}A^{(0)}_{\mu}
\big)\psi_e \nonumber\\
\hspace{-0.3in}&& + \bar{\psi}^{(0)}_p
\big(i\gamma^{\mu}\partial_{\mu} - m^{(0)}_N - (+ e)\,Z^{(p)}_1
Z^{(p)-1}_2 Z^{1/2}_3 A^{(0)}_{\mu}\big)\psi^{(0)}_p\nonumber\\
\hspace{-0.3in}&& + \big(\partial_{\mu} + i (+ e)\,Z^{(\pi)}_1
Z^{(\pi)-1}_2 Z^{1/2}_3 A^{(0)}_{\mu}\big) \pi^{(0)+}\big(\partial^{\mu} +
i (- e)\, Z^{(\pi)}_1 Z^{(\pi)-1}_2 Z^{1/2}_3 A^{(0)\mu}\big)
\pi^{(0)-} - m^{(0)2}_{\pi}\pi^{(0)+} \pi^{(0)-} \nonumber\\
\hspace{-0.3in}&& + \bar{\psi}^{(0)}_n\big(i\gamma^{\mu}\partial_{\mu}
- m^{(0)}_N\big)\psi^{(0)}_n + \frac{1}{2}\,\big(\partial_{\mu}\sigma^{(0)}
\partial^{\mu}\sigma^{(0)} - m^{(0)2}_{\sigma} \sigma^{(0)2}\big) +
\frac{1}{2}\,\big(\partial_{\mu}\pi^{(0)0} \partial^{\mu}\pi^{(0)0} -
m^{(0)2}_{\pi}(\pi^{(0)0})^2\big)\nonumber\\
\hspace{-0.3in}&& + g_{\pi N}\,Z_{M N} Z^{-1/2}_M
Z^{-1}_N\,\bar{\psi}^{(0)}_p\psi^{(0)}_p\,\sigma^{(0)} + g_{\pi
  N}\,Z_{M N} Z^{-1/2}_M
\bar{\psi}^{(0)}_n\psi^{(0)}_n\,\sigma^{(0)} + g_{\pi N}\,Z_{ M
  N} Z^{-1}_N Z^{-1/2}_M \,\bar{\psi}^{(0)}_p
i\gamma^5\psi^{(0)}_p\,\pi^{(0)0}\nonumber\\
\hspace{-0.3in}&& - g_{\pi N}\,Z_{M N} Z^{-1}_N
Z^{-1/2}_M\,\bar{\psi}^{(0)}_n i\gamma^5\psi^{(0)}_n\,\pi^{(0)0} +
\sqrt{2}\, g_{\pi N}\,Z_{M N}
Z^{-1}_NZ^{-1/2}_M\,\bar{\psi}^{(0)}_p i \gamma^5\psi^{(0)}_n\,
\pi^{(0)+}\nonumber\\
\hspace{-0.3in}&& + \sqrt{2}\,g_{\pi N}\,Z_{M N}
Z^{-1}_NZ^{-1/2}_M\,\bar{\psi}^{(0)}_n i \gamma^5
\psi^{(0)}_p\,\pi^{(0)-} + \gamma f_{\pi}\,Z_{3 M} Z^{-3/2}_M
\sigma^{(0)}\big((\sigma^{(0)})^2 + (\pi^{(0)0})^2\big) \nonumber\\
\hspace{-0.3in}&& + 2\, \gamma f_{\pi}\,Z_{3 M}Z^{-3/2}_M\,
\sigma^{(0)} \,\pi^{(0)+} \pi^{(0)-} - \frac{\gamma}{4}\,Z_{4 M}
Z^{-2}_M\big( (\sigma^{(0)})^2 + (\pi^{(0)0})^2\big)^2 \nonumber\\
\hspace{-0.3in}&&- \gamma Z_{4M} Z^{-2}_M\,\big((\sigma^{(0)})^2 +
(\pi^{(0)0})^2\big)\,\pi^{(0)+} \pi^{(0)-0} - \gamma\,Z_{4 M} Z^{-2}_M
(\pi^{(0)+} \pi^{(0)-})^2,
\end{eqnarray}
where we have used the relations
\begin{eqnarray}\label{eq:26}
\hspace{-0.3in}m^{(0)}_N = m_p + \delta m_p = m_n + \delta m_n\;,\;
m^{(0)2}_{\sigma} = m^2_{\sigma} + \delta m^2_{\sigma}\;,\;
m^{(0)2}_{\pi} = m^2_{\pi} + \delta m^2_{\pi} = m^2_{\pi^0} + \delta
m^2_{\pi^0}.
\end{eqnarray}
The Lagrangian Eq.(\ref{eq:25}) reduces to the Lagrangian
Eq.(\ref{eq:20}) if the coupling constants and renormalization
constants satisfy the relations
\begin{eqnarray}\label{eq:27}
\hspace{-0.3in} (- e_0) &=& (- e)\,Z^{(e)}_1 Z^{(e)-1}_2Z^{-1/2}_3 = (-
e)\,Z^{(\pi)}_1 Z^{(\pi)-1}_2Z^{-1/2}_3, \nonumber\\
\hspace{-0.3in} (+ e_0) &=& (+ e)\,Z^{(p)}_1 Z^{(p)-1}_2Z^{-1/2}_3 = (+
e)\,Z^{(\pi)}_1 Z^{(\pi)-1}_2Z^{-1/2}_3, \nonumber\\
\hspace{-0.3in}g^{(0)}_{\pi N} &=& g_{\pi N}\,Z_{M N} Z^{-1/2}_M
Z^{-1}_N, \nonumber\\
\hspace{-0.3in} f^{(0)}_{\pi} &=& Z_{3 M} Z^{-1}_{4 M} Z^{1/2}_M
f_{\pi}\;,\; \gamma^{(0)} = Z_{4 M}Z^{-2}_M \gamma,\nonumber\\ Z_{3 M}
&=& Z_{4 M}.
\end{eqnarray}
Because of the Ward identities $Z^{(a)}_1 = Z^{(a)}_2$ for $a = e,p$
\cite{Ward1950} (see also
\cite{Itzykson1980,Weinberg1995,Bogoliubov1959}) and $a = \pi$
\cite{Salam1952,Matthews1954,Fry1973} we get $(\mp e_0) =
Z^{-1/3}\,(\mp e)$. We would like to emphasize that for the
calculation of radiative corrections to order $O(\alpha/\pi)$ to the
neutron $\beta^-$--decay the renormalization constant $Z_3$ is equal
to unity because of the absence of closed fermion and meson loops,
i.e.  $Z_3 = 1$.  This means that in such an approximation the {\it
  bare} electric charge $e_0$ coincides with the renormalized electric
charge $e$, i.e. $e_0 = e$. This is because of local conservation of
leptonic and hadronic electromagnetic currents.

Now we may proceed to the analysis of the properties of hadronic
structure of the neutron and proton in neutron $\beta^-$ decays within
the standard $V - A$ effective theory of weak interactions, where
electromagnetic and strong low--energy interactions are described the
Lagrangian Eq.(\ref{eq:22}).

\section{Neutron beta decays in standard $V - A$ effective theory with 
QED and L$\sigma$M}
\label{sec:weak}

The neutron $\beta^-$--decay $n \to p + e^- + \bar{\nu}_e$ we describe
within the standard $V - A$ effective theory of weak interactions by
the effective Lagrangian \cite{Feynman1958,Nambu1960}
\begin{eqnarray}\label{eq:28}
\hspace{-0.3in}{\cal L}_W(x) = - G_V J^{+}_{\mu}(x)[\bar{\psi}_e(x)
  \gamma^{\mu} (1 - \gamma^5)\psi_{\nu_e}(x)],
\end{eqnarray}
where $G_V = G_FV_{ud}/\sqrt{2}$ is the vector weak coupling constant,
and $G_F$ and $V_{ud}$ are the Fermi weak coupling constant and the
matrix element of the Cabibbo--Kobayashi--Maskawa (CKM) mixing matrix
\cite{PDG2016}, respectively. The hadronic current $J^{+}_{\mu}(x)$ is
defined by $J^{+}_{\mu}(x) = V^{+}_{\mu}(x) - A^{+}_{\mu}(x)$, where
the $V^{+}_{\mu}(x)$ and $A^{+}_{\mu}(x)$ are the charged vector and
axial--vector hadronic currents. In the phase of spontaneously broken
chiral symmetry of the L$\sigma$M these currents are given by
\begin{eqnarray}\label{eq:29}
\hspace{-0.3in}&&V^{+}_{\mu}(x) = \bar{\psi}_p(x) \gamma_{\mu}
\psi_n(x) + i\,\sqrt{2}\,\Big(\pi^0(x)\big(\partial_{\mu} + i (-
e)\, A_{\mu}(x)\big) \pi^-(x) - \pi^-(x)\partial_{\mu}\pi^0
(x)\Big) +  (Z_V - 1)\, \bar{\psi}_p(x)
\gamma_{\mu} \psi_n(x),\nonumber\\ \hspace{-0.3in}&&A^{+}_{\mu}(x) =
\bar{\psi}_p (x)\gamma_{\mu} \gamma^5\psi_n(x) +
\sqrt{2}\,\Big(\sigma(x)\big(\partial_{\mu} + i (- e)\,
A_{\mu}(x)\big) \pi^-(x) -
\pi^-(x)\partial_{\mu}\sigma(x)\Big)\nonumber\\ \hspace{-0.3in}&&-
\sqrt{2}\,f_{\pi}\big(\partial_{\mu} + i (- e)\, A_{\mu}\big) \pi^-(x)
+ (Z_A - 1)\, \bar{\psi}_p (x)\gamma_{\mu} \gamma^5\psi_n(x) -
(Z^{(\pi)}_A - 1)\,\sqrt{2}\,f_{\pi}\big(\partial_{\mu} + i (- e)\,
A_{\mu}\big) \pi^-(x),
\end{eqnarray}
where $\psi_e(x)$, $\psi_p(x)$, $\psi_n(x)$, $\sigma(x)$, $\pi^0(x)$,
$\pi^-(x)$ and $A_{\mu}(x)$ are operators of the electron, proton,
neutron, $\sigma$--meson, $\pi^0$--meson, $\pi^-$--meson and photon
fields; $\psi_{\nu_e}(x)$ is the operator of the electron neutrino
(antineutrino) field. Then, $Z_V$, $Z_A$ and $Z^{(\pi)}_A$ are the
counter--terms of the vector and axial--vector baryonic and
axial--vector mesonic currents, respectively, which are enough to
remove divergent contributions in the one--hadron--loop approximation.
The amplitude of the neutron $\beta^-$--decay is defined by
\cite{Ivanov2018}
\begin{eqnarray}\label{eq:30}
M(n \to p e^-\bar{\nu}_e) = \big\langle {\rm out},
\bar{\nu}_e(\vec{k}_{\nu},+ \frac{1}{2}), e^-(\vec{k}_e, \sigma_e),
p(\vec{k}_p,\sigma_p) \big|{\cal L}_W(0)\big| n(\vec{k}_n,
\sigma_n), {\rm in}\big\rangle,
\end{eqnarray}
where $\langle {\rm out}, \chi(\vec{k}_{\chi},\sigma_{\chi})|$ and
$|{\rm in}, n(\vec{k}_n, \sigma_n)\rangle$ are the wave functions of
the free antineutrino, electron and proton ($\chi = \bar{\nu}_e, e^-,
p$) in the final state (i.e. out--state at $t \to + \infty$) and the
free neutron in the initial state (i.e. in--state at $t \to - \infty$)
\cite{Itzykson1980}. Using the relation $\langle {\rm out}, 
\prod_{\chi} \chi(\vec{k}_{\chi},\sigma_{\chi})| = \langle {\rm in},
\prod_{\chi} \chi(\vec{k}_{\chi},\sigma_{\chi})|{\mathbb S}$, where
${\mathbb S}$ is the S--matrix, we rewrite the matrix element
Eq.(\ref{eq:30}) as follows
\begin{eqnarray}\label{eq:31}
M(n \to p e^-\bar{\nu}_e) = \big\langle {\rm in},
\bar{\nu}_e(\vec{k}_{\nu},+ \frac{1}{2}), e^-(\vec{k}_e, \sigma_e),
p(\vec{k}_p,\sigma_p) \big|{\mathbb S}\,{\cal L}_W(0)\big|
n(\vec{k}_n, \sigma_n), {\rm in}\big\rangle.
\end{eqnarray}
 The corresponding S--matrix is determined by
 \cite{Itzykson1980,Ivanov2018}
\begin{eqnarray}\label{eq:32}
{\mathbb S} = {\rm T}e^{\textstyle i\int d^4x\,{\cal L}_{\rm QED +
    L\sigma M}(x)},
\end{eqnarray}
where ${\rm T}$ is a time--ordering operator and ${\cal L}_{\rm QED +
  L\sigma M}(x)$ is given by Eq.(\ref{eq:22}). Plugging
Eq.(\ref{eq:32}) into Eq.(\ref{eq:31}) we get \cite{Ivanov2018}
\begin{eqnarray}\label{eq:33}
M(n \to p e^-\bar{\nu}_e) = \big\langle {\rm in},
\bar{\nu}_e(\vec{k}_{\nu},+ \frac{1}{2}), e^-(\vec{k}_e, \sigma_e),
p(\vec{k}_p,\sigma_p) \big|{\rm T}\big(e^{\textstyle i\int d^4x\,{\cal
    L}_{\rm QED + L\sigma M}(x)} {\cal L}_W(0)\big)\big| n(\vec{k}_n,
\sigma_n), {\rm in} \big\rangle.
\end{eqnarray}
The wave functions of fermions we determine in terms of the operators
of creation (annihilation)
 \begin{eqnarray}\label{eq:34}
|n(\vec{k}_n, \sigma_n), {\rm in}\rangle &=& a^{\dagger}_{n,\rm
  in}(\vec{k}_n, \sigma_n)|0\rangle,\nonumber\\ \big\langle {\rm in},
\bar{\nu}_e(\vec{k}_{\nu},+ \frac{1}{2}), e^-(\vec{k}_e, \sigma_e),
p(\vec{k}_p,\sigma_p) &=& \langle 0|b_{\bar{\nu}_e,\rm in}(\vec{k}_{\nu},+
\frac{1}{2})a_{e,\rm in} (\vec{k}_e, \sigma_e) a_{p,\rm
  in}(\vec{k}_p,\sigma_p).
\end{eqnarray}
The operators of creation (annihilation) obey standard anticommutation
relations \cite{Itzykson1980,Ivanov2018}. 

The amplitude of the neutron radiative $\beta^-$--decay $n \to p + e^-
+ \bar{\nu}_e + \gamma$ can be defined in analogous way
\begin{eqnarray}\label{eq:35}
\hspace{-0.15in}M(n \to p e^-\bar{\nu}_e\gamma)_{\lambda} =
\big\langle {\rm in},
\gamma(\vec{k},\lambda),\bar{\nu}_e(\vec{k}_{\nu},+ \frac{1}{2}),
e^-(\vec{k}_e, \sigma_e), p(\vec{k}_p,\sigma_p) \big|{\rm
  T}\big(e^{\textstyle i\int d^4x\,{\cal L}_{\rm QED + L\sigma M}(x)}
{\cal L}_W(0)\big)\big| n(\vec{k}_n, \sigma_n), {\rm
  in}\big\rangle,
\end{eqnarray}
where $\lambda = 1,2$ characterizes physical polarization states of
the photon \cite{Ivanov2013, Ivanov2017a, Gaponov1996, Bernard2004,
  Gardner2012, Gardner2013, Ivanov2017,Ivanov2017b}.

\section{Neutron beta decay in the tree-- and one--hadron--loop  approximation 
for strong low--energy interactions in L$\sigma$M}
\label{sec:strong}

In this section we switch off electromagnetic interactions and analyse
the contributions of strong low--energy interactions, described by the
L$\sigma$M, to the amplitude of the neutron $\beta^-$--decay $n \to p
+ e^- + \bar{\nu}_e$. The amplitude of the neutron $\beta^-$--decay is
defined by \cite{Ivanov2018}
\begin{eqnarray}\label{eq:36}
\hspace{-0.15in}M(n \to p e^-\bar{\nu}_e) = - G_V\langle
p(\vec{k}_p,\sigma_p)|J^+_{\mu}(0)|n(\vec{k}_n, \sigma_n)\rangle\,
\Big[\bar{u}_e\big(\vec{k}_e, \sigma_e\big) \gamma^{\mu}\big(1 -
  \gamma^5\big) v_{\nu}\big(\vec{k}_{\nu}, + \frac{1}{2}\big)\Big],
\end{eqnarray}
where $\bar{u}_e$ and $v_{\nu}$ are Dirac wave functions of the free
electron and electron antineutrino, respectively, a momentum
transferred of the decay is equal to $q = k_p - k_n = - k_e -
k_{\nu}$. Then, since strong low--energy interactions give the
contributions to the matrix element of the charged hadronic current
only, we have denoted $\langle {\rm in}, p(\vec{k}_p,\sigma_p)|{\rm
  T}\big(e^{\textstyle i\int d^4x\,{\cal L}_{\rm L\sigma
    M}(x)}J^+_{\mu}(0)\big)|n(\vec{k}_n, \sigma_n),{\rm in}\rangle =
\langle p(\vec{k}_p,\sigma_p)|J^+_{\mu}(0)|n(\vec{k}_n,
\sigma_n)\rangle$. This matrix element describes the hadronic $n \to
p$ transition in the neutron $\beta^-$--decay
\cite{Leitner2006,Ivanov2018}.

\subsection{Neutron beta decay in the tree--approximation 
for strong low--energy interactions in L$\sigma$M}

\begin{figure}
\centering \includegraphics[height=0.12\textheight]{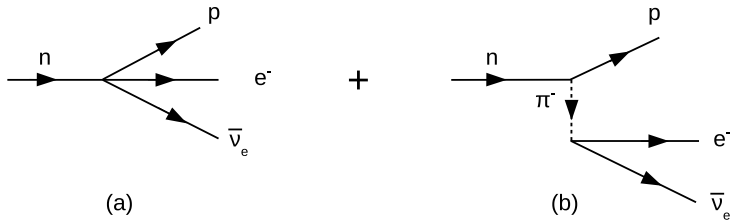}
  \caption{The Feynman diagrams, describing the amplitude of the
    neutron $\beta^-$--decay in the tree--approximation for strong
    low--energy interactions in the L$\sigma$M.}
\label{fig:fig1}
\end{figure}

In the tree--approximation for strong low-energy interactions in
the L$\sigma$M the amplitude of the neutron $\beta^-$--decay is
described by the Feynman diagrams in Fig.\,\ref{fig:fig1}.
The matrix element of the hadronic $V - A$ current $\langle
p(\vec{k}_p,\sigma_p)|J^+_{\mu}(0)|n(\vec{k}_n, \sigma_n)\rangle$,
calculated in the tree--approximation (see Fig.\,\ref{fig:fig1}), is
equal to (see also \cite{Ivanov2018})
\begin{eqnarray}\label{eq:37}
\langle p(\vec{k}_p,\sigma_p)|J^+_{\mu}(0)|n(\vec{k}_n,
\sigma_n)\rangle_{\rm Fig. \ref{fig:fig1}} = \bar{u}_p\big(\vec{k}_p,
\sigma_p\big)\Big(\gamma_{\mu}\big(1 - \gamma^5\big) - \frac{2\,g_{\pi
    N}\,f_{\pi}}{m^2_{\pi} - q^2}\,
q_{\mu}\,\gamma^5\Big)\,u_n\big(\vec{k}_n, \sigma_n\big),
\end{eqnarray}
where $\bar{u}_p$ and $u_n$ are the Dirac wave functions of the free
proton and neutron. Since in the limit $m_{\pi} \to 0$ the charged
axial--vector hadronic current $A^+_{\mu}$ is locally conserved
$\partial^{\mu}A^+_{\mu} = 0$ \cite{Feynman1958,Nambu1960}, we get
\begin{eqnarray}\label{eq:38}
\lim_{m_{\pi} \to\, 0}q^{\mu} \langle
p(\vec{k}_p,\sigma_p)|J^+_{\mu}(0)|n(\vec{k}_n, \sigma_n)\rangle_{\rm
  Fig. \ref{fig:fig1}} = \big(- 2m_N + 2g_{\pi N}
f_{\pi}\big)\,\bar{u}_p\big(\vec{k}_p, \sigma_p\big)\,\gamma^5\,
u_n\big(\vec{k}_n, \sigma_n\big) = 0,
\end{eqnarray}
where we have used local conservation of the charged vector hadronic
current $\partial^{\mu} V^+_{\mu} = 0$ \cite{Feynman1958} (see also
\cite{Ivanov2018,Leitner2006}) and the Dirac equations
$\bar{u}_p\hat{k}_p = \bar{u}_p m_N$ and $\hat{k}_n u_n = m_N u_n$ for
the free proton and neutron. Thus, from Eq.(\ref{eq:38}) we obtain the
well--known Goldberger--Treiman (GT) relation $g_{\pi N} =
m_N/f_{\pi}$ \cite{Goldberger1958} (see also
\cite{GellMann1960,Bernstein1960,Nambu1960,Strubbe1972}), which
appears naturally in the L$\sigma$M (see Eq.(\ref{eq:8}) at $b = -
f_{\pi}$), where the axial coupling constant $g_A$ is equal to $g_A =
1$. This is caused by the account for the strong low--energy
interactions in the tree--approximation. Plugging the GT--relation
$g_{\pi N} = m_N/f_{\pi}$ into Eq.(\ref{eq:37}) we arrive at the
matrix element of the charged $V - A$ hadronic current, calculated in
the tree--approximation in the L$\sigma$M \cite{Ivanov2018}
\begin{eqnarray}\label{eq:39}
\langle p(\vec{k}_p,\sigma_p)|J^+_{\mu}(0)|n(\vec{k}_n,
\sigma_n)\rangle_{\rm Fig.\,\ref{fig:fig1}} = \bar{u}_p\big(\vec{k}_p,
  \sigma_p\big)\Big(\gamma_{\mu}\big(1 - \gamma^5\big) - \frac{2\,m_N
  }{m^2_{\pi} - q^2}\, q_{\mu}\, \gamma^5\Big)\,u_n\big(\vec{k}_n,
  \sigma_n\big).
\end{eqnarray}
The matrix element of the charged hadronic current Eq.(\ref{eq:39})
has the standard Lorentz structure with the vector, axial--vector and
pseudoscalar form factors equal to unity
\cite{Nambu1960,Marshak1969,Leitner2006} (see also \cite{Ivanov2018}).

\subsection{Neutron beta decay in one--hadron--loop  approximation 
for strong low--energy interactions in L$\sigma$M}

\begin{figure}
\centering 
\includegraphics[height=0.235\textheight]{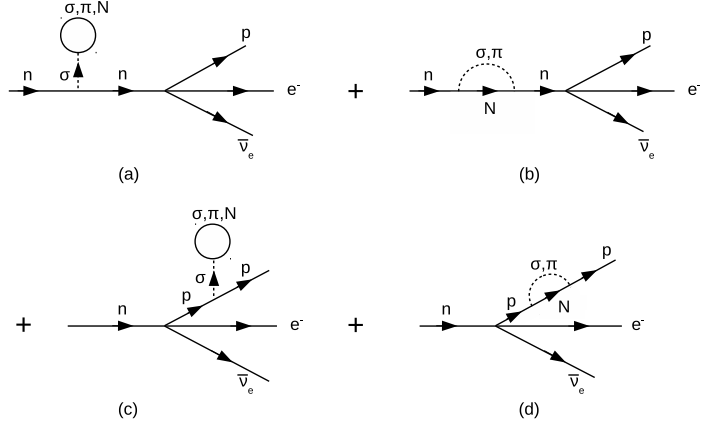}
  \caption{The Feynman diagrams, describing the contributions to the
    amplitude of the neutron $\beta^-$--decay of the self--energy
    corrections to the neutron and proton states in the
    one--hadron--loop approximation in the L$\sigma$M.}
\label{fig:fig2}
\end{figure}

\begin{figure}
\centering \includegraphics[height=0.35\textheight]{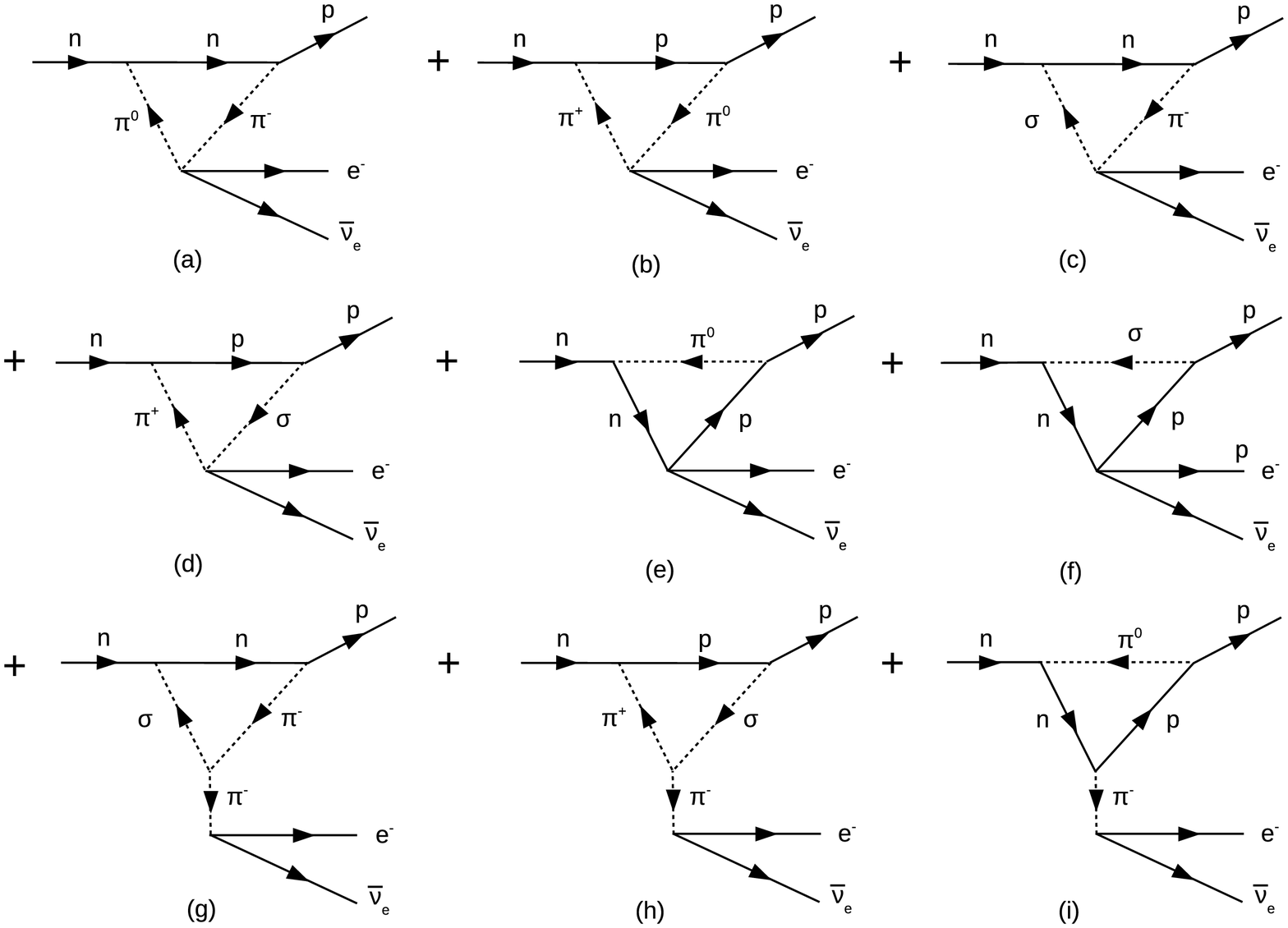}
\includegraphics[height=0.21\textheight]{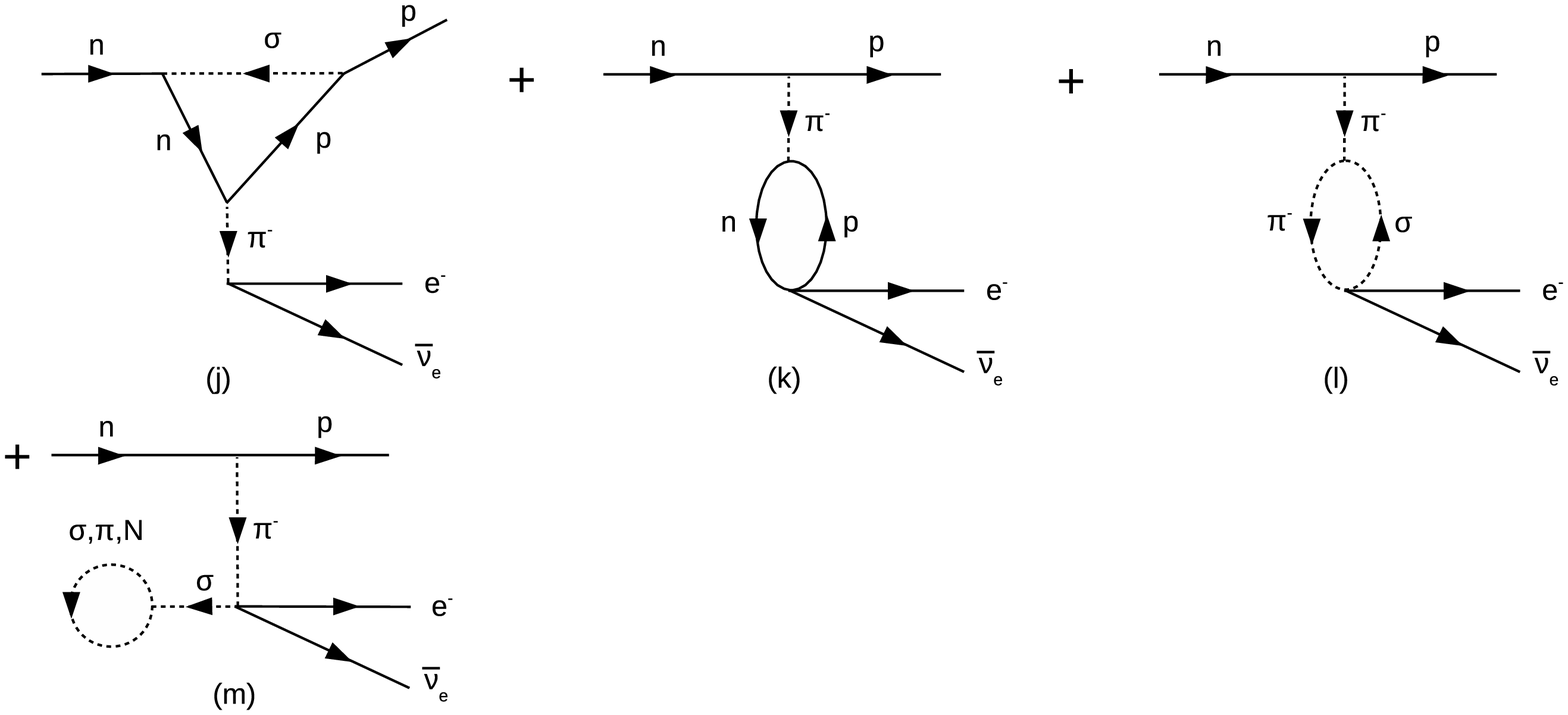}
  \caption{The Feynman diagrams, describing the contributions to the
    amplitude of the neutron $\beta^-$--decay of strong low--energy
    interactions in the one--hadron--loop approximation in the
    L$\sigma$M.}
\label{fig:fig3}
\end{figure}

\begin{figure}
\centering \includegraphics[height=0.10\textheight]{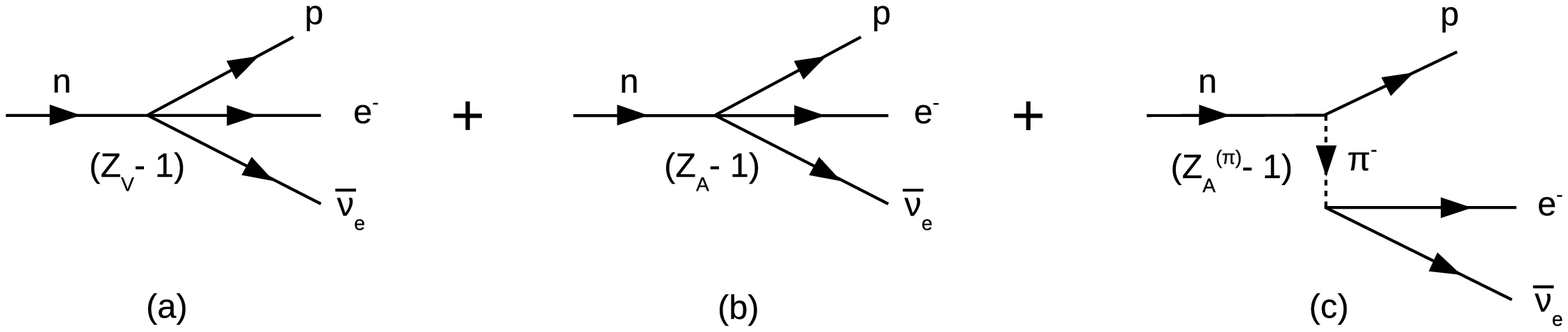}
  \caption{The Feynman diagrams, describing the contributions of the
    counter--terms of the vector (a) and axial--vector (b) charged
    baryonic and axial--vector (c) charged mesonic currents.}
\label{fig:fig4}
\end{figure}

The matrix element $\langle
p(\vec{k}_p,\sigma_p)|J^+_{\mu}(0)|n(\vec{k}_n, \sigma_n)\rangle$ of
the charged hadronic current in the one--hadron--loop approximation
acquires the contributions, given by the Feynman diagrams in
Fig.\,\ref{fig:fig2} and Fig.\,\ref{fig:fig3}, which are caused by
strong low--energy interactions described by the L$\sigma$M in the
phase of spontaneously broken chiral symmetry.  The analytical
expressions of the Feynman diagrams in Fig.\,\ref{fig:fig2} and
Fig.\,\ref{fig:fig3} are equal to
\begin{eqnarray*}
\hspace{-0.3in}&&\langle
p(\vec{k}_p,\sigma_p)|J^+_{\mu}(0)|n(\vec{k}_n, \sigma_n)\rangle_{\rm
  Fig. \ref{fig:fig2}a + Fig. \ref{fig:fig2}b} =
\bar{u}_p\big(\vec{k}_p, \sigma_p\big)\,\gamma_{\mu}(1 -
\gamma^5)\,\frac{1}{m_N - \hat{k}_n - i 0}\,
\Sigma_n(k_n)\,u_n\big(\vec{k}_n,
\sigma_n\big),\nonumber\\ \hspace{-0.3in}&&\langle
p(\vec{k}_p,\sigma_p)|J^+_{\mu}(0)|n(\vec{k}_n, \sigma_n)\rangle_{\rm
  Fig.\,\ref{fig:fig2}c + Fig.\,\ref{fig:fig2}d} =
\bar{u}_p\big(\vec{k}_p, \sigma_p\big)\,\Sigma_p(k_p)\,\frac{1}{m_N -
  \hat{k}_p - i 0}\,\gamma_{\mu}(1 - \gamma^5)\,u_n\big(\vec{k}_n,
\sigma_n\big),\nonumber\\ \hspace{-0.3in}&&\langle
p(\vec{k}_p,\sigma_p)|J^+_{\mu}(0)|n(\vec{k}_n, \sigma_n)\rangle_{\rm
  Fig.\,\ref{fig:fig3}a + Fig.\,\ref{fig:fig3}b}
=\nonumber\\ \hspace{-0.3in}&& = \bar{u}_p\big(\vec{k}_p,
\sigma_p\big)\,\Big\{4\,g^2_{\pi N}\gamma^5 \int \frac{d^4p}{(2\pi)^4
  i}\,\frac{(2 p - q)_{\mu}}{m_N - \hat{p} - \hat{k}_n - i 0}\,
\frac{1}{m^2_{\pi} - (p - q)^2 - i 0}\,\frac{1}{m^2_{\pi} - p^2 - i
  0}\,\gamma^5\,\Big\}\,u_n\big(\vec{k}_n, \sigma_n\big),
\nonumber\\ \hspace{-0.3in}&&\langle
p(\vec{k}_p,\sigma_p)|J^+_{\mu}(0)|n(\vec{k}_n, \sigma_n)\rangle_{\rm
  Fig.\,\ref{fig:fig3}c + Fig.\,\ref{fig:fig3}d}
=\nonumber\\ \hspace{-0.3in}&& = \bar{u}_p\big(\vec{k}_p,
\sigma_p\big)\,\Big\{2\,g^2_{\pi N}\gamma^5 \int \frac{d^4p}{(2\pi)^4
  i}\,\frac{(2 p - q)_{\mu}}{m_N - \hat{p} - \hat{k}_n - i 0}\,
\frac{1}{m^2_{\pi} - (p - q)^2 - i 0}\,\frac{1}{m^2_{\sigma} - p^2 - i
  0}\Big\}\,u_n\big(\vec{k}_n, \sigma_n\big)\nonumber\\ &&
-\bar{u}_p\big(\vec{k}_p, \sigma_p\big)\,\Big\{ 2\,g^2_{\pi N}\int
\frac{d^4p}{(2\pi)^4 i}\,\frac{(2 p - q)_{\mu}}{m_N - \hat{p} -
  \hat{k}_n - i 0}\, \frac{1}{m^2_{\sigma} - (p - q)^2 - i
  0}\,\frac{1}{m^2_{\pi} - p^2 - i 0}\,\gamma^5\Big\}\,
u_n\big(\vec{k}_n, \sigma_n\big),\nonumber\\ \hspace{-0.3in}&&\langle
p(\vec{k}_p,\sigma_p)|J^+_{\mu}(0)|n(\vec{k}_n, \sigma_n)\rangle_{\rm
  Fig.\,\ref{fig:fig3}e + Fig.\,\ref{fig:fig3}f}
=\nonumber\\ &&=\bar{u}_p\big(\vec{k}_p, \sigma_p\big)\,\Big\{
g^2_{\pi N}\int \frac{d^4p}{(2\pi)^4 i}\, \gamma^5\, \frac{1}{m_N -
  \hat{k}_p - \hat{p} - i0}\,\gamma_{\mu}(1 - \gamma^5)\,\frac{1}{m_N
  - \hat{k}_n - \hat{p} - i0}\,\gamma^5\,\frac{1}{m^2_{\pi} - p^2 -
  i0}\Big\}\, u_n\big(\vec{k}_n, \sigma_n\big)
\nonumber\\ \hspace{-0.3in}&&+ \bar{u}_p\big(\vec{k}_p,
\sigma_p\big)\,\Big\{ g^2_{\pi N}\int \frac{d^4p}{(2\pi)^4
  i}\,\frac{1}{m_N - \hat{k}_p - \hat{p} - i0}\,\gamma_{\mu}(1 -
\gamma^5)\,\frac{1}{m_N - \hat{k}_n - \hat{p} -
  i0}\,\frac{1}{m^2_{\sigma} - p^2 - i0}\Big\}\, u_n\big(\vec{k}_n,
\sigma_n\big),\nonumber\\ \hspace{-0.3in}&&\langle
p(\vec{k}_p,\sigma_p)|J^+_{\mu}(0)|n(\vec{k}_n, \sigma_n)\rangle_{\rm
  Fig.\,\ref{fig:fig3}g + Fig.\,\ref{fig:fig3}h +
  Fig.\,\ref{fig:fig3}i + Fig.\,\ref{fig:fig3}j +
  Fig.\,\ref{fig:fig3}k + Fig.\,\ref{fig:fig3}\ell +
  Fig.\,\ref{fig:fig3}m} = \frac{q_{\mu}}{m^2_{\pi} - q^2 -
  i0}\nonumber\\ \hspace{-0.3in}&&\times \Big(\bar{u}_p\big(\vec{k}_p,
\sigma_p\big)\,\Big\{(- 4)\,g^2_{\pi N}\gamma f^2_{\pi}\gamma^5 \int
\frac{d^4p}{(2\pi)^4 i}\,\frac{1}{m_N - \hat{p} - \hat{k}_n - i 0}\,
\frac{1}{m^2_{\pi} - (p - q)^2 - i 0}\,\frac{1}{m^2_{\sigma} - p^2 - i
  0}\Big\}\,u_n\big(\vec{k}_n, \sigma_n\big)
\nonumber\\ \hspace{-0.3in}&&+ \bar{u}_p\big(\vec{k}_p,
\sigma_p\big)\,\Big\{(- 4)\,g^2_{\pi N}\gamma f^2_{\pi} \int
\frac{d^4p}{(2\pi)^4 i}\,\frac{1}{m_N - \hat{p} - \hat{k}_n - i 0}\,
\frac{1}{m^2_{\sigma} - (p - q)^2 - i 0}\,\frac{1}{m^2_{\pi} - p^2 - i
  0}\,\gamma^5\Big\}\,u_n\big(\vec{k}_n,
\sigma_n\big)\nonumber\\ \hspace{-0.3in}&&+ \bar{u}_p\big(\vec{k}_p,
\sigma_p\big)\,\Big\{ 2g^3_{\pi N} f_{\pi}\int \frac{d^4p}{(2\pi)^4
  i}\, \gamma^5\, \frac{1}{m_N - \hat{k}_p - \hat{p} -
  i0}\,\gamma^5\,\frac{1}{m_N - \hat{k}_n - \hat{p} -
  i0}\,\gamma^5\,\frac{1}{m^2_{\pi} - p^2 - i0}\Big\}\,
u_n\big(\vec{k}_n, \sigma_n\big) \nonumber\\ 
\end{eqnarray*}
\begin{eqnarray}\label{eq:40}
\hspace{-0.3in}&&+
\bar{u}_p\big(\vec{k}_p, \sigma_p\big)\,\Big\{ 2g^3_{\pi N}f_{\pi}
\int \frac{d^4p}{(2\pi)^4 i}\, \frac{1}{m_N - \hat{k}_p - \hat{p} -
  i0}\,\gamma^5\,\frac{1}{m_N - \hat{k}_n - \hat{p} -
  i0}\,\frac{1}{m^2_{\sigma} - p^2 - i0}\Big\}\, u_n\big(\vec{k}_n,
\sigma_n\big)\Big) \nonumber\\
&& + \frac{1}{m^2_{\pi} - q^2 -
  i0}\,\Big(\bar{u}_p\big(\vec{k}_p, \sigma_p\big)\,\Big\{ (- 2)
g^2_{\pi N}\int \frac{d^4p}{(2\pi)^4 i}\,{\rm tr}\Big\{\gamma^5\,
\frac{1}{m_N - \hat{p} - i0}\,\gamma_{\mu}(1 - \gamma^5)\,
\frac{1}{m_N - \hat{p} + \hat{q} - i0}\Big\}\, u_n\big(\vec{k}_n,
\sigma_n\big) \nonumber\\  
\hspace{-0.3in}&&\bar{u}_p\big(\vec{k}_p,
\sigma_p\big)\,\Big\{(- 4) g_{\pi N} \gamma f_{\pi} \int
\frac{d^4p}{(2\pi)^4 i}\,\frac{(2 p - q)_{\mu}}{m^2_{\pi} - (p - q)^2
  - i0}\,\frac{1}{m^2_{\sigma} - p^2 - i0}\Big\}\,\gamma^5\,
u_n\big(\vec{k}_n, \sigma_n\big)\Big) - \frac{2 g_{\pi N}
  q_{\mu}}{m^2_{\pi} - q^2 - i0}\bar{u}_p\big(\vec{k}_p,
\sigma_p\big)\nonumber\\
\hspace{-0.3in}&& \times \,\Big(\frac{3\gamma f_{\pi}}{m^2_{\sigma}}\int
\frac{d^4p}{(2\pi)^4 i}\,\frac{1}{m^2_{\sigma} - p^2 - i0} +
\frac{3\gamma f_{\pi}}{m^2_{\sigma}}\int \frac{d^4p}{(2\pi)^4
  i}\,\frac{1}{m^2_{\pi} - p^2 - i0} - \frac{ g_{\pi
    N}}{m^2_{\sigma}}\int \frac{d^4p}{(2\pi)^4 i}\,{\rm
  tr}\Big\{\frac{1}{m_N - \hat{p} - i0}\Big\}\Big)\nonumber\\
\hspace{-0.3in}&& \times \gamma^5\, u_n\big(\vec{k}_n, \sigma_n\big),
\end{eqnarray}
where the self--energy correction to the neutron state is equal to
\begin{eqnarray}\label{eq:41}
\hspace{-0.3in}\Sigma_n(k_n) &=& - \delta m_n - (Z_N - 1)\,(m_N -
\hat{k}_n) + g_{\pi N}\frac{\gamma f_{\pi}}{m^2_{\sigma}}\int
\frac{d^4p}{(2\pi)^4 i}\frac{1}{m^2_{\sigma} - p^2 - i 0} + g_{\pi
  N}\frac{3 \gamma f_{\pi}}{m^2_{\sigma}} \int \frac{d^4p}{(2\pi)^4
  i}\frac{1}{m^2_{\pi} - p^2 - i 0} \nonumber\\ \hspace{-0.3in}&-&
\frac{2g^2_{\pi N}}{m^2_{\sigma}}\int \frac{d^4p}{(2\pi)^4 i}{\rm
  tr}\Big\{\frac{1}{m_N - \hat{p} - i 0}\Big\} + g^2_{\pi N}\int
\frac{d^4p}{(2\pi)^4 i}\,\frac{1}{m_N - \hat{p} - \hat{k}_n - i
  0}\,\frac{1}{m^2_{\sigma} - p^2 - i 0}
\nonumber\\ \hspace{-0.3in}&-& 3 g^2_{\pi N}\int \frac{d^4p}{(2\pi)^4
  i}\,\gamma^5\,\frac{1}{m_N - \hat{p} - \hat{k}_n - i
  0}\,\gamma^5\,\frac{1}{m^2_{\pi} - p^2 - i 0}.
\end{eqnarray}
The self--energy correction to the proton state $\Sigma_p(k_p)$ is
defined by Eq.(\ref{eq:41}) with a replacement of indices $n \to
p$. Following \cite{Ivanov2017b} one may assert that after
renormalization the Feynman diagrams in Fig.\,\ref{fig:fig2}a,
Fig.\,\ref{fig:fig2}b, Fig.\,\ref{fig:fig2}c and Fig.\,\ref{fig:fig2}d
do not contribute to the matrix element $\langle
p(\vec{k}_p,\sigma_p)|J^+_{\mu}(0)|n(\vec{k}_n,
\sigma_n)\rangle$. Skipping standard intermediate calculations
\cite{Ivanov2018} we adduce the final expressions for the momentum
integrals defining non--trivial contributions of the Feynman diagrams
in Fig\,3 to the matrix element of the charged hadronic current
$\langle p(\vec{k}_p,\sigma_p)|J^+_{\mu}(0)|n(\vec{k}_n,
\sigma_n)\rangle$
\begin{eqnarray}\label{eq:42}
\langle p(\vec{k}_p,\sigma_p)|J^+_{\mu}(0)|n(\vec{k}_n,
\sigma_n)\rangle_{\rm Fig. \ref{fig:fig3}a + Fig. \ref{fig:fig3}b} &=&
\bar{u}_p\big(\vec{k}_p, \sigma_p\big)\,\Big\{\,\frac{g^2_{\pi
    N}}{8\pi^2}\Big({\ell n}\frac{\Lambda^2}{m^2_N} +
\frac{3}{2}\Big)\,\gamma_{\mu} - \frac{g^2_{\pi
    N}}{4\pi^2}\,\frac{P_{\mu}}{2 m_N}\,\Big\}\, u_n\big(\vec{k}_n,
\sigma_n\big),
\end{eqnarray}
where $P = k_p + k_n$. 
The contribution of the Feynman diagrams in Fig.\,\ref{fig:fig3}c and
\ref{fig:fig3}d is equal to
\begin{eqnarray}\label{eq:43}
\hspace{-0.15in}&&\langle
p(\vec{k}_p,\sigma_p)|J^+_{\mu}(0)|n(\vec{k}_n, \sigma_n)\rangle_{\rm
  Fig. \ref{fig:fig3}c + Fig.\ref{fig:fig3}d}
=\bar{u}_p\big(\vec{k}_p, \sigma_p\big)\,\Big\{\frac{g^2_{\pi
    N}}{8\pi^2}\,\Big({\ell n}\frac{\Lambda^2}{m^2_N}- {\ell
  n}\frac{m^2_{\sigma}}{m^2_N}\Big)\,\gamma_{\mu}\gamma^5 -
\frac{g^2_{\pi N}}{24\pi^2}\,\frac{m_N q_{\mu}}{m^2_{\sigma} -
  m^2_{\pi}}\gamma^5\Big\}\, u_n\big(\vec{k}_n,
\sigma_n\big).\nonumber\\
\hspace{-0.15in}&&
\end{eqnarray}
In turn, for the contribution of the Feynman diagrams in
Fig.\,\ref{fig:fig3}e and \ref{fig:fig3}f we obtain the following
expression
\begin{eqnarray}\label{eq:44}
\hspace{-0.15in}\langle
p(\vec{k}_p,\sigma_p)|J^+_{\mu}(0)|n(\vec{k}_n, \sigma_n)\rangle_{\rm
  Fig. \ref{fig:fig3}e + Fig.\ref{fig:fig3}f}
&=&\bar{u}_p\big(\vec{k}_p, \sigma_p\big)\,\Big\{ \frac{g^2_{\pi
    N}}{8\pi^2}\,\Big(- \frac{1}{4}\,{\ell
  n}\frac{m^2_{\sigma}}{m^2_N} + \frac{1}{8} - \frac{1}{2}\,{\ell
  n}2\Big)\,\gamma_{\mu} + \frac{g^2_{\pi
    N}}{8\pi^2}\,\Big(\frac{1}{2} - {\ell n}
2\Big)\,\frac{P_{\mu}}{2 m_N}\nonumber\\ &-& \frac{g^2_{\pi
    N}}{8\pi^2}\,\Big(\frac{1}{4} - \frac{1}{2}{\ell n}
2\Big)\,\gamma_{\mu} \gamma^5\Big\}\, u_n\big(\vec{k}_n, \sigma_n\big)
\end{eqnarray}
Following \cite{Ivanov2018} and using the
Gordon identity \cite{Itzykson1980}
\begin{eqnarray}\label{eq:45}
\bar{u}_p(\vec{k}_p, \sigma_p)\,\frac{(k_p +
  k_n)_{\mu}}{2m_N}\,u_n(\vec{k}_n,\sigma_n) = \bar{u}_p(\vec{k}_p,
\sigma_p)\,\gamma_{\mu}\,u_n(\vec{k}_n,\sigma_n) -
\bar{u}_p(\vec{k}_p,
\sigma_p)\,\frac{i\sigma_{\mu\nu}q^{\nu}}{2m_N}\,u_n(\vec{k}_n,\sigma_n)
\end{eqnarray}
we transcribe the sum of the contributions of the Feynman diagrams
Fig.\,\ref{fig:fig3}a - Fig.\,\ref{fig:fig3}f into the form
\begin{eqnarray}\label{eq:46}
&&\langle p(\vec{k}_p,\sigma_p)|J^+_{\mu}(0)|n(\vec{k}_n,
  \sigma_n)\rangle_{\rm Fig. \ref{fig:fig3}a + Fig. \ref{fig:fig3}b +
    Fig. \ref{fig:fig3}c + Fig. \ref{fig:fig3}d + Fig. \ref{fig:fig3}e
    + Fig. \ref{fig:fig3}f} = \nonumber\\ &&=\bar{u}_p\big(\vec{k}_p,
  \sigma_p\big)\,\Big\{\,\frac{g^2_{\pi N}}{8\pi^2}\Big({\ell
    n}\frac{\Lambda^2}{m^2_N} - \frac{1}{4}\,{\ell
    n}\frac{m^2_{\sigma}}{m^2_N} + \frac{9}{8}\Big)\,\gamma_{\mu} +
  \frac{g^2_{\pi N}}{16\pi^2}\,\big(3 + 2 {\ell n}
  2\big)\,\frac{i\sigma_{\mu\nu}q^{\nu}}{2 m_N}\nonumber\\ &&-
  \frac{g^2_{\pi N}}{8\pi^2}\,\Big({\ell n}\frac{\Lambda^2}{m^2_N}-
       {\ell n}\frac{m^2_{\sigma}}{m^2_N} + \frac{1}{4} -
       \frac{1}{2}{\ell n} 2\Big)\,\gamma_{\mu}\gamma^5 -
       \frac{g^2_{\pi N}}{24\pi^2}\,\frac{m_N q_{\mu}}{m^2_{\sigma} -
         m^2_{\pi}}\gamma^5\Big\}\, u_n\big(\vec{k}_n, \sigma_n\big),
\end{eqnarray}
where $\sigma_{\mu\nu} = \frac{i}{2}(\gamma_{\mu}\gamma_{\nu} -
\gamma_{\nu}\gamma_{\mu})$ are the Dirac matrices
\cite{Itzykson1980}. The term with the Lorentz structure
$i\sigma_{\mu\nu}q^{\nu}$ describes the contribution of the weak
magnetism \cite{Bilenky1959,Bilenky1960,Wilkinson1982} with the
isovector anomalous magnetic moment of the nucleon $\kappa = (g^2_{\pi
  N}/16\pi^2) (3+ 2\,{\ell n}2)$.
The contribution of the Feynman diagrams in Fig.\,\ref{fig:fig3}g -
Fig.\,\ref{fig:fig3}m is
\begin{eqnarray}\label{eq:47}
\hspace{-0.15in}&&\langle
p(\vec{k}_p,\sigma_p)|J^+_{\mu}(0)|n(\vec{k}_n, \sigma_n)\rangle_{\rm
  Fig. \ref{fig:fig3}g + Fig. \ref{fig:fig3}h + Fig. \ref{fig:fig3}i +
  Fig. \ref{fig:fig3}j + Fig. \ref{fig:fig3}k +
  Fig. \ref{fig:fig3}\ell+ Fig. \ref{fig:fig3}m} =- \frac{
  q_{\mu}}{m^2_{\pi} - q^2 - i0}\nonumber\\ &&\times \,
\Big\{\frac{g^2_{\pi N}}{16\pi^2}\,\frac{4 m_N\gamma
  f^2_{\pi}}{m^2_{\sigma} - m^2_{\pi}}\,\Big({\ell
  n}\frac{m^2_{\sigma}}{m^2_N} + 2\Big) + m_N\frac{g^2_{\pi
    N}}{8\pi^2}\,\Big({\ell n}\frac{\Lambda^2}{m^2_N} -{\ell
  n}\frac{m^2_{\sigma}}{m^2_N} \Big)\Big\}\,\bar{u}_p\big(\vec{k}_p,
\sigma_p\big) \gamma^5 u_n\big(\vec{k}_n, \sigma_n\big).
\end{eqnarray}
For the calculation of Eq.(\ref{eq:47}) we have used dimensional
regularization and have kept only leading contributions.  From
Eq.(\ref{eq:8}) we define $\gamma = (m^2_{\sigma} -
m^2_{\pi})/2f^2_{\pi}$. Plugging this relation into Eq.(\ref{eq:47})
we get
\begin{eqnarray}\label{eq:48}
\hspace{-0.3in}\langle p(\vec{k}_p,\sigma_p)|J^+_{\mu}(0)|n(\vec{k}_n,
\sigma_n)\rangle_{\rm Fig. \ref{fig:fig3}e + Fig. \ref{fig:fig3}f} =-
\frac{2 m_N q_{\mu}}{m^2_{\pi} - q^2 - i0}\, \frac{g^2_{\pi
    N}}{16\pi^2}\,\Big({\ell n}\frac{\Lambda^2}{m^2_N} +
2\Big)\,\bar{u}_p\big(\vec{k}_p, \sigma_p\big) \gamma^5
u_n\big(\vec{k}_n, \sigma_n\big).
\end{eqnarray}
Thus, to leading order in the large $\sigma$--meson mass expansion and
to leading order in the large nucleon mass expansion we obtain the
matrix element of the hadronic $n\to p$ transition defined by the
Feynman diagrams in Fig.\,\ref{fig:fig1}, Fig.\,\ref{fig:fig2},
Fig.\,\ref{fig:fig3} and Fig.\,\ref{fig:fig4}. We get
\begin{eqnarray}\label{eq:49}
\hspace{-0.3in}&&\langle
p(\vec{k}_p,\sigma_p)|J^+_{\mu}(0)|n(\vec{k}_n, \sigma_n)\rangle =
\bar{u}_p\big(\vec{k}_p, \sigma_p\big)\Bigg\{\Big[1 + \Big(Z_V - 1 +
  \frac{g^2_{\pi N}}{8\pi^2}\Big({\ell n}\frac{\Lambda^2}{m^2_N} -
  \frac{1}{4}\,{\ell n}\frac{m^2_{\sigma}}{m^2_N} +
  \frac{9}{8}\Big)\Big)\Big]\,\gamma_{\mu} - \Big[1 + \Big(Z_A -
  1\nonumber\\
\hspace{-0.3in}&& - \frac{g^2_{\pi
      N}}{8\pi^2}\,\Big({\ell n}\frac{\Lambda^2}{m^2_N} - {\ell
    n}\frac{m^2_{\sigma}}{m^2_N} + \frac{1}{4} - \frac{1}{2}\,{\ell n}
  2\Big)\Big)\Big]\,\gamma_{\mu}\gamma^5 - \Big[1 + \Big(Z^{(\pi)}_A -
  1 + \frac{g^2_{\pi N}}{16\pi^2}\,\Big({\ell n}\frac{m^2_{\sigma}
  }{m^2_N} + 2\Big)\Big)\Big]\,\frac{2 m_N q_{\mu}}{m^2_{\pi} - q^2 -
  i0}\,\gamma^5\nonumber\\
\hspace{-0.3in}&& + \frac{g^2_{\pi N}}{16\pi^2}\,\big(3 + 2{\ell
  n}2\big)\, \frac{i\sigma_{\mu\nu}q^{\nu}}{2
  m_N}\Bigg\}\,u_n\big(\vec{k}_n, \sigma_n\big).
\end{eqnarray}
Because of conservation of the charged vector hadronic current
\cite{Feynman1958} we set
\begin{eqnarray}\label{eq:50}
Z_V = 1 + \frac{g^2_{\pi
      N}}{8\pi^2}\Big({\ell n}\frac{\Lambda^2}{m^2_N} -
  \frac{1}{4}\,{\ell n}\frac{m^2_{\sigma}}{m^2_N} +
  \frac{9}{8}\Big).
\end{eqnarray}
In turn, renormalization of the charged axial--vector hadronic current
by strong low--energy interactions
\cite{GellMann1960,Bernstein1960,DeAlfaro1973} should lead to the
axial coupling constant $g_A - 1$. Setting
\begin{eqnarray}\label{eq:51}
&& Z_A - 1 - \frac{g^2_{\pi N}}{8\pi^2}\,\Big({\ell
    n}\frac{\Lambda^2}{m^2_N} - {\ell n}\frac{m^2_{\sigma}}{m^2_N} +
  \frac{1}{4} - \frac{1}{2}\,{\ell n} 2\Big)= g_A - 1,\nonumber\\ &&
  Z^{(\pi)}_A - 1 + \frac{g^2_{\pi N}}{16\pi^2}\,\Big({\ell
    n}\frac{m^2_{\sigma} }{m^2_N} + 2\Big) = g_A - 1
\end{eqnarray}
we arrive at the matrix element of the hadronic $n \to p$ transition
\begin{eqnarray}\label{eq:52}
\langle p(\vec{k}_p,\sigma_p)|J^+_{\mu}(0)|n(\vec{k}_n,
\sigma_n)\rangle = \bar{u}_p\big(\vec{k}_p,
\sigma_p\big)\Big\{\gamma_{\mu}\big(1 - g_A\gamma^5\big)+
\frac{\kappa}{2 m_N}\, i \sigma_{\mu\nu}q^{\nu} - \frac{2\,g_A m_N
}{m^2_{\pi} - q^2}\, q_{\mu} \gamma^5 \Big\}\,u_n\big(\vec{k}_n,
\sigma_n\big).
\end{eqnarray}
We would like to notice that the pion--nucleon coupling constant
$g_{\pi N}$ in our calculations is fully defined by the GT--relation
$g_{\pi N} = m_N/f_{\pi}$.
The relations Eq.(\ref{eq:51}) are justified by the vanishing of the
matrix element
\begin{eqnarray}\label{eq:53}
\lim_{m_{\pi} \to\, 0}q^{\mu} \langle
p(\vec{k}_p,\sigma_p)|J^+_{\mu}(0)|n(\vec{k}_n, \sigma_n)\rangle = 0
\end{eqnarray}
in the limit $m_{\pi} \to 0$, caused by local conservation of the
vector and axial--vector hadronic currents
$\partial^{\mu}\vec{V}_{\mu}(x) = \partial^{\mu}\vec{A}_{\mu}(x) = 0$
\cite{Feynman1958,Nambu1960}. We would like to emphasize that we do
not claim that we have calculated the axial coupling constant
$g_A$. We assert only that in the L$\sigma$M to one--hadron--loop
approximation we have reproduced in the limit $m_{\sigma} \to \infty$
the Lorentz structure of the matrix element of the hadronic $n \to p$
transition agreeing well with the results, which can be obtained in
the current algebra approach \cite{Marshak1969}. Indeed, the term
proportional to $q_{\mu}\gamma^5$ in Eq.(\ref{eq:43}), vanishing in
the limit $m_{\sigma} \to \infty$, does not appear in the current
algebra approach.
\begin{figure}
\centering \includegraphics[height=0.23\textheight]{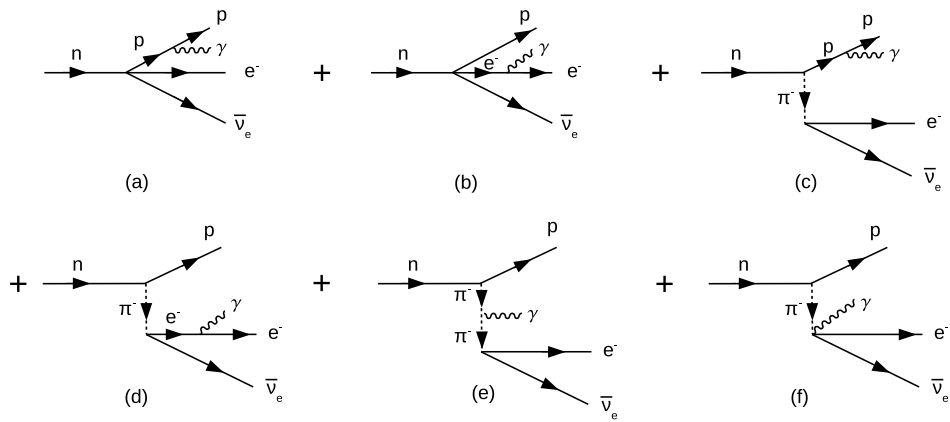}
  \caption{The Feynman diagrams, defining within the standard $V - A$
    effective theory of weak interactions the amplitude of the neutron
    radiative $\beta-$--decay, calculated in the tree--approximation
    for strong low--energy interactions in the L$\sigma$M and QED.}
\label{fig:fig5}
\end{figure}

\section{Neutron radiative beta decay in the tree-- and
one--hadron--loop approximation for strong low--energy interactions in
L$\sigma$M and QED}
\label{sec:strahlung}

In this section we analyse gauge properties of strong low--energy
interactions in the neutron radiative $\beta^-$--decay $n \to p + e^-
+ \bar{\nu}_e + \gamma$ in the tree-- and one--hadron--loop
approximation within the standard $V - A$ effective theory of weak
interactions with the L$\sigma$M and QED, describing strong
low--energy and electromagnetic interactions, respectively.

\subsection{Neutron radiative beta decay 
in the tree--approximation for strong low--energy interactions in
L$\sigma$M and QED}

The Feynman diagrams, defining the amplitude of the neutron radiative
$\beta^-$--decay within the standard $V - A$ effective theory of weak
interactions with QED and in the tree--approximation for strong
low--energy interactions in the L$\sigma$M, are shown in
Fig.\,\ref{fig:fig5}. The amplitude of the neutron radiative
$\beta^-$--decay, described by the Feynman diagrams in
Fig.\,\ref{fig:fig5}, can be written as follows
\begin{eqnarray}\label{eq:54}
\hspace{-0.3in}M_{\rm Fig. \ref{fig:fig5}}(n \to p e^- \bar{\nu}_e
\gamma)_{\lambda} &=& M_{\rm Fig. \ref{fig:fig5}a +
  Fig. \ref{fig:fig5}b}(n \to p e^- \bar{\nu}_e \gamma)_{\lambda} +
M_{\rm Fig. \ref{fig:fig5}c + Fig. \ref{fig:fig5}d}(n \to p e^-
\bar{\nu}_e \gamma)_{\lambda} \nonumber\\
\hspace{-0.3in}&+& M_{\rm Fig. \ref{fig:fig5}e +
  Fig. \ref{fig:fig5}f}(n \to p e^- \bar{\nu}_e \gamma)_{\lambda}.
\end{eqnarray}
The amplitude $M_{\rm Fig. \ref{fig:fig5}a + Fig. \ref{fig:fig5}b}(n
\to p e^- \bar{\nu}_e \gamma)_{\lambda}$, defined by the Feynman
diagrams in Fig.\ref{fig:fig5}a and Fig.\ref{fig:fig5}a, is equal to
\cite{Ivanov2013,Ivanov2017,Ivanov2017b}
\begin{eqnarray}\label{eq:55}
\hspace{-0.3in}&& M_{\rm Fig. \ref{fig:fig5}a +
  Fig. \ref{fig:fig5}b}(n \to p e^- \bar{\nu}_e \gamma)_{\lambda} = e
G_V\nonumber\\
\hspace{-0.3in}&&\times \Big\{\Big[\bar{u}_p(\vec{k}_p, \sigma_p)
  \gamma^{\mu}(1 - \gamma^5) u_n(\vec{k}_n, \sigma_n)\Big]
\Big[\bar{u}_e(\vec{k}_e,\sigma_e)\,\frac{1}{2k_e\cdot k}\,Q_{e,
    \lambda}\,\gamma_{\mu} (1 - \gamma^5) v_{\nu}(\vec{k}_{\nu}, +
  \frac{1}{2})\Big]\nonumber\\
\hspace{-0.3in}&&- \Big[\bar{u}_p(\vec{k}_p,
  \sigma_p)\,Q_{p, \lambda} \,\frac{1}{2k_p \cdot k}\,\gamma^{\mu}(1 -
  \gamma^5) u_n(\vec{k}_n,
  \sigma_n)\Big]\Big[\bar{u}_e(\vec{k}_e,\sigma_e) \gamma^{\mu} (1 -
  \gamma^5) v_{\nu}(\vec{k}_{\nu}, + \frac{1}{2})\Big]\Big\},
\end{eqnarray}
where $Q_{e,\lambda}$ and $Q_{p, \lambda}$ are given by
\cite{Ivanov2013,Ivanov2017,Ivanov2017b}
\begin{eqnarray}\label{eq:56}
\hspace{-0.3in}Q_{e, \lambda} = 2 \varepsilon^{*}_{\lambda}(k)\cdot k_e +
\hat{\varepsilon}^*_{\lambda}(k)\hat{k}\;,\; Q_{p,\lambda} = 2
\varepsilon^{*}_{\lambda}(k)\cdot k_p +
\hat{\varepsilon}^*_{\lambda}(k)\hat{k}.
\end{eqnarray}
Here $\varepsilon^*_{\lambda}(k)$ is the polarization vector of the
photon with the 4--momentum $k$ and in two polarization states
$\lambda = 1,2$, obeying the constraint $k\cdot
\varepsilon^*_{\lambda}(k) = 0$. For the derivation of
Eq.(\ref{eq:55}) we have used the Dirac equations for the free proton
and electron. Replacing $\varepsilon^{*}_{\lambda}(k) \to k$ and using
$k^2 = 0$ we get \cite{Ivanov2017,Ivanov2017b} (see also
\cite{Ivanov2013b})
\begin{eqnarray}\label{eq:57}
\hspace{-0.3in}M_{\rm Fig. \ref{fig:fig5}a + Fig. \ref{fig:fig5}b}(n
\to p e^- \bar{\nu}_e \gamma)_{\lambda}\Big|_{\varepsilon^{*}_{\lambda}(k) \to k} = 0.
\end{eqnarray}
This confirms invariance of the Feynman diagrams in
Fig.\,\ref{fig:fig5}a and Fig.\,\ref{fig:fig5}b  under a
gauge transformation $\varepsilon^*_{\lambda'}(k) \to
\varepsilon^*_{\lambda'}(k) + c\,k$, where $c$ is an arbitrary
constant.

In turn, the contribution of the Feynman diagrams in
Fig.\,\ref{fig:fig5}c, Fig.\,\ref{fig:fig5}d, Fig.\,\ref{fig:fig5}e
and Fig.\,\ref{fig:fig5}f goes beyond the previous analysis of the
neutron radiative $\beta^-$--decay \cite{Gaponov1996,Bernard2004,
  Gardner2012, Gardner2013} (see also
\cite{Ivanov2013,Ivanov2017,Ivanov2017a,Ivanov2017b}). The
contribution of the Feynman diagrams in Fig.\,\ref{fig:fig5}c and
Fig.\,\ref{fig:fig5}d to the amplitude of the neutron radiative
$\beta^-$--decay takes the form
\begin{eqnarray}\label{eq:58}
\hspace{-0.3in}&& M_{\rm Fig. \ref{fig:fig5}c +
  Fig. \ref{fig:fig5}d}(n \to p e^- \bar{\nu}_e \gamma)_{\lambda} = e
G_V\nonumber\\
\hspace{-0.3in}&&\times\,\Big\{\frac{2 m_N (q - k)_{\mu}}{m^2_{\pi} -
  (q - k)^2 - i 0}\,[\bar{u}_p(\vec{k}_p, \sigma_p)\gamma^5
  u_n(\vec{k}_n, \sigma_n)] \,\Big[\bar{u}_e(\vec{k}_e,\sigma_e)Q_{e,
    \lambda}\frac{1}{2 k_e\cdot k} \gamma^{\mu} (1 - \gamma^5)
  v_{\nu}(\vec{k}_{\nu}, +
  \frac{1}{2})\Big]\nonumber\\ \hspace{-0.3in}&& - \frac{2 m_N
  q_{\mu}}{m^2_{\pi} - q^2 - i 0}\,\Big[\bar{u}_p(\vec{k}_p,
  \sigma_p)\,Q_{p, \lambda} \,\frac{1}{2k_p \cdot k}\,\gamma^5
  u_n(\vec{k}_n, \sigma_n)\Big]\Big[\bar{u}_e(\vec{k}_e,\sigma_e)
  \gamma^{\mu} (1 - \gamma^5) v_{\nu}(\vec{k}_{\nu}, +
  \frac{1}{2})\Big]\Big\},
\end{eqnarray}
where we have used the GT--relation $g_{\pi N} = m_N/f_{\pi}$. The
contribution of the Feynman diagrams in Fig\,\ref{fig:fig5}e and
Fig\,\ref{fig:fig5}f to the amplitude of the neutron radiative
$\beta^-$--decay is defined by the analytical expression
\begin{eqnarray}\label{eq:59}
\hspace{-0.3in}&& M_{\rm Fig. \ref{fig:fig5}e +
  Fig. \ref{fig:fig5}f}(n \to p e^- \bar{\nu}_e \gamma)_{\lambda} = e
G_V\nonumber\\
\hspace{-0.3in}&&\times\,\Big\{\frac{2 m_N q_{\mu}}{m^2_{\pi} - q^2 -
  i 0}\,\frac{(2 q - k)\cdot \varepsilon^*_{\lambda}(k)}{m^2_{\pi} -
  (q - k)^2 - i0}\,[\bar{u}_p(\vec{k}_p, \sigma_p)\gamma^5
  u_n(\vec{k}_n, \sigma_n)]\Big[\bar{u}_e(\vec{k}_e,\sigma_e)
  \gamma^{\mu} (1 - \gamma^5) v_{\nu}(\vec{k}_{\nu}, +
  \frac{1}{2})\Big]\nonumber\\\hspace{-0.3in}&& + \frac{2
  m_N}{m^2_{\pi} - (q - k)^2 - i0}\,[\bar{u}_p(\vec{k}_p,
  \sigma_p)\gamma^5 u_n(\vec{k}_n, \sigma_n)]\,
\Big[\bar{u}_e(\vec{k}_e,\sigma_e)\hat{\varepsilon}^*_{\lambda}(k) (1
  - \gamma^5) v_{\nu}(\vec{k}_{\nu}, + \frac{1}{2})\Big]\Big\}.
\end{eqnarray}
 Replacing $\varepsilon^{*}_{\lambda}(k) \to k$ one may show the sum
 of the Feynman diagrams in Fig.\,\ref{fig:fig5}c,
 Fig.\,\ref{fig:fig5}d, Fig\,\ref{fig:fig5}e and Fig\,\ref{fig:fig5}f,
 defined by Eq.(\ref{eq:58}) and Eq.(\ref{eq:59}), vanishes
\begin{eqnarray}\label{eq:60}
\hspace{-0.3in}M_{\rm Fig. \ref{fig:fig5}c +
  Fig. \ref{fig:fig5}d}(n \to p e^- \bar{\nu}_e \gamma)_{\lambda} +
M_{\rm Fig. \ref{fig:fig5}e + Fig. \ref{fig:fig5}f}(n \to p e^-
\bar{\nu}_e \gamma)_{\lambda}\Big|_{\varepsilon^{*}_{\lambda}(k)
  \to k} = 0.
\end{eqnarray}
This confirms invariance of the amplitude of the neutron radiative
$\beta^-$--decay, defined by the sum of Eq.(\ref{eq:58}) and
Eq.(\ref{eq:59}), under a gauge transformation
$\varepsilon^*_{\lambda}(k) \to \varepsilon^*_{\lambda}(k) + c\,k$.
We would like to notice that the Feynman diagrams in
Fig.\,\ref{fig:fig5}c, Fig.\,\ref{fig:fig5}d, Fig\,\ref{fig:fig5}e and
Fig\,\ref{fig:fig5}f describe the contribution of strong low--energy
interactions, which is fully caused by mesonic parts of the charged
axial--vector hadronic current, defined in the L$\sigma$M in the phase
of spontaneously broken chiral symmetry. It should be also noticed
that such a contribution does not appear in previous calculations of
the neutron radiative $\beta^-$--decay \cite{Gaponov1996,Bernard2004,
  Gardner2012, Gardner2013} (see also \cite{Ivanov2013, Ivanov2017,
  Ivanov2017a, Ivanov2017b}).

\subsection{Neutron radiative beta decay in the one--hadron--loop  
approximation for strong low--energy interactions in L$\sigma$M and QED}

The amplitude of the neutron radiative $\beta^-$--decay in the
one--hadron--loop approximation in the L$\sigma$M is defined by the
Feynman diagrams in Fig.\,\ref{fig:fig6}, Fig.\,\ref{fig:fig7} and
Fig.\,\ref{fig:fig8}
\begin{eqnarray}\label{eq:61}
\hspace{-0.3in}M(n \to p e^- \bar{\nu}_e \gamma)_{\lambda} = M_{\rm
  Fig. \ref{fig:fig6}}(n \to p e^- \bar{\nu}_e \gamma)_{\lambda} +
M_{\rm Fig. \ref{fig:fig7}}(n \to p e^- \bar{\nu}_e \gamma)_{\lambda}
+ M_{\rm Fig. \ref{fig:fig8}}(n \to p e^- \bar{\nu}_e
\gamma)_{\lambda}.
\end{eqnarray}
The analytical expressions and properties of the amplitudes $M_{\rm
  Fig\,j}(n \to p e^- \bar{\nu}_e \gamma)_{\lambda}$ for $j = 6,7,8$
with respect a gauge transformation $\varepsilon^*_{\lambda}(k) \to
\varepsilon^*_{\lambda}(k) + c\,k$ are given and discussed below.

\begin{figure}
\centering \includegraphics[height=0.21\textheight]{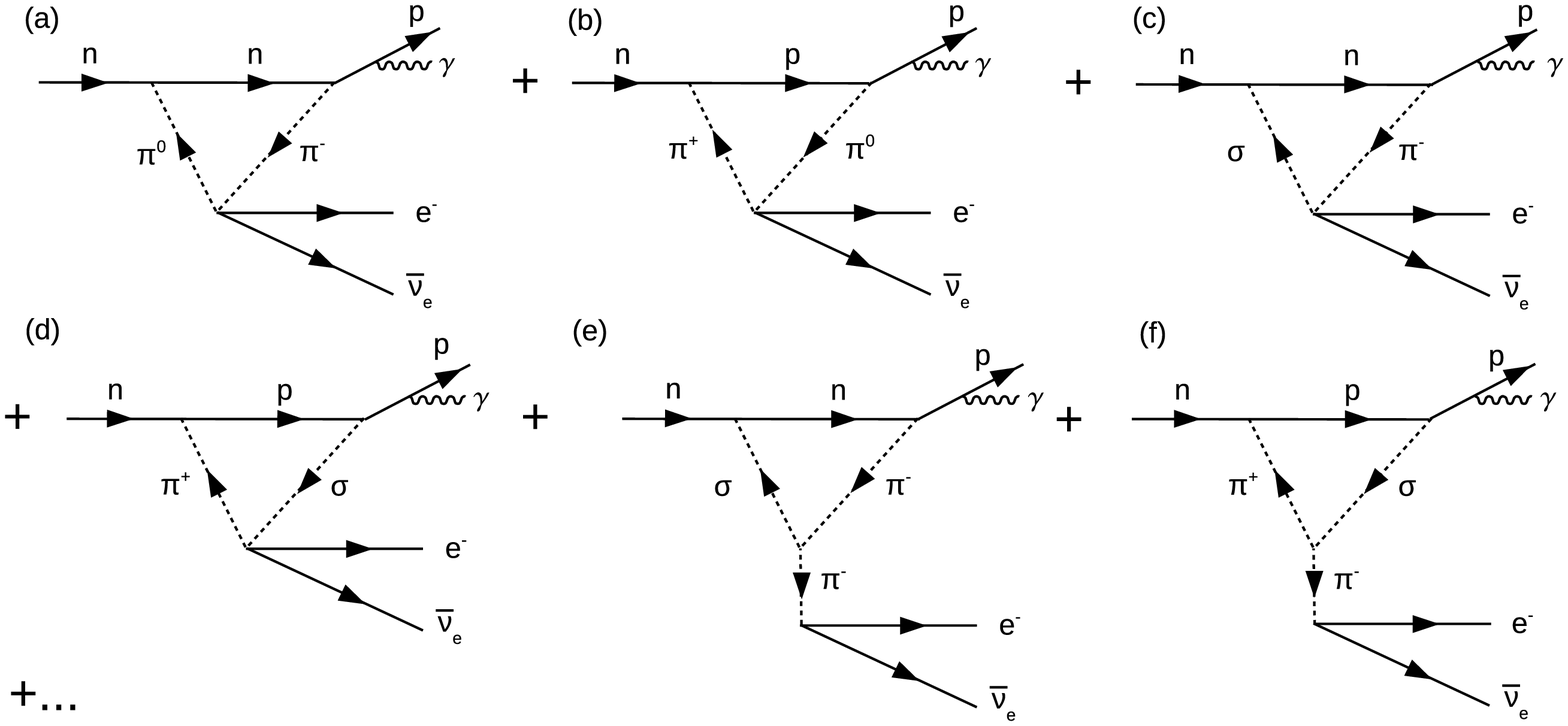}
\includegraphics[height=0.24\textheight]{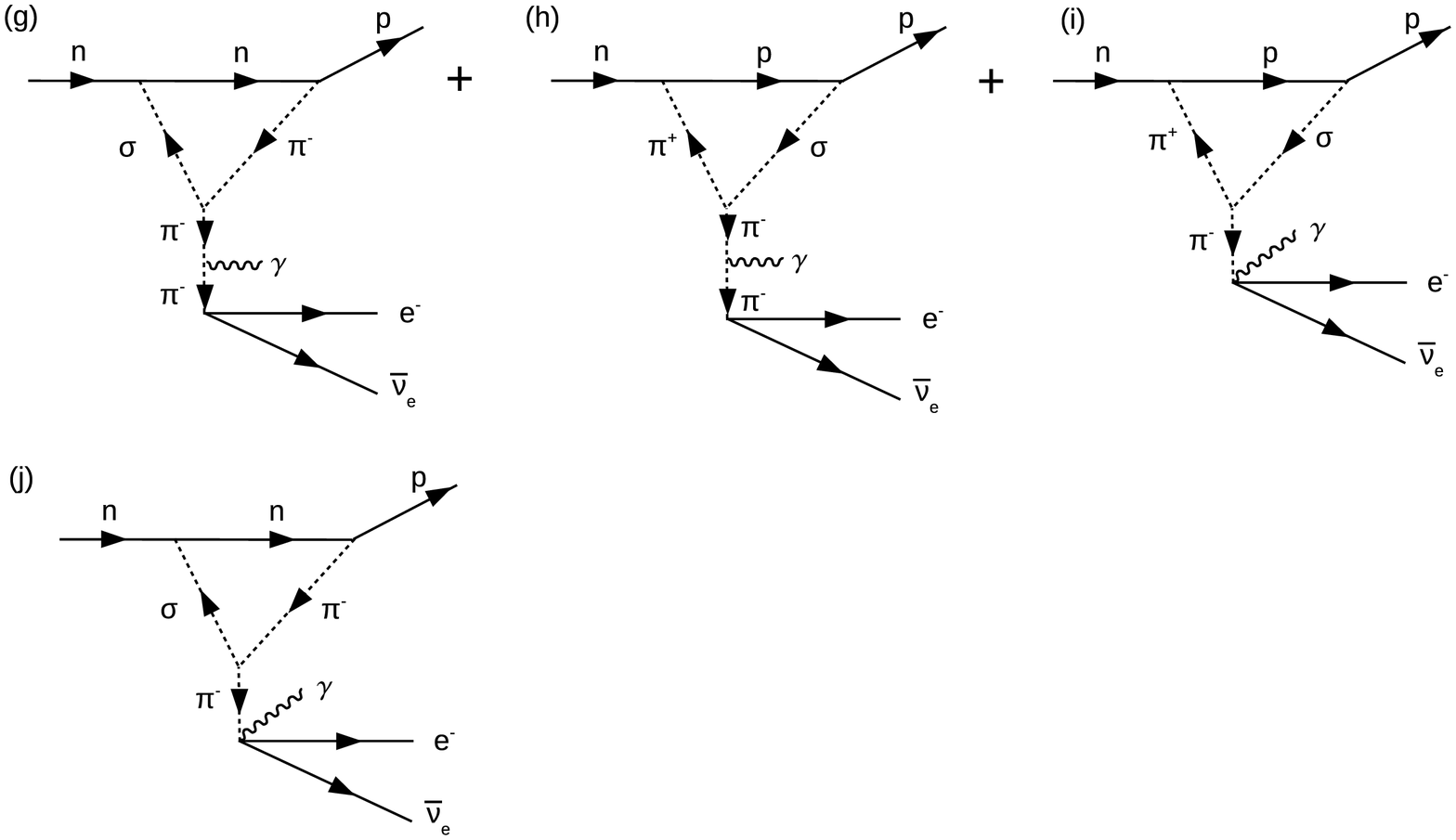}
  \caption{The Feynman diagrams, defining the contributions of strong
    low--energy interactions to the amplitude of the neutron radiative
    $\beta-$--decay in the one--hadron--loop approximation in the
    L$\sigma$M. They are obtained from the Feynman diagrams in
    Fig.\,\ref{fig:fig3} by emission of a photon from external proton
    and electron lines, and from the $\pi^-$--meson line of the
    one--pion--pole diagrams. We show only some of the complete set of
    Feynman diagrams. The other diagrams can be trivially added using
    a complete set of Feynman diagrams in Fig.\,\ref{fig:fig3} and
    inserting a photon line into external proton and electron lines,
    and the $\pi^-$--meson line of the one--pion--pole diagrams.}
\label{fig:fig6}
\end{figure}

\subsubsection*{\bf 1. The contribution to the amplitude of the neutron 
radiative $\beta^-$--decay caused by the Feynman diagrams in
Fig.\ref{fig:fig6}}

In Fig.\,\ref{fig:fig6} we show a set of Feynman diagrams, describing
one--hadron--loop contributions of strong low--energy interactions in
the L$\sigma$M to the amplitude of the neutron radiative
$\beta^-$--decay. The complete set of the Feynman diagrams in
Fig.\,\ref{fig:fig6} can be obtained from the Feynman diagrams in
Fig.\,\ref{fig:fig3} with a photon emitted from external proton and
electron lines, and from the $\pi^-$--meson line of the
one--pion--pole diagrams. The complete set of the Feynman diagrams,
obtained in such a way, is invariant under a gauge transformation
$\varepsilon^*_{\lambda}(k) \to \varepsilon^*_{\lambda}(k) +
c\,k$. After renormalization, which is similar to renormalization of
the Feynman diagrams in Fig.\,\ref{fig:fig3} with the same set of
counter--terms, this set of Feynman diagrams together with the Feynman
diagrams in Fig.\,\ref{fig:fig5} define the following contribution to
the amplitude of the neutron radiative $\beta^-$--decay
\begin{eqnarray*}
\hspace{-0.3in}&& M_{\rm Fig. \ref{fig:fig6}}(n \to p e^- \bar{\nu}_e
\gamma)_{\lambda} = e G_V\nonumber\\
\hspace{-0.3in}&&\times \Big\{\Big[\bar{u}_p(\vec{k}_p, \sigma_p)
  \gamma^{\mu}(1 - g_A\gamma^5) u_n(\vec{k}_n, \sigma_n)\Big]
\Big[\bar{u}_e(\vec{k}_e,\sigma_e)\,\frac{1}{2k_e\cdot k}\,Q_{e,
    \lambda}\,\gamma_{\mu} (1 - \gamma^5) v_{\nu}(\vec{k}_{\nu}, +
  \frac{1}{2})\Big] \nonumber\\
\end{eqnarray*}
\begin{eqnarray}\label{eq:62}
\hspace{-0.3in}&&- \Big[\bar{u}_p(\vec{k}_p,
  \sigma_p)\,Q_{p, \lambda} \,\frac{1}{2k_p \cdot k}\,\gamma^{\mu}(1 -
  g_A\gamma^5) u_n(\vec{k}_n,
  \sigma_n)\Big]\Big[\bar{u}_e(\vec{k}_e,\sigma_e) \gamma^{\mu} (1 -
  \gamma^5) v_{\nu}(\vec{k}_{\nu}, + \frac{1}{2})\Big]\nonumber\\
\hspace{-0.3in}&&+ \frac{2 g_A m_N (q - k)_{\mu}}{m^2_{\pi} - (q -
  k)^2 - i 0}\,\Big[\bar{u}_p(\vec{k}_p, \sigma_p)\gamma^5 u_n(\vec{k}_n,
  \sigma_n)\Big] \,\Big[\bar{u}_e(\vec{k}_e,\sigma_e)Q_{e,
    \lambda}\frac{1}{2 k_e\cdot k} \gamma^{\mu} (1 - \gamma^5)
  v_{\nu}(\vec{k}_{\nu}, + \frac{1}{2})\Big]\nonumber\\
\hspace{-0.3in}&& - \frac{2 g_A m_N
  q_{\mu}}{m^2_{\pi} - q^2 - i 0}\,\Big[\bar{u}_p(\vec{k}_p,
  \sigma_p)\,Q_{p, \lambda} \,\frac{1}{2k_p \cdot k}\,\gamma^5
  u_n(\vec{k}_n, \sigma_n)\Big]\Big[\bar{u}_e(\vec{k}_e,\sigma_e)
  \gamma^{\mu} (1 - \gamma^5) v_{\nu}(\vec{k}_{\nu}, +
  \frac{1}{2})\Big]\nonumber\\
\hspace{-0.3in}&&+ \frac{2 g_A m_N q_{\mu}}{m^2_{\pi} - q^2 - i
  0}\,\frac{(2 q - k)\cdot \varepsilon^*_{\lambda}(k)}{m^2_{\pi} - (q
  - k)^2 - i0}\,\Big[\bar{u}_p(\vec{k}_p, \sigma_p)\gamma^5
  u_n(\vec{k}_n, \sigma_n)\Big]\Big[\bar{u}_e(\vec{k}_e,\sigma_e)
  \gamma^{\mu} (1 - \gamma^5) v_{\nu}(\vec{k}_{\nu}, +
  \frac{1}{2})\Big]\nonumber\\\hspace{-0.3in}&& + \frac{2 g_A
  m_N}{m^2_{\pi} - (q - k)^2 - i0}\,\Big[\bar{u}_p(\vec{k}_p,
  \sigma_p)\gamma^5 u_n(\vec{k}_n, \sigma_n)\Big]\,
\Big[\bar{u}_e(\vec{k}_e,\sigma_e) \hat{\varepsilon}^*_{\lambda}(k) (1
  - \gamma^5) v_{\nu}(\vec{k}_{\nu}, + \frac{1}{2})\Big]\Big\},
\end{eqnarray}
 where the contribution of strong low--energy interactions is given in
 terms of the axial coupling constant $g_A$. It is obvious that the
 amplitude Eq.(\ref{eq:62}) is invariant under a gauge transformation
 $\varepsilon^*_{\lambda}(k) \to \varepsilon^*_{\lambda}(k) + c\,k$.
 The first two terms in Eq.(\ref{eq:62}) describe the amplitude of the
 neutron radiative $\beta^-$--decay to leading order in the large
 proton mass expansion in the previous analysis of such a decay
 \cite{Gaponov1996,Bernard2004, Gardner2012, Gardner2013} (see also
 \cite{Ivanov2013,Ivanov2017a,Ivanov2017,Ivanov2017b}), where strong
 low--energy interactions contribute only in terms of the axial
 couping constant $g_A$. The contributions of the last four terms in
 Eq.(\ref{eq:62}) go beyond the previous analysis of the neutron
 radiative $\beta^-$--decay. They are specific for the L$\sigma$M,
 since they are fully caused by the contribution of the mesonic part
 of the charged axial--vector hadronic current.

\begin{figure}
\centering \includegraphics[height=0.34\textheight]{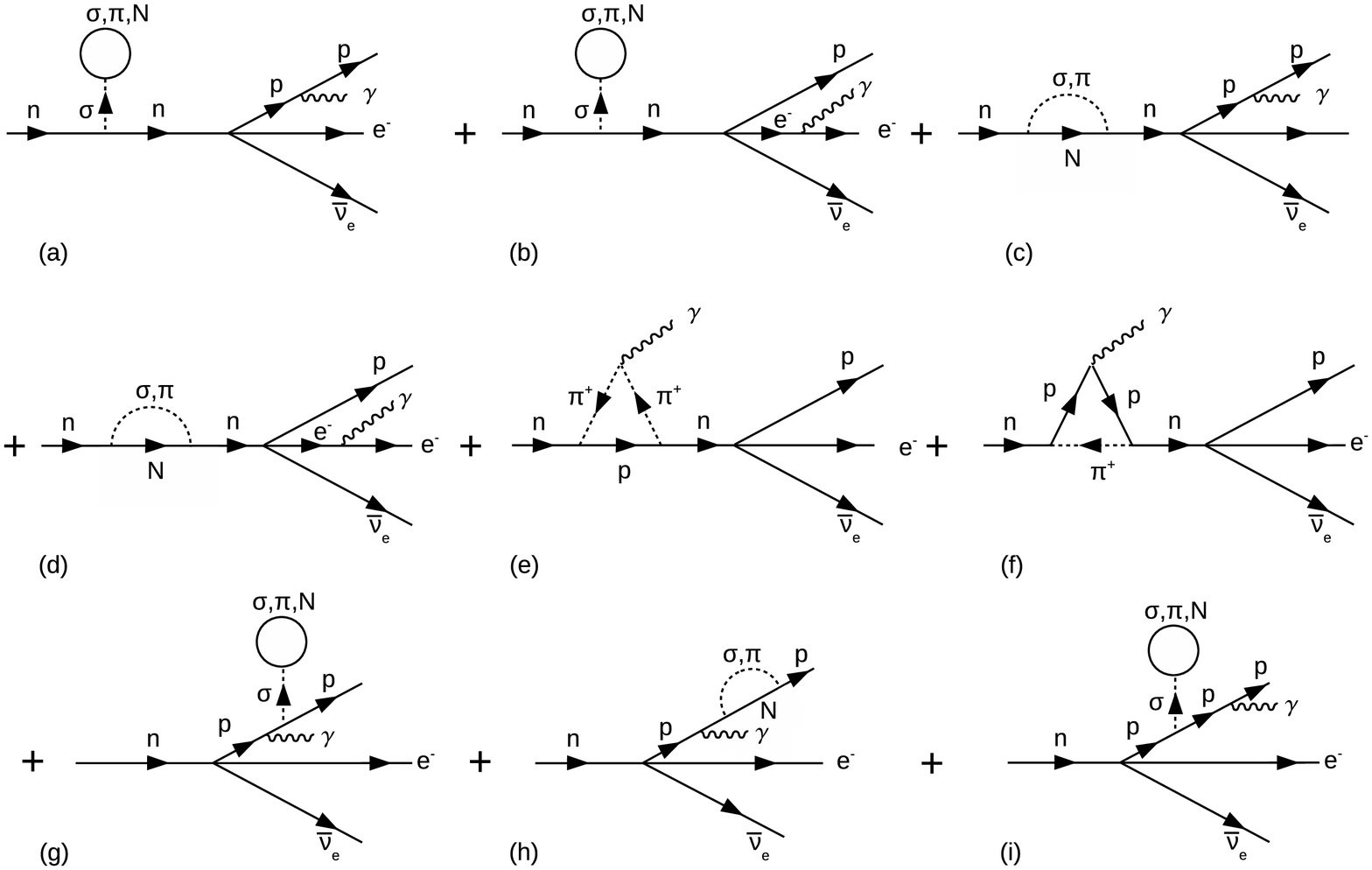}
\includegraphics[height=0.23\textheight]{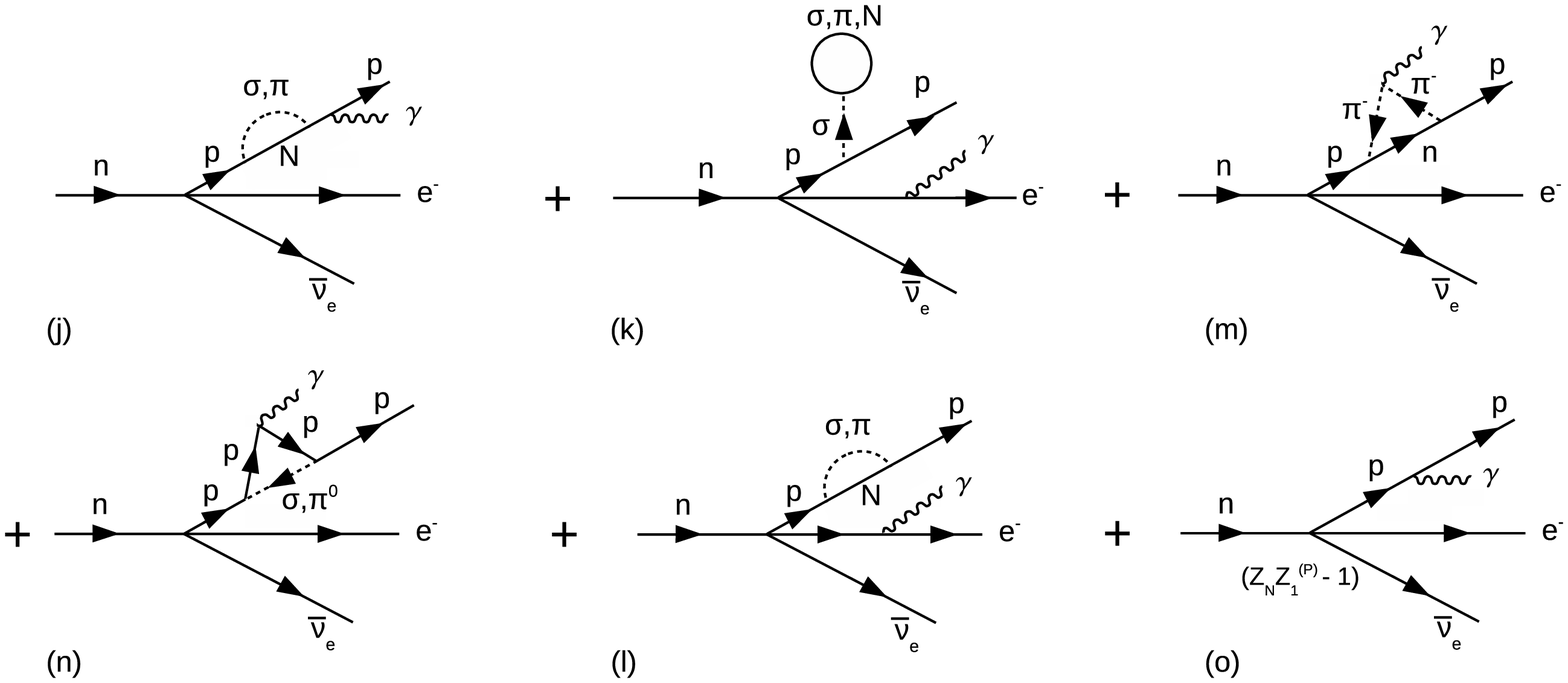}
  \caption{The Feynman diagrams, describing the contributions to the
    amplitude of the neutron radiative $\beta^-$--decay in the
    one--hadron--loop approximation for strong low--energy
    interactions in the L$\sigma$M and QED.  These Feynman diagrams
    can be obtained from the Feynman diagrams in Fig.\,\ref{fig:fig2}
    by emitting a photon from all lines of charged particles.}
\label{fig:fig7}
\end{figure}

\subsubsection*{\bf 2. The contribution to the amplitude of the neutron 
radiative $\beta^-$--decay caused by the Feynman diagrams in
Fig.\ref{fig:fig7}}

The set of the Feynman diagrams in Fig.\ref{fig:fig7} describes in the
L$\sigma$M the contribution of hadronic structure of the neutron and
proton to the neutron radiative $\beta^-$--decay. It is obtained from
the self--energy Feynman diagrams in Fig.\ref{fig:fig2} with photon
emitted from the lines of all charged particles.

The contribution of hadronic structure of the neutron is described by
the Feynman diagrams in Fig.\,\ref{fig:fig7}a - Fig.\,\ref{fig:fig7}f.
Following \cite{Ivanov2017b} one may show that the Feynman diagrams in
Fig.\,\ref{fig:fig7}a - Fig.\,\ref{fig:fig7}d are invariant under a
gauge transformation $\varepsilon^*_{\lambda'}(k) \to
\varepsilon^*_{\lambda'}(k) + c\,k$ and vanish after renormalization.
Thus, after renormalization of the mass and wave function of the
neutron we may write
\begin{eqnarray}\label{eq:63}
\hspace{-0.3in} M_{\rm Fig. \ref{fig:fig7}a} (n \to p e^- \bar{\nu}_e
\gamma)_{\lambda} + M_{\rm Fig. \ref{fig:fig7}b} (n \to p e^-
\bar{\nu}_e \gamma)_{\lambda} + M_{\rm Fig. \ref{fig:fig7}c} (n \to p
e^- \bar{\nu}_e \gamma)_{\lambda} + M_{\rm Fig. \ref{fig:fig7}d} (n
\to p e^- \bar{\nu}_e \gamma)_{\lambda} = 0.
\end{eqnarray}
The contributions of the Feynman diagrams in Fig.\,\ref{fig:fig7}e and
Fig.\,\ref{fig:fig7}f are given by the analytical expressions
\begin{eqnarray*}
\hspace{-0.3in}&&M_{\rm Fig. \ref{fig:fig7}e} (n \to p e^- \bar{\nu}_e
 \gamma)_{\lambda} = eG_V\,(- 2g^2_{\pi N})\,\Big[\bar{u}_p(\vec{k}_p, \sigma_p)
  \gamma_{\mu} (1 - \gamma^5)\,\frac{1}{m_N - \hat{k}_n + \hat{k} -
    i0}\nonumber\\
\hspace{-0.3in}&&\times\int \frac{d^4p}{(2\pi)^4
  i}\,\gamma^5\,\frac{1}{m_N - \hat{k}_n - \hat{p} -
  i0}\,\gamma^5\,\frac{(2p + k)\cdot
  \varepsilon^*_{\lambda}(k)}{m^2_{\pi} - (p + k)^2 - i
  0}\,\frac{1}{m^2_{\pi} - p^2 - i 0}\,u_n(\vec{k}_n, \sigma_n)\Big]
 \nonumber\\
\end{eqnarray*}
\begin{eqnarray}\label{eq:64}
\hspace{-0.3in}&&\times\, \Big[\bar{u}_e(\vec{k}_e,\sigma_e)
  \gamma^{\mu} (1 - \gamma^5) v_{\nu}(\vec{k}_{\nu}, +
  \frac{1}{2})\Big],\nonumber\\
\hspace{-0.3in}&&M_{\rm Fig. \ref{fig:fig7}f} (n \to p e^- \bar{\nu}_e
 \gamma)_{\lambda} = eG_V\,(- 2g^2_{\pi N})\,\Big[\bar{u}_p(\vec{k}_p, \sigma_p)
  \gamma_{\mu} (1 - \gamma^5)\,\frac{1}{m_N - \hat{k}_n + \hat{k} -
    i0}\nonumber\\
\hspace{-0.3in}&&\times\int \frac{d^4p}{(2\pi)^4
  i}\,\gamma^5\,\frac{1}{m_N - \hat{k}_n - \hat{p}+ \hat{k} -
  i0}\,\hat{\varepsilon}^*_{\lambda}(k)\,\frac{1}{m_N - \hat{k}_n -
  \hat{p} - i0}\,\gamma^5\,\frac{1}{m^2_{\pi} - p^2 - i
  0}\,u_n(\vec{k}_n, \sigma_n)\Big] \nonumber\\
\hspace{-0.3in}&&\times\, \Big[\bar{u}_e(\vec{k}_e,\sigma_e)
  \gamma^{\mu} (1 - \gamma^5) v_{\nu}(\vec{k}_{\nu}, +
  \frac{1}{2})\Big].
\end{eqnarray}
The sum of the Feynman diagrams in Fig.\ref{fig:fig7}e and
Fig.\ref{fig:fig7}f is invariant under a gauge transformation
$\varepsilon^*_{\lambda'}(k) \to \varepsilon^*_{\lambda'}(k) +
c\,k$. Indeed, summing up the amplitudes in Eq.(\ref{eq:64}) and
replacing $\varepsilon^*_{\lambda'}(k) \to k$ we get

\begin{eqnarray}\label{eq:65}
\hspace{-0.3in}&&M_{\rm Fig. \ref{fig:fig7}e} (n \to p e^- \bar{\nu}_e
\gamma)_{\lambda} + M_{\rm Fig. \ref{fig:fig7}f} (n \to p e^-
\bar{\nu}_e \gamma)_{\lambda}\Big|_{\varepsilon^*_{\lambda}(k) \to k}
= e G_V(- 2 g^2_{\pi N})\Big[\bar{u}_p(\vec{k}_p, \sigma_p)
  \gamma_{\mu} (1 - \gamma^5)\,\frac{1}{m_N - \hat{k}_n + \hat{k} -
    i0}\nonumber\\
\hspace{-0.3in}&&\times\Big\{\int
\frac{d^4p}{(2\pi)^4 i}\,\gamma^5\,\frac{1}{m_N - \hat{k}_n - \hat{p} -
    i0}\,\gamma^5\,\frac{1}{m^2_{\pi} - (p + k)^2 - i0} - \int
  \frac{d^4p}{(2\pi)^4 i}\,\gamma^5\,\frac{1}{m_N - \hat{k}_n - \hat{p} -
      i0}\,\gamma^5\,\frac{1}{m^2_{\pi} - p^2 - i0}\nonumber\\
\hspace{-0.3in}&&+ \int \frac{d^4p}{(2\pi)^4
  i}\,\gamma^5\,\frac{1}{m_N - \hat{k}_n - \hat{p} -
  i0}\,\gamma^5\,\frac{1}{m^2_{\pi} - p^2 - i0} - \int
\frac{d^4p}{(2\pi)^4 i}\,\gamma^5\,\frac{1}{m_N - \hat{k}_n - \hat{p}
  + \hat{k} - i0}\,\gamma^5\,\frac{1}{m^2_{\pi} - p^2 -
  i0}\Big\}\nonumber\\
\hspace{-0.3in}&&\times\, \Big[\bar{u}_e(\vec{k}_e,\sigma_e)
  \gamma^{\mu} (1 - \gamma^5) v_{\nu}(\vec{k}_{\nu}, +
  \frac{1}{2})\Big].
\end{eqnarray}
Making a shift of variables $p \to p + k$ in the last integral in
curly brackets, which is allowed within dimensional regularization of
divergent integrals \cite{Itzykson1980,Sirlin1975a}, one may see that
the r.h.s. of Eq.(\ref{eq:65}) vanishes. This confirms invariance of
the sum of the Feynman diagrams in Fig.\,\ref{fig:fig7}e and
Fig.\,\ref{fig:fig7}f under a gauge transformation
$\varepsilon^*_{\lambda}(k) \to \varepsilon^*_{\lambda}(k) +
c\,k$. Using dimensional regularization of the momentum integrals one
may show that the sum of the amplitude Eq.(\ref{eq:64}) vanishes in
the accepted approximation, i.e. to leading order in the large nucleon
mass expansion. Thus, the Feynman diagrams in Fig.\,\ref{fig:fig7}a -
Fig.\,\ref{fig:fig7}f, taking into account in the L$\sigma$M the
contribution of hadronic structure of the neutron to the neutron
radiative $\beta^-$--decay, are gauge invariant, but vanish after
renormalisation and do not contribute to the amplitude of the neutron
radiative $\beta^-$--decay.

The contribution of hadronic structure of the proton is described by
the Feynman diagrams in Fig.\,\ref{fig:fig7}g - Fig.\,\ref{fig:fig7}o,
where the Feynman diagram in Fig.\,\ref{fig:fig7}o is caused by the
counter--term of renormalization of the proton--proton--photon vertex
by strong low--energy interactions. The analytical expressions for the
Feynman diagrams Fig.\,\ref{fig:fig7}g - Fig.\,\ref{fig:fig7}$\ell$
are given by
\begin{eqnarray}\label{eq:66}
\hspace{-0.3in}&&M_{\rm Fig. \ref{fig:fig7}g} (n \to p e^- \bar{\nu}_e
\gamma)_{\lambda} + M_{\rm Fig. \ref{fig:fig7}h} (n \to p e^-
\bar{\nu}_e \gamma)_{\lambda} =\nonumber\\
\hspace{-0.3in}&&=  eG_V\Big[\bar{u}_p(\vec{k}_p,
    \sigma_p) \Sigma_p(k_p)\,\frac{1}{m_N - \hat{k}_p -
      i0}\,\hat{\varepsilon}^*_{\lambda}(k)\,\frac{1}{m_N - \hat{k}_p
      - \hat{k} - i0}\,\gamma^{\mu}(1 - \gamma^5)\,u_n(\vec{k}_n,
    \sigma_n)\Big]\nonumber\\
\hspace{-0.3in}&&\times\, \Big[\bar{u}_e(\vec{k}_e,\sigma_e)
  \gamma_{\mu} (1 - \gamma^5) v_{\nu}(\vec{k}_{\nu}, +
  \frac{1}{2})\Big],\nonumber\\
\hspace{-0.3in}&&M_{\rm Fig. \ref{fig:fig7}i} (n \to p e^- \bar{\nu}_e
\gamma)_{\lambda} + M_{\rm Fig. \ref{fig:fig7}j} (n \to p e^-
\bar{\nu}_e \gamma)_{\lambda} = \nonumber\\
\hspace{-0.3in}&&=  eG_V\Big[\bar{u}_p(\vec{k}_p,
  \sigma_p)\,\hat{\varepsilon}^*_{\lambda}(k)\,\frac{1}{m_N -
    \hat{k}_p - \hat{k} - i0}\,\Sigma_p(k_p + k)\,\frac{1}{m_N -
    \hat{k}_p - \hat{k} - i0}\,\gamma^{\mu}(1 -
  \gamma^5)\,u_n(\vec{k}_n, \sigma_n)\Big]\nonumber\\
\hspace{-0.3in}&&\times\, \Big[\bar{u}_e(\vec{k}_e,\sigma_e)
  \gamma^{\mu} (1 - \gamma^5) v_{\nu}(\vec{k}_{\nu}, +
  \frac{1}{2})\Big],\nonumber\\
\hspace{-0.3in}&&M_{\rm Fig. \ref{fig:fig7}k} (n \to p e^- \bar{\nu}_e
\gamma)_{\lambda} + M_{\rm Fig. \ref{fig:fig7}\ell} (n \to p e^-
\bar{\nu}_e \gamma)_{\lambda} = - eG_V \Big[\bar{u}_p(\vec{k}_p,
  \sigma_p) \Sigma_p(k_p)\,\frac{1}{m_N - \hat{k}_p -
    i0}\,\gamma^{\mu}(1 - \gamma^5)\,u_n(\vec{k}_n,
  \sigma_n)\Big]\nonumber\\
\hspace{-0.3in}&&\times \Big[\bar{u}_e(\vec{k}_e,\sigma_e)\,
  \hat{\varepsilon}^*_{\lambda}(k)\, \frac{1}{m_e - \hat{k}_e -
    \hat{k} - i0}\, \gamma_{\mu} (1 - \gamma^5) v_{\nu}(\vec{k}_{\nu},
  + \frac{1}{2})\Big].
\end{eqnarray}
The analytical expressions for the Feynman diagrams in
Fig.\,\ref{fig:fig7}m and Fig.\,\ref{fig:fig7}n we give in the
following form
\begin{eqnarray}\label{eq:67}
\hspace{-0.3in}M_{\rm Fig. \ref{fig:fig7}m + Fig. \ref{fig:fig7}n}
(n \to p e^- \bar{\nu}_e \gamma)_{\lambda} &=& eG_V
\Big[\bar{u}_p(\vec{k}_p, \sigma_p)\,\varepsilon^*_{\lambda}(k)\cdot
  \Lambda_p(k_p, k)\, \frac{1}{m_N - \hat{k}_p - \hat{k} -
    i0}\,\gamma^{\mu}(1 - \gamma^5) \,u_n(\vec{k}_n,
  \sigma_n)\Big]\nonumber\\
\hspace{-0.3in}&&\times 
\Big[\bar{u}_e(\vec{k}_e,\sigma_e)\, \gamma_{\mu} (1 - \gamma^5)
  v_{\nu}(\vec{k}_{\nu}, + \frac{1}{2})\Big],
\end{eqnarray}
where $\Lambda^{\alpha}_p(k_p, k)$ is the vertex function, defined by
the momentum integrals
\begin{eqnarray}\label{eq:68}
\Lambda^{\alpha}_p(k_p, k) &=& 2g^2_{\pi
  N}\int\frac{d^4p}{(2\pi)^4i}\,\gamma^5\,\frac{1}{m_N - \hat{k}_p -
  \hat{k} - \hat{p} - i 0}\,\gamma^5\,\frac{(2p +
  k)^{\alpha}}{m^2_{\pi} - (p + k)^2 - i0}\,\frac{1}{m^2_{\pi} - p^2 -
  i 0}\nonumber\\
\hspace{-0.3in}&-& g^2_{\pi
  N}\int\frac{d^4p}{(2\pi)^4i}\,\gamma^5\,\frac{1}{m_N - \hat{k}_p -
  \hat{p} - i 0}\,\gamma^{\alpha}\,\frac{1}{m_N - \hat{k}_p - \hat{k}
  - \hat{p} - i 0}\,\gamma^5\,\frac{1}{m^2_{\pi} - p^2 - i0}\nonumber\\
\hspace{-0.3in}&+& g^2_{\pi N}
\int\frac{d^4p}{(2\pi)^4i}\,\frac{1}{m_N - \hat{k}_p - \hat{p} - i
  0}\,\gamma^{\alpha}\,\frac{1}{m_N - \hat{k}_p -
  \hat{k} - \hat{p} - i 0}\,\frac{1}{m^2_{\sigma} - p^2 - i0}.
\end{eqnarray}
The self--energy corrections $\Sigma_p(k_p)$ and $\Sigma_p(k_p + k)$
are defined by Eq.(\ref{eq:41}) with a replacement $k_n \to k_p$ and
$k_n \to k_p + k$, respectively. The Feynman diagram in
Fig.\,\ref{fig:fig7}o has the following analytical expression
\begin{eqnarray}\label{eq:69}
\hspace{-0.3in}&&M_{\rm Fig. \ref{fig:fig7}o} (n \to p e^- \bar{\nu}_e
\gamma)_{\lambda} = eG_V (Z_N Z^{(p)}_1 - 1)\nonumber\\
\hspace{-0.3in}&&\times \Big[\bar{u}_p(\vec{k}_p, \sigma_p)
  \,\hat{\varepsilon}^*_{\lambda}(k)\,\frac{1}{m_N - \hat{k}_p -
    \hat{k} - i0}\,\gamma^{\mu}(1 - \gamma^5)\,u_n(\vec{k}_n,
  \sigma_n)\Big]\Big[\bar{u}_e(\vec{k}_e,\sigma_e) \gamma_{\mu} (1 -
  \gamma^5) v_{\nu}(\vec{k}_{\nu}, + \frac{1}{2})\Big].
\end{eqnarray}
Since the proton--proton--photon vertex is not renormalized by
electromagnetic interactions, we set $Z^{(p)}_1 = 1$. In order to show
that the Feynman diagrams in Fig.\,\ref{fig:fig7}g -
Fig.\,\ref{fig:fig7}o are gauge invariant we propose to sum up these
diagrams and replace $\varepsilon^*_{\lambda}(k) \to k$. This gives
\begin{eqnarray}\label{eq:70}
\hspace{-0.3in}&&M_{\rm Fig. \ref{fig:fig7}g + Fig. \ref{fig:fig7}h +
  Fig. \ref{fig:fig7}i + Fig. \ref{fig:fig7}j + Fig. \ref{fig:fig7}k +
  Fig. \ref{fig:fig7}\ell + Fig. \ref{fig:fig7}m +
  Fig. \ref{fig:fig7}n + Fig. \ref{fig:fig7}o} (n \to p e^-
\bar{\nu}_e \gamma)_{\lambda}\Big|_{\varepsilon^*_{\lambda}(k) \to k}
= eG_V\nonumber\\
\hspace{-0.3in}&&\times\Big\{\Big[\bar{u}_p(\vec{k}_p,
  \sigma_p)\,\Sigma_p(k_p)\,\frac{1}{m_N - \hat{k}_p - \hat{k} -
    i0}\,\gamma^{\mu}(1 - \gamma^5) \,u_n(\vec{k}_n, \sigma_n)\Big]
\Big[\bar{u}_e(\vec{k}_e,\sigma_e)\, \gamma_{\mu} (1 - \gamma^5)
  v_{\nu}(\vec{k}_{\nu}, + \frac{1}{2})\Big]\nonumber\\
\hspace{-0.3in}&&- \Big[\bar{u}_p(\vec{k}_p,
\sigma_p)\,\Sigma_p(k_p)\,\frac{1}{m_N - \hat{k}_p -
  i0}\,\gamma^{\mu}(1 - \gamma^5) \,u_n(\vec{k}_n, \sigma_n)\Big]
\Big[\bar{u}_e(\vec{k}_e,\sigma_e)\, \gamma_{\mu} (1 - \gamma^5)
  v_{\nu}(\vec{k}_{\nu}, + \frac{1}{2})\Big]\nonumber\\
\hspace{-0.3in}&&- \Big[\bar{u}_p(\vec{k}_p, \sigma_p)\,\Sigma_p(k_p +
  k)\,\frac{1}{m_N - \hat{k}_p - \hat{k} - i0}\,\gamma^{\mu}(1 -
  \gamma^5) \,u_n(\vec{k}_n, \sigma_n)\Big]
\Big[\bar{u}_e(\vec{k}_e,\sigma_e)\, \gamma_{\mu} (1 - \gamma^5)
  v_{\nu}(\vec{k}_{\nu}, + \frac{1}{2})\Big]\nonumber\\
\hspace{-0.3in}&&+ \Big[\bar{u}_p(\vec{k}_p,
  \sigma_p)\,\Sigma_p(k_p)\,\frac{1}{m_N - \hat{k}_p -
    i0}\,\gamma^{\mu}(1 - \gamma^5) \,u_n(\vec{k}_n, \sigma_n)\Big]
\Big[\bar{u}_e(\vec{k}_e,\sigma_e)\, \gamma_{\mu} (1 - \gamma^5)
  v_{\nu}(\vec{k}_{\nu}, + \frac{1}{2})\Big]\nonumber\\
\hspace{-0.3in}&& + \Big[\bar{u}_p(\vec{k}_p, \sigma_p)\,k \cdot
  \Lambda_p(k_p, k)\, \frac{1}{m_N - \hat{k}_p - \hat{k} -
    i0}\,\gamma^{\mu}(1 - \gamma^5) \,u_n(\vec{k}_n, \sigma_n)\Big]
\Big[\bar{u}_e(\vec{k}_e,\sigma_e)\, \gamma_{\mu} (1 - \gamma^5)
  v_{\nu}(\vec{k}_{\nu}, + \frac{1}{2})\Big]\nonumber\\
\hspace{-0.3in}&& + (Z_N -
1)\Big[\bar{u}_p(\vec{k}_p, \sigma_p)\,\hat{k}\,\frac{1}{m_N -
    \hat{k}_p - \hat{k} - i0}\,\gamma^{\mu}(1 - \gamma^5)
  \,u_n(\vec{k}_n, \sigma_n)\Big] \Big[\bar{u}_e(\vec{k}_e,\sigma_e)\,
  \gamma_{\mu} (1 - \gamma^5) v_{\nu}(\vec{k}_{\nu}, +
  \frac{1}{2})\Big].
\end{eqnarray}
The r.h.s. of Eq.(\ref{eq:70}) we transcribe into the form
\begin{eqnarray}\label{eq:71}
\hspace{-0.3in}&&M_{\rm Fig. \ref{fig:fig7}g + Fig. \ref{fig:fig7}h +
  Fig. \ref{fig:fig7}i + Fig. \ref{fig:fig7}j + Fig. \ref{fig:fig7}k +
  Fig. \ref{fig:fig7}\ell + Fig. \ref{fig:fig7}m +
  Fig. \ref{fig:fig7}n + Fig. \ref{fig:fig7}o} (n \to p e^-
\bar{\nu}_e \gamma)_{\lambda}\Big|_{\varepsilon^*_{\lambda}(k) \to k}
= eG_V\nonumber\\
\hspace{-0.3in}&&\times\Big\{\Big[\bar{u}_p(\vec{k}_p,
  \sigma_p)\,\Sigma'_p(k_p)\,\frac{1}{m_N - \hat{k}_p - \hat{k} -
    i0}\,\gamma^{\mu}(1 - \gamma^5) \,u_n(\vec{k}_n, \sigma_n)\Big]
\Big[\bar{u}_e(\vec{k}_e,\sigma_e)\, \gamma_{\mu} (1 - \gamma^5)
  v_{\nu}(\vec{k}_{\nu}, + \frac{1}{2})\Big]\nonumber\\
\hspace{-0.3in}&&- \Big[\bar{u}_p(\vec{k}_p, \sigma_p)\,\Sigma'_p(k_p
  + k)\,\frac{1}{m_N - \hat{k}_p - \hat{k} - i0}\,\gamma^{\mu}(1 -
  \gamma^5) \,u_n(\vec{k}_n, \sigma_n)\Big]
\Big[\bar{u}_e(\vec{k}_e,\sigma_e)\, \gamma_{\mu} (1 - \gamma^5)
  v_{\nu}(\vec{k}_{\nu}, + \frac{1}{2})\Big]\nonumber\\
\hspace{-0.3in}&& + \Big[\bar{u}_p(\vec{k}_p, \sigma_p)\,k \cdot
  \Lambda_p(k_p, k)\, \frac{1}{m_N - \hat{k}_p - \hat{k} -
    i0}\,\gamma^{\mu}(1 - \gamma^5) \,u_n(\vec{k}_n, \sigma_n)\Big]
\Big[\bar{u}_e(\vec{k}_e,\sigma_e)\, \gamma_{\mu} (1 - \gamma^5)
  v_{\nu}(\vec{k}_{\nu}, + \frac{1}{2})\Big]\nonumber\\
\hspace{-0.3in}&& + (Z_N - 1)\Big[\bar{u}_p(\vec{k}_p,
  \sigma_p)\,\hat{k}\,\frac{1}{m_N - \hat{k}_p - \hat{k} -
    i0}\,\gamma^{\mu}(1 - \gamma^5) \,u_n(\vec{k}_n, \sigma_n)\Big]
\Big[\bar{u}_e(\vec{k}_e,\sigma_e)\, \gamma_{\mu} (1 - \gamma^5)
  v_{\nu}(\vec{k}_{\nu}, + \frac{1}{2})\Big],
\end{eqnarray}
where in Eq.(\ref{eq:70}) the second term in curly brackets is
cancelled by the Feynman diagrams in Fig.\,\ref{fig:fig7}k and
Fig.\,\ref{fig:fig7}$\ell$, and in Eq.(\ref{eq:71}) the self--energy
corrections $\Sigma'_p(k_p)$ and $\Sigma'_p(k_p + k)$ have no
contributions of the tadpole Feynman diagrams, i.e the Feynman
diagrams in Fig.\,\ref{fig:fig7}g and Fig.\,\ref{fig:fig7}i. Such
tadpole contributions are mutually cancelled from the first two terms
in curly brackets in Eq.(\ref{eq:71}). For the scalar product $k\cdot
\Lambda_p(k_p, k)$ we obtain the following expression
\begin{eqnarray*}
k\cdot \Lambda^{\alpha}_p(k_p, k) &=& 2g^2_{\pi
  N}\int\frac{d^4p}{(2\pi)^4i}\,\gamma^5\,\frac{1}{m_N - \hat{k}_p -
  \hat{k}- \hat{p} - i 0}\,\gamma^5\,\frac{1}{m^2_{\pi} - (p + k)^2 -
  i 0}\nonumber\\
\hspace{-0.3in}&-& 2g^2_{\pi
  N}\int\frac{d^4p}{(2\pi)^4i}\,\gamma^5\,\frac{1}{m_N - \hat{k}_p -
  \hat{k}- \hat{p} - i 0}\,\gamma^5\,\frac{1}{m^2_{\pi} - p^2 - i
  0}\nonumber\\
\hspace{-0.3in}&+& g^2_{\pi
  N}\int\frac{d^4p}{(2\pi)^4i}\,\gamma^5\,\frac{1}{m_N - \hat{k}_p -
  \hat{p} - i 0}\,\gamma^5\,\frac{1}{m^2_{\pi} - p^2 - i 0}\nonumber\\
\hspace{-0.3in}&-& g^2_{\pi
  N}\int\frac{d^4p}{(2\pi)^4i}\,\gamma^5\,\frac{1}{m_N - \hat{k}_p -
  \hat{k} - \hat{p} - i 0}\,\gamma^5\,\frac{1}{m^2_{\pi} - p^2 - i
  0}\nonumber\\
\end{eqnarray*}
\begin{eqnarray}\label{eq:72}
\hspace{-0.3in}&-& g^2_{\pi
  N}\int\frac{d^4p}{(2\pi)^4i}\,\gamma^5\,\frac{1}{m_N - \hat{k}_p -
  \hat{p} - i 0}\,\gamma^5\,\frac{1}{m^2_{\sigma} - p^2 - i
  0}\nonumber\\
\hspace{-0.3in}&+& g^2_{\pi
  N}\int\frac{d^4p}{(2\pi)^4i}\,\gamma^5\,\frac{1}{m_N - \hat{k}_p -
  \hat{k}- \hat{p} - i 0}\,\gamma^5\,\frac{1}{m^2_{\sigma} - p^2 - i
  0}.
\end{eqnarray}
Making a shift of variables $p + k \to p$ in the first momentum
integral we may transcribe the r.h.s. of Eq.(\ref{eq:72}) into the
form
\begin{eqnarray}\label{eq:73}
k\cdot \Lambda^{\alpha}_p(k_p, k) &=&\big( - \delta m'_p - (Z_N -
1)\,(m_N - \hat{k}_p ) - \Sigma'_p(k_p)\big) - \big( - \delta m'_p -
(Z_N - 1)\,(m_N - \hat{k}_p - \hat{k}) - \Sigma'_p(k_p + k)\big)
\nonumber\\ &=& \Sigma'_p(k_p + k) - \Sigma'_p(k_p) - (Z_N -
1)\,\hat{k},
\end{eqnarray}
where $\delta m'_p$ is a counter--term of the proton mass with the
excluded contribution of the tadpole Feynman diagrams.  Plugging
Eq.(\ref{eq:73}) into Eq.(\ref{eq:71}) we get
\begin{eqnarray}\label{eq:74}
\hspace{-0.3in}&&M_{\rm Fig. \ref{fig:fig7}g + Fig. \ref{fig:fig7}h +
  Fig. \ref{fig:fig7}i + Fig. \ref{fig:fig7}j + Fig. \ref{fig:fig7}k +
  Fig. \ref{fig:fig7}\ell + Fig. \ref{fig:fig7}m +
  Fig. \ref{fig:fig7}n + Fig. \ref{fig:fig7}o} (n \to p e^- \bar{\nu}_e
\gamma)_{\lambda}\Big|_{\varepsilon^*_{\lambda}(k) \to k} = 0.
\end{eqnarray}
This confirms invariance of the Feynman diagrams in
Fig.\,\ref{fig:fig7}g - Fig.\,\ref{fig:fig7}o under a gauge
transformation $\varepsilon^*_{\lambda}(k) \to
\varepsilon^*_{\lambda}(k) + c\,k$. The relation Eq.(\ref{eq:73}) is
the Ward identity for the proton--proton--photon vertex renormalized
by strong low--energy interactions. The direct calculation of the
momentum integrals gives one
\begin{eqnarray}\label{eq:75}
\hspace{-0.3in}\Sigma'_p(k_p) &=& - \delta m'_p - (Z_N - 1)\,(m_N -
\hat{k}_p ) - m_N\, \frac{g^2_{\pi N}}{32\pi^2}\,\Big(3\,{\ell
  n}\frac{m^2_{\sigma}}{m^2_N} + \frac{1}{2}\Big)\nonumber\\
\hspace{-0.3in}&&- (m_N - \hat{k}_p)\,\frac{g^2_{\pi
    N}}{32\pi^2}\,\Big(4\,{\ell n}\frac{\Lambda^2}{m^2_N} - {\ell
  n}\frac{m^2_{\sigma}}{m^2_N} + \frac{11}{2}\Big).
\end{eqnarray}
The self--energy correction $\Sigma'_p(k_p)$ vanishes after
renormalization \cite{Ivanov2017b}
\begin{eqnarray}\label{eq:76}
\hspace{-0.3in}\delta m'_p &=& - m_N \,\frac{g^2_{\pi
    N}}{32\pi^2}\,\Big(3\,{\ell n}\frac{m^2_{\sigma}}{m^2_N} +
\frac{1}{2}\Big),\nonumber\\
\hspace{-0.3in} Z_N &=&1 - \frac{g^2_{\pi
    N}}{32\pi^2}\,\Big(4\,{\ell n}\frac{\Lambda^2}{m^2_N} - {\ell
    n}\frac{m^2_{\sigma}}{m^2_N} + \frac{11}{2}\Big).
\end{eqnarray}
Denoting $\bar{\Sigma}_p(k_p)$ as a renormalized self--energy
correction $\Sigma_p(k_p)$ we get $\bar{\Sigma}_p(k_p) = 0$
\cite{Ivanov2017b}.  For the self--energy correction $\Sigma'_p(k_p +
k)$ we obtain the following expression
\begin{eqnarray}\label{eq:77}
\Sigma'_p(k_p + k) &=& - \delta m'_p - (Z_N - 1)\,(m_N - \hat{k}_p -
\hat{k}) - m_N\,\frac{g^2_{\pi N}}{32\pi^2}\,\Big(3\,{\ell
  n}\frac{m^2_{\sigma}}{m^2_N} + \frac{1}{2}\Big)\nonumber\\
\hspace{-0.3in}&& - (m_N - \hat{k}_p - \hat{k})\,\frac{g^2_{\pi
    N}}{32\pi^2}\,\Big(4\,{\ell n}\frac{\Lambda^2}{m^2_N} - {\ell
  n}\frac{m^2_{\sigma}}{m^2_N} + \frac{11}{2}\Big)\nonumber\\ && +
\frac{3g^2_{\pi N}}{16\pi^2}\,\Big[m_N \,F_1\Big( \frac{2k\cdot
    k_p}{m^2_N}\Big) + (m_N - \hat{k}_p - \hat{k})\,F_2\Big(
  \frac{2k\cdot k_p}{m^2_N}\Big)\Big].
\end{eqnarray}
After renormalization the self--energy correction $\Sigma'_p(k_p + k)$
takes the form
\begin{eqnarray}\label{eq:78}
\bar{\Sigma}_p(k_p + k) = \frac{3g^2_{\pi N}}{16\pi^2}\,\Big[m_N
  \,F_1\Big( \frac{2k\cdot k_p}{m^2_N}\Big) + (m_N - \hat{k}_p -
  \hat{k})\,F_2\Big( \frac{2k\cdot k_p}{m^2_N}\Big)\Big].
\end{eqnarray}
The functions $F_1(z)$ and $F_2(z)$, where $z = 2k\cdot k_p/m^2_N$,
are defined by
\begin{eqnarray}\label{eq:79}
F_1(z) &=& \int^1_0dx\,x\,{\ell n}\Big(1 - z\,\frac{1-x}{x}\Big) =
\frac{1}{2}\,\frac{z}{(1 + z)^2}\big(-1 - z + z\,{\ell
  n}(-z)\big),\nonumber\\ F_2(z) &=& \int^1_0dx\,(1- x)\,{\ell
  n}\Big(1 - z\,\frac{1-x}{x}\Big) = \frac{z}{1 + z}\,{\ell n}(-z) -
F_1(z) = \frac{1}{2}\,\frac{z}{(1 + z)^2}\big(1 + z +(2 + z)\,{\ell
  n}(-z)\big).
\end{eqnarray}
In the limit $m_{\sigma} \to \infty$ and after renormalization the
vertex function $\Lambda^{\alpha}_p(k_p, k)$ takes the form
\begin{eqnarray}\label{eq:80}
\bar{\Lambda}^{\alpha}_p(k_p, k) &=& \frac{g^2_{\pi
    N}}{8\pi^2}\,\Big\{\gamma^{\alpha}\Big[\frac{1}{4} +
  \frac{2}{z}\,F_1(z) + \Big(- \frac{3}{2} + \frac{1}{2z}\Big)\,F_2(z)
  - \frac{1}{2z}\,\Big({\ell n}(1 + z)\,{\ell n}(-z) + {\rm
    Li}_2(-z)\Big)\Big]\nonumber\\ &+& \frac{k^{\alpha}_p
  \hat{k}}{m^2_N}\,\Big[ - \frac{1}{2z} - \frac{4}{z^2}\,F_1(z) -
  \frac{1}{z^2}\,F_2(z) + \frac{1}{z^2}\,\Big({\ell n}(1 + z)\,{\ell
    n}(-z) + {\rm Li}_2(-z)\Big)\Big] +
\frac{k^{\alpha}_p}{m_N}\,\Big[
  \frac{3}{z}\,F_1(z)\Big]\nonumber\\ &+&
\frac{i\sigma^{\alpha\beta}k_{\beta}}{2m_N}\,
\Big[\frac{1}{z}\,\Big(F_1(z) + F_2(z)\Big) - \frac{1}{2z}\,\Big({\ell
    n}(1 + z)\,{\ell n}(-z) + {\rm Li}_2(-z)\Big)\Big]\Big\},
\end{eqnarray}
where ${\rm Li}_2(-z)$ is the Polylogarithmic function. The
renormalized vertex function Eq.(\ref{eq:80}) is fully defined by the
Feynman diagrams with virtual $\pi$--meson exchanges. The contribution
of the Feynman diagrams with the $\sigma$--meson exchanges is absorbed
in the limit $m _{\sigma} \to \infty$ by the counter--term only.

The renormalized amplitude of the neutron radiative $\beta^-$--decay,
caused by the contributions of the Feynman diagrams in
Fig.\,\ref{fig:fig7}, takes the form
\begin{eqnarray}\label{eq:81}
M_{\rm Fig. \ref{fig:fig7}} (n \to p e^- \bar{\nu}_e \gamma)_{\lambda}
&=& eG_V \Big\{\Big[\bar{u}_p(\vec{k}_p,
  \sigma_p)\,\varepsilon^*_{\lambda}(k)\cdot \bar{\Lambda}_p(k_p,
  k)\,\frac{1}{m_N - \hat{k}_p - \hat{k} - i0}\,\gamma^{\mu}(1 -
  \gamma^5)\,u_n(\vec{k}_n, \sigma_n)\Big]\nonumber\\
\hspace{-0.3in}&+& \Big[\bar{u}_p(\vec{k}_p,
  \sigma_p)\,\hat{\varepsilon}^*_{\lambda}(k)\,\frac{1}{m_N -
    \hat{k}_p - \hat{k} - i0}\,\Sigma_p(k_p + k)\,\frac{1}{m_N -
    \hat{k}_p - \hat{k} - i0}\,\gamma^{\mu}(1 -
  \gamma^5)\,u_n(\vec{k}_n, \sigma_n)\Big]\Big\}\nonumber\\
\hspace{-0.3in}&&\times\, \Big[\bar{u}_e(\vec{k}_e,\sigma_e)
  \gamma^{\mu} (1 - \gamma^5) v_{\nu}(\vec{k}_{\nu}, +
  \frac{1}{2})\Big].
\end{eqnarray}
Making a replacement $\varepsilon^*_{\lambda}(k) \to k$ we arrive at
the amplitude
\begin{eqnarray}\label{eq:82}
&&M_{\rm Fig. \ref{fig:fig7}} (n \to p e^- \bar{\nu}_e
  \gamma)_{\lambda}\Big|_{\varepsilon^*_{\lambda}(k) \to k}=
  eG_V\,\Big[\bar{u}_e(\vec{k}_e,\sigma_e)
  \gamma^{\mu} (1 - \gamma^5) v_{\nu}(\vec{k}_{\nu}, +
  \frac{1}{2})\Big]\nonumber\\ &&\times \,\Big[\bar{u}_p(\vec{k}_p,
    \sigma_p)\,\Big(k\cdot \bar{\Lambda}_p(k_p, k) -
    \bar{\Sigma}_p(k_p + k)\Big)\,\frac{1}{m_N - \hat{k}_p - \hat{k} -
      i0}\,\gamma^{\mu}(1 - \gamma^5)\,u_n(\vec{k}_n,
    \sigma_n)\Big],
\end{eqnarray}
which vanishes because of the Ward identity \cite{Ivanov2017b}
\begin{eqnarray}\label{eq:83}
\bar{u}_p(\vec{k}_p, \sigma_p)\,\Big(k\cdot \bar{\Lambda}_p(k_p, k) -
\bar{\Sigma}_p(k_p + k)\Big) = 0.
\end{eqnarray}
Using Eq.(\ref{eq:80}) and Eq.(\ref{eq:78}) the relation
Eq.(\ref{eq:83}) can be verified by a direct calculation.

\begin{figure}
\centering \includegraphics[height=0.32\textheight]{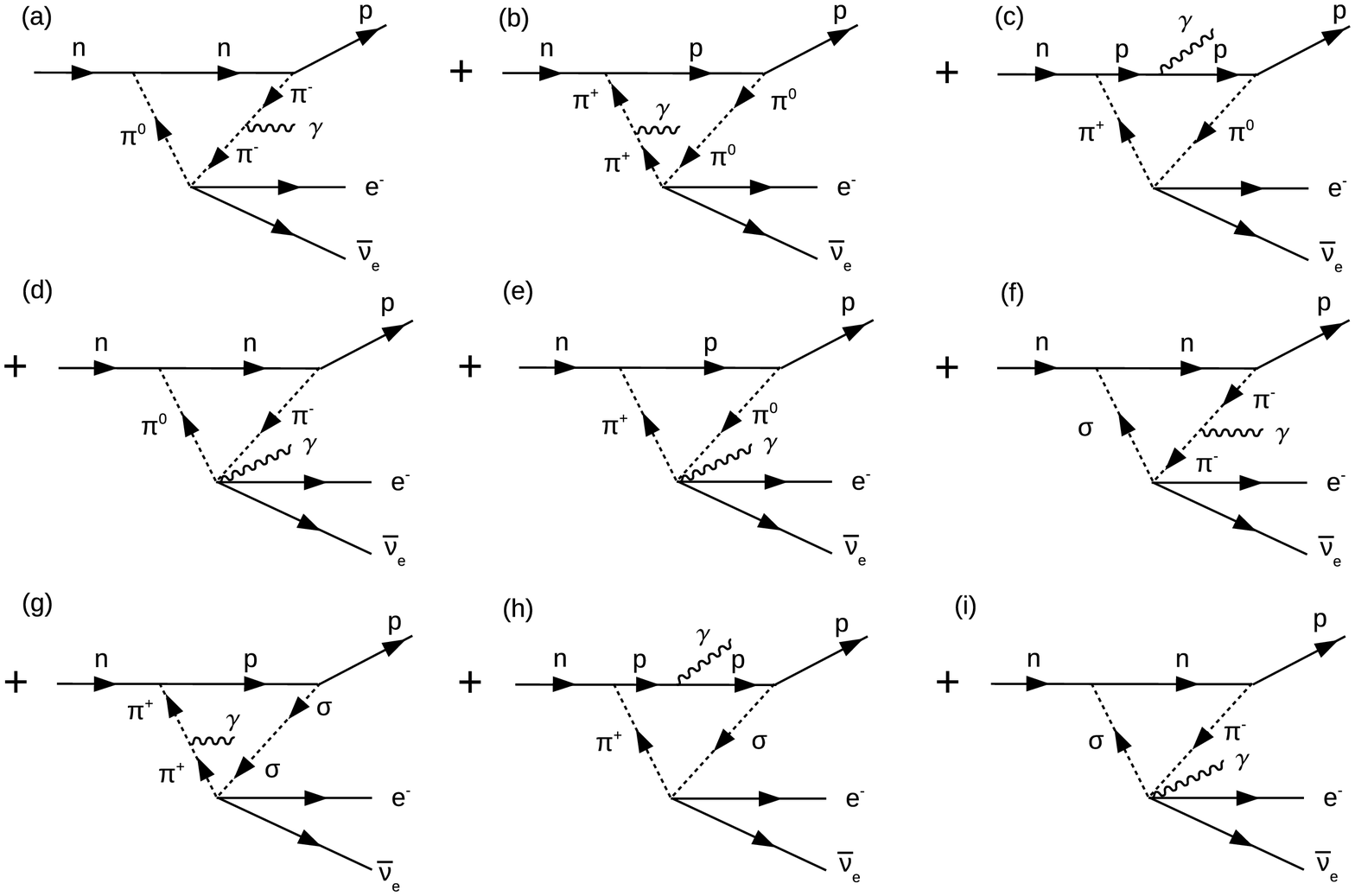}
\includegraphics[height=0.32\textheight]{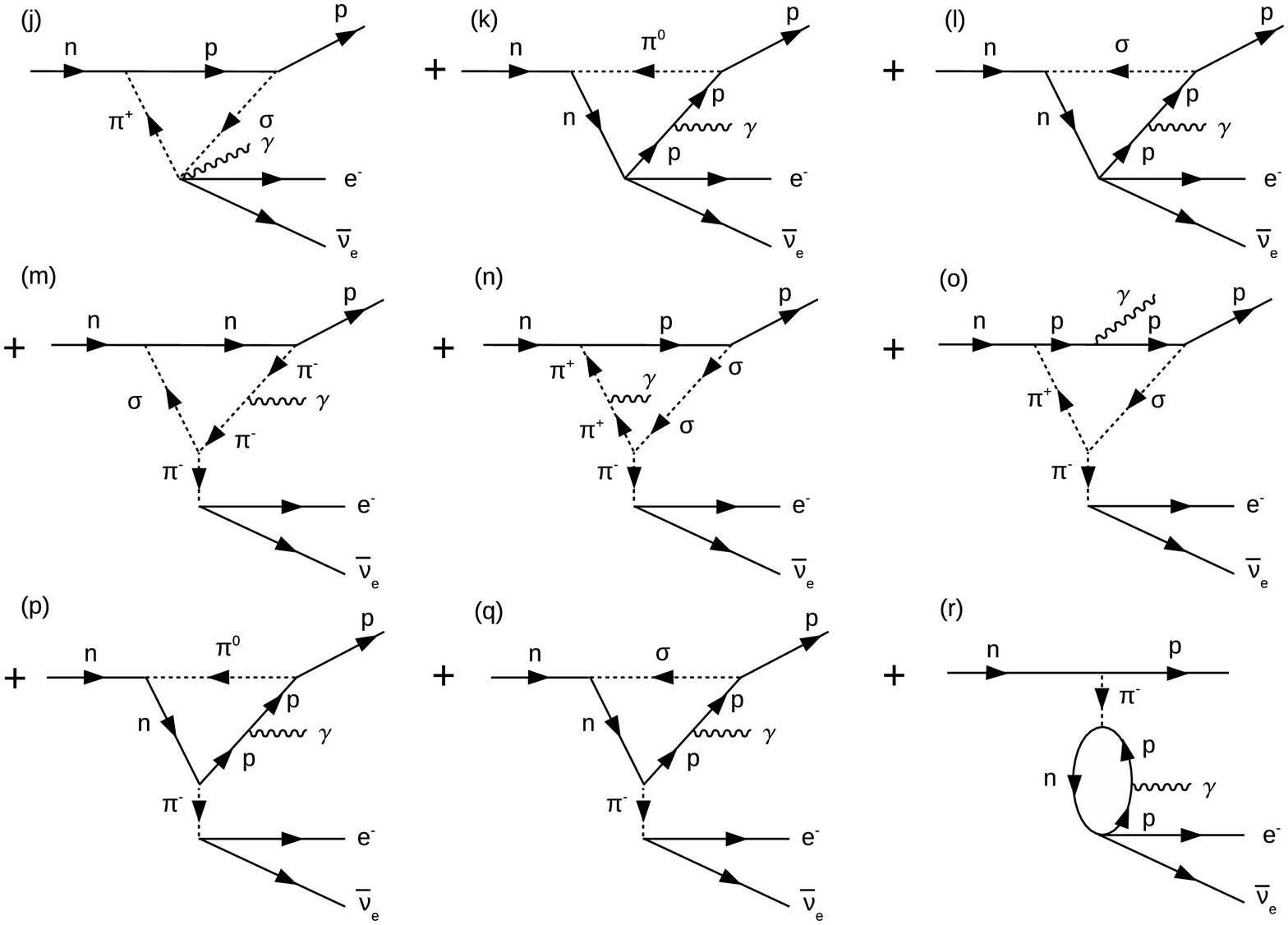}
\includegraphics[height=0.115\textheight]{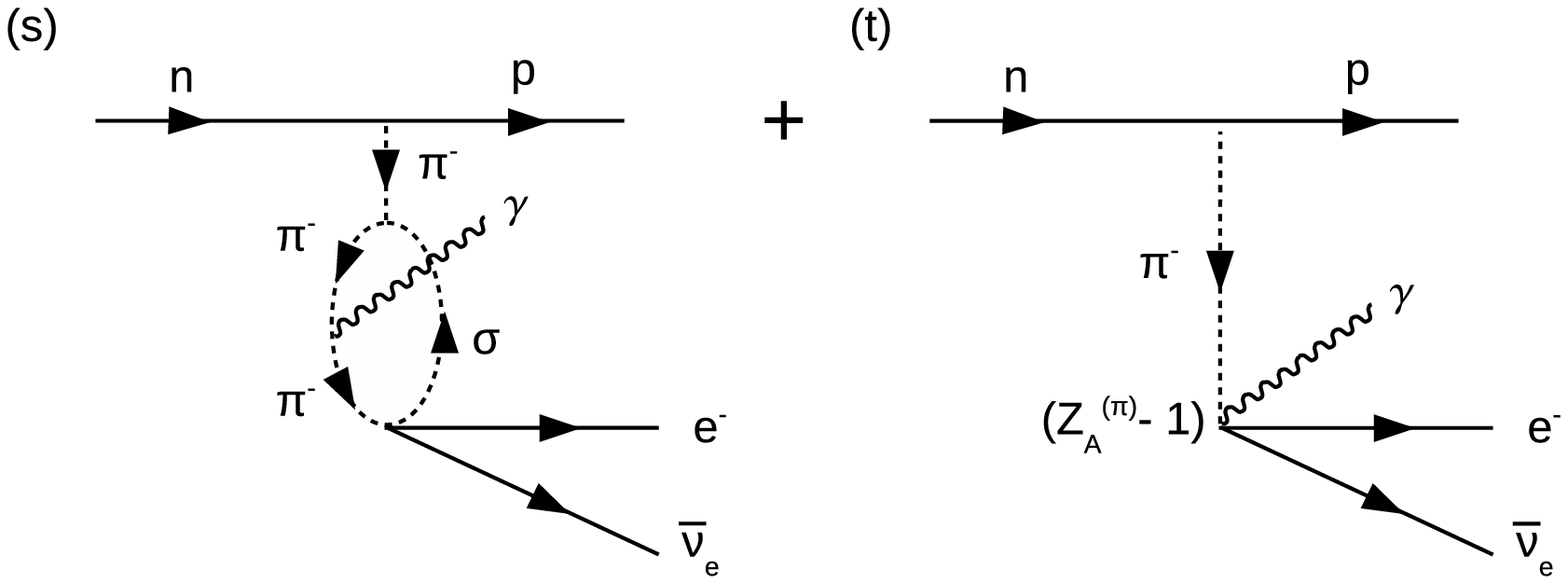}
  \caption{The Feynman diagrams, defining contributions to the
    amplitude of the neutron radiative $\beta-$--decay in the
    one--hadron--loop approximation for strong low--energy
    interactions in the L$\sigma$M and QED, where a photon is emitted
    by virtual charged hadrons in the one--hadron--loops.}
\label{fig:fig8}
\end{figure}

\subsubsection*{\bf 3. The contribution to the amplitude of the neutron 
radiative $\beta^-$--decay caused by the Feynman diagrams in
Fig.\ref{fig:fig8}}

The set of the Feynman diagrams in Fig.\,\ref{fig:fig8} can be
obtained from the Feynman diagrams in Fig.\,\ref{fig:fig3} by emitting
a photon from all lines of virtual charged hadrons.
In the limit of the infinite mass of the $\sigma$--meson $m_{\sigma}
\to \infty$ non--trivial contributions are given by the Feynman
diagrams without virtual $\sigma$--meson exchanges and the Feynman
diagrams in Fig.\,\ref{fig:fig8}$\ell$, Fig.\,\ref{fig:fig8}m,
Fig.\,\ref{fig:fig8}n, Fig.\,\ref{fig:fig8}o and
Fig.\,\ref{fig:fig8}s. The contributions of these diagrams with
virtual $\sigma$--meson exchanges do not vanish in the limit
$m_{\sigma} \to \infty$ because of the $\sigma \pi^+\pi^-$ coupling
constant equal to $\gamma f_{\pi} = (m^2_{\sigma} -
m^2_{\pi})/f_{\pi}$.

The analytical expressions of the Feynman diagrams in
Fig.\,\ref{fig:fig8}a - Fig.\,\ref{fig:fig8}e, caused by the mesonic
part of the charged vector hadronic current, are given by
\begin{eqnarray}\label{eq:84}
\hspace{-0.3in}&&M_{\rm Fig.\ref{fig:fig8}a}(n\to
p\,e^-\,\bar{\nu}_e\gamma)_{\lambda} = eG_V (- 2g^2_{\pi N})
\Big[\bar{u}_p(\vec{k}_p, \sigma_p) \int
  \frac{d^4p}{(2\pi)^4i}\,\gamma^5\,\frac{1}{m_N - \hat{k}_p - \hat{k}
    - \hat{p} - i0}\,\gamma^5\,\frac{(2p + k)\cdot
    \varepsilon^*_{\lambda}(k)}{m^2_{\pi} - (p + k)^2 - i0}\nonumber\\
\hspace{-0.3in}&&\times\,\frac{(2p + q)^{\mu}}{m^2_{\pi} - (p + q)^2 -
  i0}\,\frac{1}{m^2_{\pi} - p^2 - i0}\,u_n(\vec{k}_n, \sigma_n)\Big]
\Big[\bar{u}_e(\vec{k}_e,\sigma_e) \gamma_{\mu} (1 - \gamma^5)
  v_{\nu}(\vec{k}_{\nu}, + \frac{1}{2})\Big],\nonumber\\
\hspace{-0.3in}&&M_{\rm Fig.\ref{fig:fig8}b}(n\to
p\,e^-\,\bar{\nu}_e\gamma)_{\lambda} = eG_V (+ 2g^2_{\pi N})
\Big[\bar{u}_p(\vec{k}_p, \sigma_p) \int
  \frac{d^4p}{(2\pi)^4i}\,\gamma^5\,\frac{1}{m_N - \hat{k}_p - \hat{p}
    - i0}\,\gamma^5\,\frac{(2(p + q) - k)\cdot
    \varepsilon^*_{\lambda}(k)}{m^2_{\pi} - (p + q - k)^2 -
    i0}\nonumber\\
\hspace{-0.3in}&&\times\,\frac{(2p + q)^{\mu}}{m^2_{\pi} - (p + q)^2 -
  i0}\,\frac{1}{m^2_{\pi} - p^2 - i0}\,u_n(\vec{k}_n, \sigma_n)\Big]
\Big[\bar{u}_e(\vec{k}_e,\sigma_e) \gamma_{\mu} (1 - \gamma^5)
  v_{\nu}(\vec{k}_{\nu}, + \frac{1}{2})\Big],\nonumber\\
\hspace{-0.3in}&&M_{\rm Fig.\ref{fig:fig8}c}(n\to
p\,e^-\,\bar{\nu}_e\gamma)_{\lambda} = eG_V (- 2g^2_{\pi N})
\Big[\bar{u}_p(\vec{k}_p, \sigma_p) \int
  \frac{d^4p}{(2\pi)^4i}\,\gamma^5\,\frac{1}{m_N - \hat{k}_p - \hat{p}
    - i0}\,\hat{\varepsilon}^*_{\lambda}(k)\,\frac{1}{m_N - \hat{k}_p
    - \hat{k} - \hat{p} - i0}\,\gamma^5 \nonumber\\
\hspace{-0.3in}&&\times\,\frac{(2p + q)^{\mu}}{m^2_{\pi} - (p + q)^2 -
  i0}\,\frac{1}{m^2_{\pi} - p^2 - i0}\,u_n(\vec{k}_n, \sigma_n)\Big]
\Big[\bar{u}_e(\vec{k}_e,\sigma_e) \gamma_{\mu} (1 - \gamma^5)
  v_{\nu}(\vec{k}_{\nu}, + \frac{1}{2})\Big],\nonumber\\
\hspace{-0.3in}&&M_{\rm Fig.\ref{fig:fig8}d}(n\to
p\,e^-\,\bar{\nu}_e\gamma)_{\lambda} = eG_V (+ 2g^2_{\pi N})
\Big[\bar{u}_p(\vec{k}_p, \sigma_p) \int
  \frac{d^4p}{(2\pi)^4i}\,\gamma^5\,\frac{1}{m_N - \hat{k}_p - \hat{k}
    - \hat{p} - i0}\,\gamma^5\,\frac{1}{m^2_{\pi} - (p + q)^2 - i0}
  \nonumber\\
\hspace{-0.3in}&&\times\,\frac{1}{m^2_{\pi} - (p + k)^2 - i0}\,u_n(\vec{k}_n, \sigma_n)\Big]
\Big[\bar{u}_e(\vec{k}_e,\sigma_e) \hat{\varepsilon}^*_{\lambda}(k) (1
  - \gamma^5) v_{\nu}(\vec{k}_{\nu}, + \frac{1}{2})\Big],\nonumber\\
\hspace{-0.3in}&&\nonumber\\
\hspace{-0.3in}&&M_{\rm Fig.\ref{fig:fig8}e}(n\to
p\,e^-\,\bar{\nu}_e\gamma)_{\lambda} = eG_V (+ 2g^2_{\pi N})
\Big[\bar{u}_p(\vec{k}_p, \sigma_p) \int
  \frac{d^4p}{(2\pi)^4i}\,\gamma^5\,\frac{1}{m_N - \hat{k}_p - \hat{k}
    - \hat{p} - i0}\,\gamma^5\,\frac{1}{m^2_{\pi} - (p + q)^2 - i0}
  \nonumber\\
\hspace{-0.3in}&&\times\,\frac{1}{m^2_{\pi} - (p + k)^2 - i0}\,
u_n(\vec{k}_n, \sigma_n)\Big] \Big[\bar{u}_e(\vec{k}_e,\sigma_e)
  \hat{\varepsilon}^*_{\lambda}(k) (1 - \gamma^5)
  v_{\nu}(\vec{k}_{\nu}, + \frac{1}{2})\Big].
\end{eqnarray}
Summing up the Feynman diagrams in Fig.\,\ref{fig:fig8}a -
Fig.\,\ref{fig:fig8}e and making a replacement
$\varepsilon^*_{\lambda}(k) \to k$ we obtain
\begin{eqnarray}\label{eq:85}
\hspace{-0.3in}&&M_{\rm Fig.\ref{fig:fig8}a + Fig.\ref{fig:fig8}b +
  Fig.\ref{fig:fig8}c + Fig.\ref{fig:fig8}d + Fig.\ref{fig:fig8}e
}(n\to
p\,e^-\,\bar{\nu}_e\gamma)_{\lambda}\Big|_{\varepsilon^*_{\lambda}(k)
  \to k} = eG_V (- 2g^2_{\pi N})\Big[\bar{u}_p(\vec{k}_p,
  \sigma_p)\nonumber\\
\hspace{-0.3in}&&\times \Big\{\int
\frac{d^4p}{(2\pi)^4i}\,\gamma^5\,\frac{1}{m_N - \hat{k}_p - \hat{k} -
  \hat{p} - i0}\,\gamma^5\,\frac{1}{m^2_{\pi} - (p + k)^2 -
  i0}\,\frac{(2p + q)^{\mu}}{m^2_{\pi} - (p + q)^2 - i0}\nonumber\\
\hspace{-0.3in}&&- \int
  \frac{d^4p}{(2\pi)^4i}\,\gamma^5\,\frac{1}{m_N - \hat{k}_p - \hat{k}
    - \hat{p} - i0}\,\gamma^5\,\frac{1}{m^2_{\pi} - p^2 -
    i0}\,\frac{(2p + q)^{\mu}}{m^2_{\pi} - (p + q)^2 - i0}\nonumber\\
\hspace{-0.3in}&&+ \int \frac{d^4p}{(2\pi)^4i}\,\gamma^5\,\frac{1}{m_N
  - \hat{k}_p - \hat{p} - i0}\,\gamma^5\,\frac{(2p +
  q)^{\mu}}{m^2_{\pi} - (p + q - k)^2 - i0}\,\frac{1}{m^2_{\pi} - p^2
  - i0}\nonumber\\
\hspace{-0.3in}&&- \int \frac{d^4p}{(2\pi)^4i}\,\gamma^5\,\frac{1}{m_N
  - \hat{k}_p - \hat{p} - i0}\,\gamma^5\,\frac{(2p +
  q)^{\mu}}{m^2_{\pi} - (p + q)^2 - i0}\,\frac{1}{m^2_{\pi} - p^2 -
  i0}\nonumber\\
\hspace{-0.3in}&&- \int \frac{d^4p}{(2\pi)^4i}\,\gamma^5\,\frac{1}{m_N
  - \hat{k}_p - \hat{k} - \hat{p} - i0}\,\gamma^5\,\frac{(2p +
  q)^{\mu}}{m^2_{\pi} - (p + q)^2 - i0}\,\frac{1}{m^2_{\pi} - p^2
  - i0}\nonumber\\
\hspace{-0.3in}&&+ \int \frac{d^4p}{(2\pi)^4i}\,\gamma^5\,\frac{1}{m_N
  - \hat{k}_p - \hat{k} - \hat{p} - i0}\,\gamma^5\,\frac{(2p +
  q)^{\mu}}{m^2_{\pi} - (p + q)^2 - i0}\,\frac{1}{m^2_{\pi} - p^2
  - i0}\nonumber\\
\hspace{-0.3in}&&+ \int \frac{d^4p}{(2\pi)^4i}\,\gamma^5\,\frac{1}{m_N
  - \hat{k}_p - \hat{k} - \hat{p} -
  i0}\,\gamma^5\,\frac{2k^{\mu}}{m^2_{\pi} - (p + q)^2 -
  i0}\,\frac{1}{m^2_{\pi} - p^2 - i0}\Big\}\,u_n(\vec{k}_n,
\sigma_n)\Big]\nonumber\\
\hspace{-0.3in}&&\times \Big[\bar{u}_e(\vec{k}_e,\sigma_e)
  \gamma_{\mu} (1 - \gamma^5) v_{\nu}(\vec{k}_{\nu}, +
  \frac{1}{2})\Big].
\end{eqnarray}
The last integral in curly brackets is defined by the contributions of
the Feynman diagrams in Fig.\,\ref{fig:fig8}d and
Fig.\,\ref{fig:fig8}e. Making a change of variables $p + k \to p$ in
the first integral in curly brackets one may show that the r.h.s. of
Eq.(\ref{eq:85}) vanishes. This confirms invariance of the sum of
Feynman diagrams Fig.\,\ref{fig:fig8}a - Fig.\,\ref{fig:fig8}e under a
gauge transformation $\varepsilon^*_{\lambda}(k) \to
\varepsilon^*_{\lambda}(k) + c\,k$.

The analytical expressions for the Feynman diagrams in
Fig.\,\ref{fig:fig8}k - Fig.\,\ref{fig:fig8}s, which survive in the
limit $m_{\sigma} \to \infty$, are given by
\begin{eqnarray}\label{eq:86}
\hspace{-0.3in}&&M_{\rm Fig. \ref{fig:fig8}k} (n \to p e^- \bar{\nu}_e
\gamma)_{\lambda} = eG_V\Big[\bar{u}_e(\vec{k}_e,\sigma_e)
  \gamma^{\mu} (1 - \gamma^5) v_{\nu}(\vec{k}_{\nu}, +
  \frac{1}{2})\Big] \Big[\bar{u}_p(\vec{k}_p, \sigma_p) \Big\{g^2_{\pi
    N}\int \frac{d^4p}{(2\pi)^4i}\,\gamma^5\,\frac{1}{m_N - \hat{k}_p
    - \hat{p} - i0}\nonumber\\
\hspace{-0.3in}&&\times
\,\hat{\varepsilon}^*_{\lambda}(k)\,\frac{1}{m_N - \hat{k}_p - \hat{k}
  - \hat{p} - i0}\,\gamma_{\mu}(1 - \gamma^5)\,\frac{1}{m_N -
  \hat{k}_n - \hat{p} - i0}\,\gamma^5\,\frac{1}{m^2_{\pi} - p^2 -
  i0}\Big\}\,u_n(\vec{k}_n, \sigma_n)\Big],\nonumber\\
\hspace{-0.3in}&&M_{\rm Fig. \ref{fig:fig8}m} (n \to p e^- \bar{\nu}_e
\gamma)_{\lambda} = eG_V\,\frac{q_{\mu}}{m^2_{\pi} - q^2 -
  i0}\,\Big[\bar{u}_e(\vec{k}_e,\sigma_e)
  \gamma_{\mu} (1 - \gamma^5) v_{\nu}(\vec{k}_{\nu}, +
  \frac{1}{2})\Big]\nonumber\\
\hspace{-0.3in}&&\times \,\Big[\bar{u}_p(\vec{k}_p, \sigma_p) \Big\{ - 2 g^2_{\pi N}\int
  \frac{d^4p}{(2\pi)^4i}\,\gamma^5\,\frac{1}{m_N - \hat{k}_n - \hat{p}
    - i0}\,\frac{(2(p - q) + k)\cdot
  \varepsilon^*_{\lambda}(k)}{m^2_{\pi} - (p - q + k)^2 -
  i0}\,\frac{1}{m^2_{\pi} - (p - q)^2 - i0}\Big\}\, u_n(\vec{k}_n,
\sigma_n)\Big]\nonumber\\
\hspace{-0.3in}&&M_{\rm Fig. \ref{fig:fig8}n} (n \to p e^- \bar{\nu}_e
\gamma)_{\lambda} = eG_V\,\frac{q_{\mu}}{m^2_{\pi} - q^2 -
  i0}\,\Big[\bar{u}_e(\vec{k}_e,\sigma_e) \gamma_{\mu} (1 -
  \gamma^5) v_{\nu}(\vec{k}_{\nu}, + \frac{1}{2})\Big]\nonumber\\
\hspace{-0.3in}&&\times \,\Big[\bar{u}_p(\vec{k}_p, \sigma_p) \Big\{ 
  2 g^2_{\pi N}\int \frac{d^4p}{(2\pi)^4i}\,\frac{1}{m_N - \hat{k}_n -
    \hat{p} - i0}\,\gamma^5 \,\frac{(2 p + k)\cdot
    \varepsilon^*_{\lambda}(k)}{m^2_{\pi} - (p + k)^2 -
    i0}\,\frac{1}{m^2_{\pi} - p^2 - i0}\Big\}\, u_n(\vec{k}_n,
  \sigma_n)\Big],\nonumber\\
\hspace{-0.3in}&&M_{\rm Fig. \ref{fig:fig8}o} (n \to p e^- \bar{\nu}_e
\gamma)_{\lambda} = eG_V\,\frac{q_{\mu}}{m^2_{\pi} - q^2 -
  i0}\,\Big[\bar{u}_e(\vec{k}_e,\sigma_e) \gamma_{\mu} (1 - \gamma^5)
  v_{\nu}(\vec{k}_{\nu}, + \frac{1}{2})\Big]\nonumber\\
\hspace{-0.3in}&&\times \Big[\bar{u}_p(\vec{k}_p, \sigma_p) \Big\{ 
  2 g^2_{\pi N}\int \frac{d^4p}{(2\pi)^4i}\,\frac{1}{m_N - \hat{k}_p -
    \hat{p} - i0}\,\hat{\varepsilon}^*_{\lambda}(k)\,\frac{1}{m_N -
    \hat{k}_p - \hat{k} - \hat{p} - i0}\,\gamma^5\,\frac{1}{m^2_{\pi}
    - (p + q)^2 - i0}\Big\}\, u_n(\vec{k}_n, \sigma_n)\Big],\nonumber\\
\hspace{-0.3in}&&M_{\rm Fig. \ref{fig:fig8}p} (n \to p e^- \bar{\nu}_e
\gamma)_{\lambda} = eG_V\,\frac{q_{\mu}}{m^2_{\pi} - q^2 -
  i0}\,\Big[\bar{u}_e(\vec{k}_e,\sigma_e) \gamma_{\mu} (1 - \gamma^5)
  v_{\nu}(\vec{k}_{\nu}, + \frac{1}{2})\Big]\nonumber\\
\hspace{-0.3in}&&\times \Big[\bar{u}_p(\vec{k}_p, \sigma_p) \Big\{ 2
  g^3_{\pi N}f_{\pi} \int \frac{d^4p}{(2\pi)^4i}\,\gamma^5\,
  \frac{1}{m_N - \hat{k}_p - \hat{p} -
    i0}\,\hat{\varepsilon}^*_{\lambda}(k)\,\frac{1}{m_N - \hat{k}_p -
    \hat{k} - \hat{p} - i0}\,\gamma^5\,\frac{1}{m_N - \hat{k}_n -
    \hat{p} - i0}\,\gamma^5\nonumber\\
\hspace{-0.3in}&&\times \,\frac{1}{m^2_{\pi} - p^2
    - i0}\Big\}\, u_n(\vec{k}_n, \sigma_n)\Big],\nonumber\\
\hspace{-0.3in}&&M_{\rm Fig. \ref{fig:fig8}r} (n \to p e^- \bar{\nu}_e
\gamma)_{\lambda} = eG_V\,\frac{1}{m^2_{\pi} - (k - q)^2 -
  i0}\,\Big[\bar{u}_e(\vec{k}_e,\sigma_e) \gamma_{\mu} (1 - \gamma^5)
  v_{\nu}(\vec{k}_{\nu}, + \frac{1}{2})\Big]\nonumber\\
\hspace{-0.3in}&&\times \Big[\bar{u}_p(\vec{k}_p, \sigma_p)\gamma^5
  \Big\{ - 2 g^2_{\pi N} \int \frac{d^4p}{(2\pi)^4i}\,{\rm
    tr}\Big\{\gamma^5\, \frac{1}{m_N - \hat{p} -
    i0}\,\hat{\varepsilon}^*_{\lambda}(k)\,\frac{1}{m_N - \hat{p} -
    \hat{k} - i0}\, \gamma_{\mu} (1 - \gamma^5)\nonumber\\
\hspace{-0.3in}&&\times \,\frac{1}{m_N - \hat{p} - \hat{k} + \hat{q}-
  i0}\Big\}\Big\}\, u_n(\vec{k}_n, \sigma_n)\Big],\nonumber\\
\hspace{-0.3in}&&M_{\rm Fig. \ref{fig:fig8}s} (n \to p e^- \bar{\nu}_e
\gamma)_{\lambda} = eG_V\,\frac{1}{m^2_{\pi} - (k - q)^2 -
  i0}\,\Big[\bar{u}_e(\vec{k}_e,\sigma_e) \gamma_{\mu} (1 - \gamma^5)
  v_{\nu}(\vec{k}_{\nu}, + \frac{1}{2})\Big]\nonumber\\
\hspace{-0.3in}&&\times \Big[\bar{u}_p(\vec{k}_p, \sigma_p)\gamma^5
  \Big\{- 2 \,\frac{g^2_{\pi N}}{m_N} \int
  \frac{d^4p}{(2\pi)^4i}\,\frac{(2(p - q) + k)\cdot
    \varepsilon^*_{\lambda}(k)}{m^2_{\pi} - (p - q + k)^2 -
    i0}\,\frac{(2p - q)_{\mu}}{m^2_{\pi} - (p - q )^2 - i0}\Big\}\,
  u_n(\vec{k}_n, \sigma_n)\Big],
\end{eqnarray}
where we have used the GT--relation $g_{\pi N} = m_N/f_{\pi}$. Now we
may classify the contributions of the Feynman diagrams, which survived
in the limit $m_{\sigma} \to \infty$ and are given by the analytical
expressions in Eq.(\ref{eq:86}), according to their properties with
respect to a gauge transformation $\varepsilon^*_{\lambda}(k) \to
\varepsilon^*_{\lambda}(k) + c\,k$. We  sum up the
contributions of the Feynman diagrams in Fig.\,\ref{fig:fig8}n and
Fig.\,\ref{fig:fig8}o and get
\begin{eqnarray}\label{eq:87}
\hspace{-0.3in}&&M_{\rm Fig. \ref{fig:fig8}n + Fig. \ref{fig:fig8}o}
(n \to p e^- \bar{\nu}_e \gamma)_{\lambda} =
eG_V\,\frac{q_{\mu}}{m^2_{\pi} - q^2 -
  i0}\,\Big[\bar{u}_e(\vec{k}_e,\sigma_e) \gamma_{\mu} (1 - \gamma^5)
  v_{\nu}(\vec{k}_{\nu}, + \frac{1}{2})\Big]\nonumber\\
\hspace{-0.3in}&&\times \,\Big[\bar{u}_p(\vec{k}_p, \sigma_p) 2
  g^2_{\pi N} \Big\{ \int \frac{d^4p}{(2\pi)^4i}\,\frac{1}{m_N -
    \hat{k}_n - \hat{p} - i0}\,\gamma^5 \,\frac{(2 p + k)\cdot
    \varepsilon^*_{\lambda}(k)}{m^2_{\pi} - (p + k)^2 -
    i0}\,\frac{1}{m^2_{\pi} - p^2 - i0}\nonumber\\
\hspace{-0.3in}&& +\int \frac{d^4p}{(2\pi)^4i}\,\frac{1}{m_N -
  \hat{k}_p - \hat{p} -
  i0}\,\hat{\varepsilon}^*_{\lambda}(k)\,\frac{1}{m_N - \hat{k}_p -
  \hat{k} - \hat{p} - i0}\,\gamma^5\,\frac{1}{m^2_{\pi} - (p + q)^2 -
  i0}\Big\}\, u_n(\vec{k}_n, \sigma_n)\Big]
\end{eqnarray}
and to replace $\varepsilon^*_{\lambda}(k) \to k$
\begin{eqnarray*}
\hspace{-0.3in}&&M_{\rm Fig. \ref{fig:fig8}n +
  Fig. \ref{fig:fig8}o}\Big|_{\varepsilon^*_{\lambda}(k) \to k} (n \to
p e^- \bar{\nu}_e \gamma)_{\lambda} = eG_V\,\frac{q_{\mu}}{m^2_{\pi} -
  q^2 - i0}\,\Big[\bar{u}_e(\vec{k}_e,\sigma_e) \gamma_{\mu} (1 -
  \gamma^5) v_{\nu}(\vec{k}_{\nu}, +
  \frac{1}{2})\Big]\,\Big[\bar{u}_p(\vec{k}_p, \sigma_p) 2 g^2_{\pi
    N}\nonumber\\
\end{eqnarray*}
\begin{eqnarray}\label{eq:88}
\hspace{-0.3in}&&\times \Big\{ \int
\frac{d^4p}{(2\pi)^4i}\,\frac{1}{m_N - \hat{k}_n - \hat{p} -
  i0}\,\gamma^5 \,\frac{1}{m^2_{\pi} - (p + k)^2 - i0} - \int
\frac{d^4p}{(2\pi)^4i}\,\frac{1}{m_N - \hat{k}_n - \hat{p} -
  i0}\,\gamma^5 \, \frac{1}{m^2_{\pi} - p^2 - i0}\nonumber\\
\hspace{-0.3in}&& + \int \frac{d^4p}{(2\pi)^4i}\,\frac{1}{m_N -
  \hat{k}_p - \hat{k} - \hat{p} - i0}\,\gamma^5\,\frac{1}{m^2_{\pi} -
  (p + q)^2 - i0} - \int \frac{d^4p}{(2\pi)^4i}\,\frac{1}{m_N -
  \hat{k}_p - \hat{p} - i0}\,\gamma^5\,\frac{1}{m^2_{\pi} - (p + q)^2
  - i0}\Big\}\nonumber\\
\hspace{-0.3in}&&\times \, u_n(\vec{k}_n, \sigma_n)\Big].
\end{eqnarray}
Making a change of variables $p + q \to p$ in the third integral and
$p + q \to p + k$ in the fourth integral in curly brackets one may
show that the r.h.s. of Eq.(\ref{eq:88}) vanishes. This confirms
invariance of the sum of the Feynman diagrams in Fig.\,\ref{fig:fig8}n
and Fig.\,\ref{fig:fig8}o  under a gauge transformation
$\varepsilon^*_{\lambda}(k) \to \varepsilon^*_{\lambda}(k) +
c\,k$. Removing the contributions of these diagrams from
Eq.(\ref{eq:86}) we are left with the following analytical expressions
of the Feynman diagrams, which are not invariant under a gauge
transformation $\varepsilon^*_{\lambda}(k) \to
\varepsilon^*_{\lambda}(k) + c\,k$. They are
\begin{eqnarray}\label{eq:89}
\hspace{-0.3in}&&M_{\rm Fig. \ref{fig:fig8}k} (n \to p e^- \bar{\nu}_e
\gamma)_{\lambda} = eG_V\Big[\bar{u}_e(\vec{k}_e,\sigma_e)
  \gamma^{\mu} (1 - \gamma^5) v_{\nu}(\vec{k}_{\nu}, +
  \frac{1}{2})\Big] \Big[\bar{u}_p(\vec{k}_p, \sigma_p) \Big\{g^2_{\pi
    N}\int \frac{d^4p}{(2\pi)^4i}\,\gamma^5\,\frac{1}{m_N - \hat{k}_p
    - \hat{p} - i0}\nonumber\\
\hspace{-0.3in}&&\times
\,\hat{\varepsilon}^*_{\lambda}(k)\,\frac{1}{m_N - \hat{k}_p - \hat{k}
  - \hat{p} - i0}\,\gamma_{\mu}(1 - \gamma^5)\,\frac{1}{m_N -
  \hat{k}_n - \hat{p} - i0}\,\gamma^5\,\frac{1}{m^2_{\pi} - p^2 -
  i0}\Big\}\,u_n(\vec{k}_n, \sigma_n)\Big],\nonumber\\
\hspace{-0.3in}&&M_{\rm Fig. \ref{fig:fig8}m} (n \to p e^- \bar{\nu}_e
\gamma)_{\lambda} = eG_V\,\frac{q_{\mu}}{m^2_{\pi} - q^2 -
  i0}\,\Big[\bar{u}_e(\vec{k}_e,\sigma_e)
  \gamma_{\mu} (1 - \gamma^5) v_{\nu}(\vec{k}_{\nu}, +
  \frac{1}{2})\Big]\nonumber\\
\hspace{-0.3in}&&\times \,\Big[\bar{u}_p(\vec{k}_p, \sigma_p) \Big\{ - 2 g^2_{\pi N}\int
  \frac{d^4p}{(2\pi)^4i}\,\gamma^5\,\frac{1}{m_N - \hat{k}_n - \hat{p}
    - i0}\,\frac{(2(p - q) + k)\cdot
  \varepsilon^*_{\lambda}(k)}{m^2_{\pi} - (p - q + k)^2 -
  i0}\,\frac{1}{m^2_{\pi} - (p - q)^2 - i0}\Big\}\, u_n(\vec{k}_n,
\sigma_n)\Big]\nonumber\\
\hspace{-0.3in}&&M_{\rm Fig. \ref{fig:fig8}p} (n \to p e^- \bar{\nu}_e
\gamma)_{\lambda} = eG_V\,\frac{q_{\mu}}{m^2_{\pi} - q^2 -
  i0}\,\Big[\bar{u}_e(\vec{k}_e,\sigma_e) \gamma_{\mu} (1 - \gamma^5)
  v_{\nu}(\vec{k}_{\nu}, + \frac{1}{2})\Big]\nonumber\\
\hspace{-0.3in}&&\times \Big[\bar{u}_p(\vec{k}_p, \sigma_p) \Big\{ 2
  g^3_{\pi N}f_{\pi} \int \frac{d^4p}{(2\pi)^4i}\,\gamma^5\,
  \frac{1}{m_N - \hat{k}_p - \hat{p} -
    i0}\,\hat{\varepsilon}^*_{\lambda}(k)\,\frac{1}{m_N - \hat{k}_p -
    \hat{k} - \hat{p} - i0}\,\gamma^5\,\frac{1}{m_N - \hat{k}_n -
    \hat{p} - i0}\,\gamma^5\nonumber\\
\hspace{-0.3in}&&\times \,\frac{1}{m^2_{\pi} - p^2
    - i0}\Big\}\, u_n(\vec{k}_n, \sigma_n)\Big],\nonumber\\
\hspace{-0.3in}&&M_{\rm Fig. \ref{fig:fig8}r} (n \to p e^- \bar{\nu}_e
\gamma)_{\lambda} = eG_V\,\frac{1}{m^2_{\pi} - (k - q)^2 -
  i0}\,\Big[\bar{u}_e(\vec{k}_e,\sigma_e) \gamma_{\mu} (1 - \gamma^5)
  v_{\nu}(\vec{k}_{\nu}, + \frac{1}{2})\Big]\nonumber\\
\hspace{-0.3in}&&\times \Big[\bar{u}_p(\vec{k}_p, \sigma_p)\gamma^5
  \Big\{ - 2 g^2_{\pi N} \int \frac{d^4p}{(2\pi)^4i}\,{\rm
    tr}\Big\{\gamma^5\, \frac{1}{m_N - \hat{p} -
    i0}\,\hat{\varepsilon}^*_{\lambda}(k)\,\frac{1}{m_N - \hat{p} -
    \hat{k} - i0}\, \gamma_{\mu} (1 - \gamma^5)\nonumber\\
\hspace{-0.3in}&&\times \,\frac{1}{m_N - \hat{p} - \hat{k} + \hat{q}-
  i0}\Big\}\Big\}\, u_n(\vec{k}_n, \sigma_n)\Big],\nonumber\\
\hspace{-0.3in}&&M_{\rm Fig. \ref{fig:fig8}s} (n \to p e^- \bar{\nu}_e
\gamma)_{\lambda} = eG_V\,\frac{1}{m^2_{\pi} - (k - q)^2 -
  i0}\,\Big[\bar{u}_e(\vec{k}_e,\sigma_e) \gamma_{\mu} (1 - \gamma^5)
  v_{\nu}(\vec{k}_{\nu}, + \frac{1}{2})\Big]\nonumber\\
\hspace{-0.3in}&&\times \Big[\bar{u}_p(\vec{k}_p, \sigma_p)\gamma^5
  \Big\{- 2 \,\frac{g^2_{\pi N}}{m_N} \int
  \frac{d^4p}{(2\pi)^4i}\,\frac{(2(p - q) + k)\cdot
    \varepsilon^*_{\lambda}(k)}{m^2_{\pi} - (p - q + k)^2 -
    i0}\,\frac{(2p - q)_{\mu}}{m^2_{\pi} - (p - q )^2 - i0}\Big\}\,
  u_n(\vec{k}_n, \sigma_n)\Big],
\end{eqnarray}
It is obvious that the analytical expressions Eq.(\ref{eq:89}) are not
invariant under a gauge transformation $\varepsilon^*_{\lambda}(k) \to
\varepsilon^*_{\lambda}(k) + c\,k$. However, in order to understand an
influence of such a non--invariance on the amplitude of the neutron
radiative $\beta^-$--decay we have to calculate them. Using
dimensional regularization for divergent integrals we get
\begin{eqnarray*}
\hspace{-0.3in}&&M_{\rm Fig. \ref{fig:fig8}k} (n \to p e^- \bar{\nu}_e
\gamma)_{\lambda} = eG_V\,\Big[\bar{u}_p(\vec{k}_p, \sigma_p)
  \Big\{\frac{g^2_{\pi N}}{32\pi^2}\,\frac{\pi}{2}\,{\ell
    n}2\,\frac{\hat{\varepsilon}^*_{\lambda}(k)}{m_N}\, \gamma_{\mu}
  (1 - \gamma^5) + O\Big(\frac{1}{m^2_N}\Big)\Big\} \,u_n(\vec{k}_n,
  \sigma_n)\Big]\nonumber\\
\hspace{-0.3in}&&\times\,\Big[\bar{u}_e(\vec{k}_e,\sigma_e)
  \gamma^{\mu} (1 - \gamma^5) v_{\nu}(\vec{k}_{\nu}, +
  \frac{1}{2})\Big],\nonumber\\
\hspace{-0.3in}&&M_{\rm Fig. \ref{fig:fig8}m} (n \to p e^- \bar{\nu}_e
\gamma)_{\lambda} = eG_V\,\frac{2 m_N q_{\mu}}{m^2_{\pi} - q^2 -
  i0}\nonumber\\
\hspace{-0.3in}&&\times \,\Big[\bar{u}_p(\vec{k}_p, \sigma_p)
  \gamma^5\Big\{ - \frac{g^2_{\pi N}}{32\pi^2}\Big({\ell
    n}\frac{\Lambda^2}{m^2_N} -
  \frac{1}{2}\Big)\,\frac{\hat{\varepsilon}^*_{\lambda}(k)}{m_N} +
  \frac{3g^2_{\pi N}}{16\pi^2}\,\frac{k_n\cdot
    \varepsilon^*_{\lambda}(k)}{m^2_N} +
  O\Big(\frac{1}{m^3_N}\Big)\Big\} u_n(\vec{k}_n,
  \sigma_n)\Big]\nonumber\\
\hspace{-0.3in}&&\times \,\Big[\bar{u}_e(\vec{k}_e,\sigma_e)
  \gamma_{\mu} (1 - \gamma^5) v_{\nu}(\vec{k}_{\nu}, +
  \frac{1}{2})\Big],\nonumber\\
\end{eqnarray*}
\begin{eqnarray}\label{eq:90}
\hspace{-0.3in}&&M_{\rm Fig. \ref{fig:fig8}p} (n \to p e^- \bar{\nu}_e
\gamma)_{\lambda} = eG_V\,\frac{2 m_N q_{\mu}}{m^2_{\pi} - q^2 -
  i0}\nonumber\\
\hspace{-0.3in}&&\times \,\Big[\bar{u}_p(\vec{k}_p, \sigma_p)\, \gamma^5\,
  \Big\{ \frac{g^2_{\pi N}}{32\pi^2}\,(2 - {\ell
    n}2)\,\frac{\hat{\varepsilon}^*_{\lambda}(k)}{m_N} +
  \frac{g^2_{\pi N}}{12\pi^2}\,\frac{k_n\cdot
    \varepsilon^*_{\lambda}(k)}{m^2_N} +
  O\Big(\frac{1}{m^3_N}\Big)\Big\}\, u_n(\vec{k}_n,
  \sigma_n)\Big]\nonumber\\
\hspace{-0.3in}&&\times \,\Big[\bar{u}_e(\vec{k}_e,\sigma_e)
  \gamma_{\mu} (1 - \gamma^5) v_{\nu}(\vec{k}_{\nu}, +
  \frac{1}{2})\Big],\nonumber\\
\hspace{-0.3in}&&M_{\rm Fig. \ref{fig:fig8}r} (n \to p e^- \bar{\nu}_e
\gamma)_{\lambda} = eG_V\,\frac{2 g_{\pi N} f_{\pi}}{m^2_{\pi} - (k -
  q)^2 - i0}\nonumber\\
\hspace{-0.3in}&&\times\, \Big[\bar{u}_e(\vec{k}_e,\sigma_e) \Big\{ -
  \frac{g^2_{\pi N}}{4\pi^2}\Big({\ell n}\frac{\Lambda^2}{m^2_N} -
  2\Big)\,\hat{\varepsilon}^*_{\lambda}(k) +
  O\Big(\frac{1}{m^2_N}\Big)\Big\}\,(1 - \gamma^5)
  v_{\nu}(\vec{k}_{\nu}, +
  \frac{1}{2})\Big]\,\Big[\bar{u}_p(\vec{k}_p, \sigma_p)\,\gamma^5\,
  u_n(\vec{k}_n, \sigma_n)\Big],\nonumber\\
\hspace{-0.3in}&&M_{\rm Fig. \ref{fig:fig8}s} (n \to p e^- \bar{\nu}_e
\gamma)_{\lambda} =  eG_V\,\frac{2 g_{\pi N} f_{\pi}}{m^2_{\pi} - (k -
  q)^2 - i0}\nonumber\\
\hspace{-0.3in}&&\times\, \Big[\bar{u}_e(\vec{k}_e,\sigma_e) \Big\{
  \frac{g^2_{\pi N}}{8\pi^2}\,\Big(\frac{m^2_{\pi}}{m^2_N}\,
  \hat{\varepsilon}^*_{\lambda}(k) + \frac{m_e}{m^2_N}\,q\cdot
  \varepsilon^*_{\lambda}(k)\Big)\,\Big({\ell
    n}\frac{\Lambda^2}{m^2_N} - 1\Big) +
  O\Big(\frac{1}{m^2_N}\Big)\Big\}\,(1 - \gamma^5)
  v_{\nu}(\vec{k}_{\nu}, + \frac{1}{2})\Big]\nonumber\\
\hspace{-0.3in}&&\times\,\Big[\bar{u}_p(\vec{k}_p,
  \sigma_p)\,\gamma^5\, u_n(\vec{k}_n, \sigma_n)\Big],
\end{eqnarray}
where we have used the GT--relation $g_{\pi N} = m_N/f_{\pi}$. The
divergent contribution of the Feynman diagram in Fig.\,\ref{fig:fig8}r
can be absorbed by the counter--term of the Feynman diagram in
Fig.\,\ref{fig:fig8}t. Then, the contributions of other diagrams are
of order $O(1/m_N)$ or even smaller and can be omitted to leading
order in the large nucleon mass expansion. This implies that gauge
non--invariant Feynman diagrams in Fig.\,\ref{fig:fig8}k,
Fig.\,\ref{fig:fig8}m, Fig.\,\ref{fig:fig8}p, Fig.\,\ref{fig:fig8}r,
Fig.\,\ref{fig:fig8}s and Fig.\,\ref{fig:fig8}t do not contribute to
the amplitude of the neutron radiative $\beta^-$--decay in the limit
$m_{\sigma} \to \infty$ and after renormalization, and to leading
order in the large nucleon mass expansion.

Thus, in the limit $m_{\sigma} \to \infty$ after renormalization and
to leading order in the large nucleon mass expansion the L$\sigma$M
and QED allow to describe only gauge invariant contributions of
hadronic structure of the neutron and proton to the neutron radiative
$\beta^-$--decay, where the main contributions come from the axial
coupling constant $g_A$. This agrees well with Sirlin's analysis of
contributions of strong low--energy interactions to the radiative
corrections of order $O(\alpha/\pi)$ \cite{Sirlin1967} and with
previous calculations of the rate and correlation coefficients of the
neutron radiative $\beta^-$--decay \cite{Gaponov1996,Bernard2004,
  Gardner2012, Gardner2013} (see also \cite{Ivanov2013,Ivanov2017a,
  Ivanov2017, Ivanov2017b}).

However, we would like to emphasize that divergences, induced by the
Feynman diagrams in Fig.\,\ref{fig:fig8}m and Fig.\,\ref{fig:fig8}s to
order $O(1/m_N)$ and $O(1/m^2_N)$, respectively, can be removed only
by the counter--terms, which are not from the set of counter--terms
defined by ${\cal L}^{(\rm CT)}_{\rm L\sigma M + QED}$ in
Eq.(\ref{eq:23}) and the counter--terms of the charged hadronic
axial--vector current Eq.(\ref{eq:29}). Hence, these diagrams violate
not only gauge invariance but also renormalizability of the amplitude
of the neutron radiative $\beta^-$--decay to orders $O(1/m_N)$ and
$O(1/m^2_N)$ in the large nucleon mass expansion.

\subsubsection*{\bf 4. Self--energy corrections to the $\pi^-$--meson 
and their contributions to the amplitude of the neutron radiative 
$\beta^-$--decay}

\begin{figure}
\centering 
\includegraphics[height=0.20\textheight]{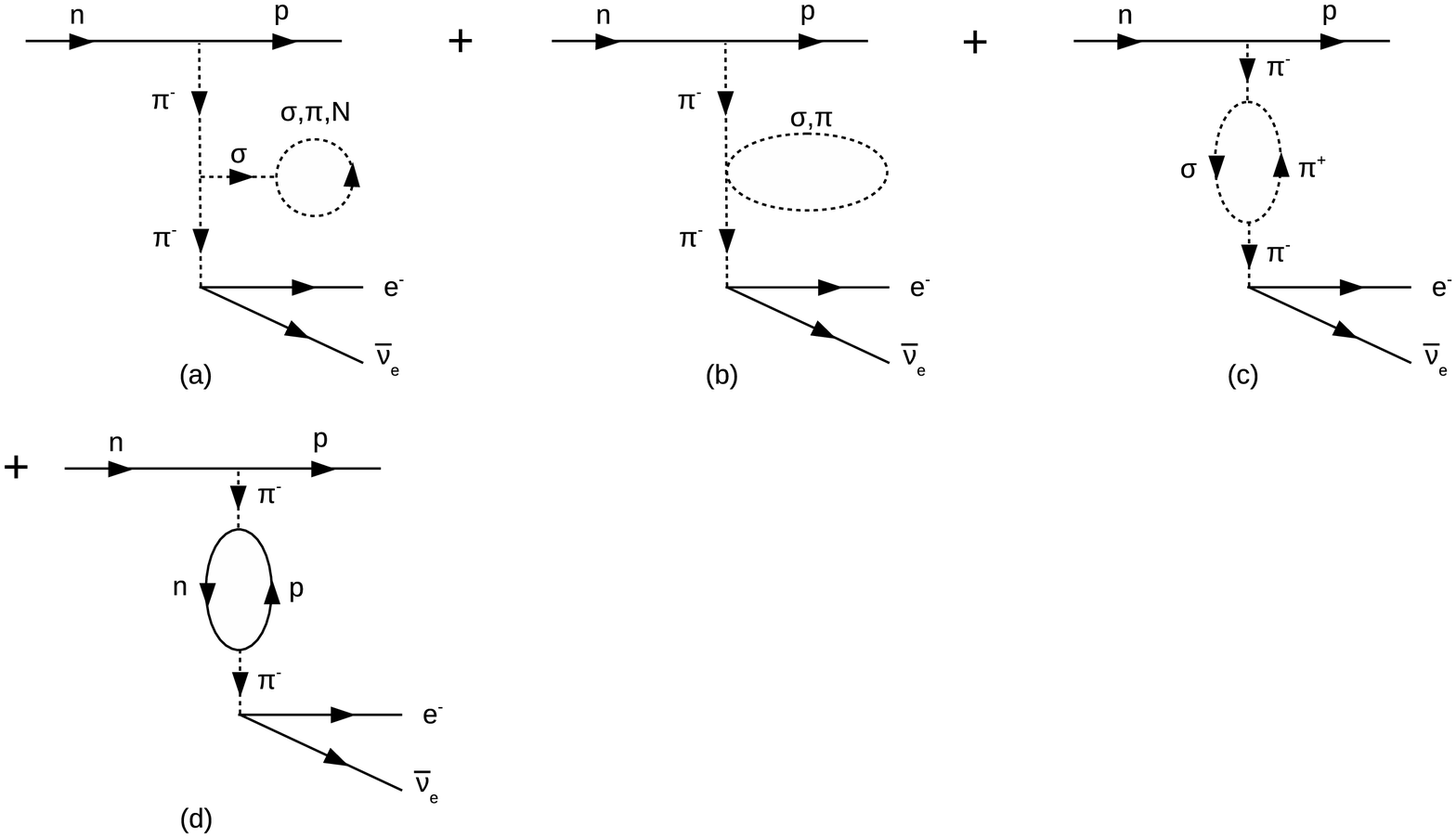}
  \caption{The Feynman diagrams, describing self--energy corrections
    for the $\pi^-$--mesons, which can be removed by renormalization
    of the mass and wave function of the $\pi^-$--meson.}
\label{fig:figM}
\end{figure}

The self--energy corrections to the $\pi^-$--meson state are defined
by the Feynman diagrams in Fig.\,\ref{fig:figM}.
In the limit $m_{\sigma} \to \infty$ and to leading order in the large
nucleon mass expansion the self--energy corrections to the
$\pi^-$--meson, given by the Feynman diagrams in Fig.\,\ref{fig:figM},
can be absorbed by mass and wave function renormalization of the
$\pi^-$--meson \cite{Matthews1954}. After renormalization the
contribution of the Feynman diagrams in Fig.\,\ref{fig:figM} to the
matrix element of the hadronic $n \to p$ transition vanishes.

Using the set of Feynman diagrams in Fig.\,\ref{fig:figM} and emitting
a photon from all lines of charged particles we determine the
contributions to the amplitude of the neutron radiative
$\beta^-$--decay. The analysis of these diagrams is similar to that
which we have done above in this section. As a result, we may argue
that the contribution of such a set of Feynman diagrams, taken in the
limit $m_{\sigma} \to \infty$ and to leading order in the large
nucleon mass expansion, to the amplitude of the neutron radiative
$\beta-$--decay vanishes after renormalization of the mass and wave
function of the $\pi^-$--meson and the $\pi^+\pi^-\gamma$--vertex.
Renormalization of the wave function of the $\pi^-$--meson state leads
to renormalization of the coupling constant $f_{\pi}$ (see
Eq.(\ref{eq:27})). However, such a renormalization of the coupling
constant $f_{\pi}$ can be taken into account only in the
two--hadron--loop approximation. In other words, in our calculations
in the one--hadron--loop approximation the coupling constant
$f_{\pi}$, describing the pion--nucleon coupling constant $g_{\pi N} =
m_N/f_{\pi}$, is {\it bare}.

\section{Conclusive discussion}
\label{sec:schluss}

We have analysed some properties of hadronic structure of the neutron
and proton in the neutron $\beta^-$--decay and neutron radiative
$\beta^-$--decay within the standard $V - A$ effective theory of weak
interactions, the linear $\sigma$--model with $SU(2)\times SU(2)$
chiral symmetry (the L$\sigma$M), describing strong low--energy
meson--nucleon interactions, and QED.  We have calculated the matrix
element of the hadronic $n \to p$ transition of the neutron
$\beta^-$--decay in the tree-- and one--hadron--loop approximation for
strong low--energy interactions in the L$\sigma$M. We have shown that
in the one--hadron--loop approximation, in the infinite limit of the
$\sigma$--meson mass, i.e. $m_{\sigma} \to \infty$, to leading order
in the large nucleon mass expansion and after renormalization the
L$\sigma$M reproduces well the standard Lorentz structure of the
matrix element of the hadronic $n \to p$ transition in the neutron
$\beta^-$--decay. A possibility to reproduce correct Lorentz structure
of the matrix element of the hadronic $n \to p$ transition in the
L$\sigma$M has been pointed out in \cite{Ivanov2018}, where the matrix
element of the hadronic $n \to p$ transition has been calculated in
Yukawa's theory of strong low--energy pion--nucleon interactions. The
term with the Lorentz structure $i\sigma_{\mu\nu}q^{\nu}/2m_N$, which
we have obtained in the matrix element of the hadronic $n \to p$
transition in the one--hadron--loop approximation, has been identified
with the contribution of the isovector anomalous magnetic moment of
the nucleon $\kappa = (g^2_{\pi N}/16 \pi^2)(3 + 2\,{\ell n}2)$ or the
contribution of the weak magnetism
\cite{Bilenky1959,Wilkinson1982}. The experimental value of the
isovector anomalous magnetic moment of the nucleon is equal to $\kappa
= \kappa_p - \kappa_n = 3.70589$ with $\kappa_p = 1.7928473$ and
$\kappa_n = - 1.9130427$, where $\kappa_p$ and $\kappa_n$ are
anomalous magnetic moments of the proton and neutron, respectively
\cite{PDG2016}. Setting $\kappa = 3.70589$ one may estimate the value
of the pion--nucleon coupling constant $g_{\pi N} =
\sqrt{\kappa\,16\pi^2/(3 + 2\,{\ell n}2)} = 11.55$.
From the GT--relation $g_{\pi N} = m_N/f_{\pi} = 11.55$ and $m_N =
(m_n + m_p)/2 \simeq 939\,{\rm MeV}$ \cite{PDG2016}, where $m_n =
939.5654 \,{\rm MeV}$ and $m_p = 938.2721\,{\rm MeV}$ are the neutron
and proton masses, one may estimate the value of the {\it bare} pion
decay constant $f_{\pi}$, i.e. $f_{\pi} = 81.3\,{\rm MeV}$, which
differs from the observable one $f_{\pi} \simeq 92.3\,{\rm MeV}$
\cite{PDG2016} by of about $12\,\%$. The observable value of the pion
decay constant $f_{\pi}$ can be obtained from the matrix element
$\langle 0|{\rm T}\big(e^{\,i\int d^4x\,{\cal L}_{\rm L\sigma
    M}(x)}J^+_{\mu}(0)\big)|\pi^-(q),{\rm in}\rangle$ after
renormalization.  Such a matrix element defines the hadronic $\pi^-
\to {\rm vacuum}$ transition in the weak $\pi^-$--meson leptonic decay
$\pi^- \to e^- + \bar{\nu}_e$. We would like to remind that in the
calculation of the amplitudes of neutron $\beta^-$ decays in the
one--hadron--loop approximation the pion decay constant $f_{\pi}$ is
{\it bare}.

Using the results, obtained for the amplitude of the neutron
$\beta^-$--decay, we have proceeded to the analysis of gauge
properties of hadronic structure of the neutron and proton in the
neutron radiative $\beta^-$--decay. We have shown that the amplitude
of the neutron radiative $\beta^-$--decay, calculated in the standard
$V - A$ effective theory of weak interactions and in the
tree--approximation for strong low--energy interactions in the
L$\sigma$M and QED, is invariant under a gauge transformation
$\varepsilon^*_{\lambda}(k) \to \varepsilon^*_{\lambda}(k) + c k$ of
the photon wave function.  The contribution of hadronic structure of
the neutron and proton appears in the form of the $\pi^-$--meson--pole
exchange, induced by the pion--nucleon strong low--energy interaction
and the mesonic part of the charged axial--vector hadronic current,
defined in the L$\sigma$M in the phase of spontaneously broken chiral
symmetry.

Then, we have considered the contributions to the amplitude of the
neutron radiative $\beta^-$--decay in the one--hadron--loop
approximation for strong low--energy interactions in the L$\sigma$M
and QCD, and analysed gauge properties of hadronic structure of the
neutron and proton in such an approximation. We have found that the
complete set of Feynman diagrams, describing in the one--hadron--loop
approximation the contributions to the amplitude of the neutron
radiative $\beta^-$--decay, is not invariant under a gauge
transformation $\varepsilon^*_{\lambda}(k) \to
\varepsilon^*_{\lambda}(k) + c\,k$. Gauge invariant Feynman diagrams
are shown in Fig.\,\ref{fig:fig6} and Fig.\,\ref{fig:fig7}. The
Feynman diagrams in Fig.\,\ref{fig:fig6} together with the Feynman
diagrams in Fig.\,\ref{fig:fig5} define the main contributions to the
amplitude of the neutron radiative $\beta^-$--decay, where strong
low--energy interactions are represented in the standard form of the
axial coupling constant and in the form of the $\pi^-$--meson--pole
exchange. The contribution of the $\pi^-$--meson--pole exchange is an
additional one, which has not been taken into account in previous
calculations of the neutron radiative $\beta^-$--decay
\cite{Gaponov1996,Bernard2004, Gardner2012, Gardner2013} (see also
\cite{Ivanov2013,Ivanov2017a,Ivanov2017b}). 

A gauge invariant contribution beyond the previous analysis of the
neutron radiative $\beta^-$--decay comes also from the Feynman
diagrams in Fig.\,\ref{fig:fig7}, which are obtained from the
self--energy corrections, caused by strong low--energy interactions in
the L$\sigma$M, to the neutron and proton states. We have found that
the main contribution comes from hadronic structure of the proton,
where the $pp\gamma$ vertex and self--energy corrections to the proton
state, renormalized by strong low--energy interactions in the
L$\sigma$M, obey the Ward identity (see Eq.(\ref{eq:73}) and
Eq.(\ref{eq:83})).

Then, in Fig.\,\ref{fig:fig8} gauge invariant contributions come from
the Feynman diagrams in Fig.\,\ref{fig:fig8}a -
Fig.\,\ref{fig:fig8}e. Other Feynman diagrams in Fig.\,\ref{fig:fig8}
are not invariant under a gauge transformation
$\varepsilon^*_{\lambda}(k) \to \varepsilon^*_{\lambda}(k) +
c\,k$. However, it is important to emphasize that some gauge
non--invariant contributions become gauge invariant in the limit
$m_{\sigma} \to \infty$. This concerns the sum of the Feynman diagrams
in Fig.\,\ref{fig:fig8}n and Fig.\,\ref{fig:fig8}o.

Excluding the contributions of the Feynman diagrams in
Fig.\,\ref{fig:fig8}n and Fig.\,\ref{fig:fig8}o, taken in the limit
$m_{\sigma} \to \infty$, we are left with the Feynman diagrams
Fig.\,\ref{fig:fig8}k, Fig.\,\ref{fig:fig8}m, Fig.\,\ref{fig:fig8}p,
Fig.\,\ref{fig:fig8}r, Fig.\,\ref{fig:fig8}s and
Fig.\,\ref{fig:fig8}t, which are not gauge invariant.  After the
calculation of the analytical expressions of the gauge non--invariant
Feynman diagrams in Fig.\,\ref{fig:fig8}k, Fig.\,\ref{fig:fig8}m,
Fig.\,\ref{fig:fig8}p, Fig.\,\ref{fig:fig8}r, Fig.\,\ref{fig:fig8}s
and Fig.\,\ref{fig:fig8}t we have found that removing the divergent
contribution of the Feynman diagram in Fig.\,\ref{fig:fig8}r by the
counter--term in Fig.\,\ref{fig:fig8}t the rest of the contributions
becomes of order $O(1/m_N)$ or even smaller with respect to the
contributions of the Feynman diagrams in Fig.\,\ref{fig:fig6}. As a
result, they can be omitted to leading order in the large nucleon mass
expansion. This result agrees well with Sirlin's analysis of strong
low--energy interactions in the radiative corrections to neutron
$\beta^-$ decays, carried out within the current algebra approach
\cite{Sirlin1967,Sirlin1978}.

Thus, we have obtained that to leading order in the large nucleon mass
expansion the contribution of hadronic structure of the neutron and
proton to the amplitude of the neutron radiative $\beta^-$--decay,
calculated in the standard $V - A$ effective theory of weak
interactions with the L$\sigma$M, describing strong low--energy
interactions, and QED, is gauge invariant, and moreover the main
contribution comes from the axial coupling constant.

However, It is important to emphasize that the Feynman diagrams in
Fig.\,\ref{fig:fig8}m and Fig.\,\ref{fig:fig8}s, violating gauge
invariance, induce also divergent contributions to order $O(1/m_N)$
and $O(1/m^2_N)$, respectively, which can be removed by the
counter--terms, which do not enter to the set of counter--terms
defined by ${\cal L}^{(\rm CT)}_{\rm L\sigma M + QED}$ in
Eq.(\ref{eq:23}) and the counter--terms of the axial--vector hadronic
current Eq.(\ref{eq:29}). Thus, these diagrams violate not only gauge
invariance but also renormalizability of the amplitude of the neutron
radiative $\beta^-$--decay. Hence, it is obvious that such a
theoretical approach to the analysis of the neutron radiative
$\beta^-$--decay, based on the standard $V - A$ effective theory of
weak interaction with the L$\sigma$M, describing strong low--energy
interactions, and QED, does not admit in principle the calculation of
the neutron radiative $\beta^-$--decay to order $O(1/m_N)$ and,
correspondingly, the radiative corrections to the neutron
$\beta^-$--decay to order $O(\alpha E_e/m_N)$.

The problem of the appearance of gauge non--invariant contributions
and contributions, violating renormalizability of the amplitude of the
neutron radiative $\beta^-$--decay, to order $O(1/m_N)$ and even
smaller, can be explained as follows. Indeed, the effective $V - A$
vertex of weak interactions is not the vertex of the combined quantum
field theory including the L$\sigma$M and QED. This implies that
correct gauge invariant contributions to the amplitude of the neutron
radiative $\beta^-$--decay and as well as to the neutron
$\beta^-$--decay can be obtained in any loop approximation and without
violation of renormalizability \cite{Ivanov1973,Ivanov1973a} only in
the Standard Electroweak Model (SEM) with strong low--energy
interactions, described by the L$\sigma$M . In such a combined quantum
field theory the vertex of the effective $V - A$ weak interactions is
defined by the $W^-$--electroweak boson exchange. This should result
in the complete set of gauge invariant Feynman diagrams including
electroweak bosons and photons coupled to the neutron and proton, and
hadrons from hadronic structure of the neutron and proton states,
described by strong low--energy interactions in the
L$\sigma$M. According to Weinberg \cite{Weinberg1967,Weinberg1971},
contributions of strong low--energy interactions to neutron
$\beta^-$--decays, obtained in such a combined quantum field theory in
the limit $m_{\sigma} \to \infty$, should reproduce after
renormalization contributions, obtained within the current algebra
approach. This should confirm the analysis of strong low--energy
interactions in neutron $\beta^-$ decays within the combined quantum
field theory with the SEM and L$\sigma$M at the confidence level of
Sirlin's analysis of strong low--energy interactions in the radiative
corrections to neutron $\beta^-$ decays \cite{Sirlin1967,Sirlin1978}.

We are planning to realize such an analysis of contributions of strong
low--energy interactions to the neutron radiative $\beta^-$--decay and
to the radiative corrections of order $O(\alpha/\pi)$ to the neutron
$\beta^-$--decay in our forthcoming publications.

We would like to emphasize that the results, which we are planning to
obtain to order $O(\alpha/\pi)$ for neutron $\beta^-$ decays in the
combined quantum field theory with the SEM and L$\sigma$M, should be
very important for the analysis of the SM corrections of order
$10^{-5}$, including i) the radiative corrections of order $O(\alpha
E_e/m_N)$ to next--to--leading order in the large proton mass
expansion, the radiative corrections of order $O(\alpha^2/\pi^2)$ to
leading order in the large proton mass expansion, where gauge
invariance of hadron structure of the neutron and proton plays an
important role, and iii) the weak magnetism and proton recoil
corrections of order $O(E^2_e/m^2_N)$.  An importance of these
corrections together with Wilkinson's corrections of order $10^{-5}$
\cite{Wilkinson1982}, which are caused by i) the proton recoil in the
Coulomb electron--proton final--state interaction, ii) the finite
proton radius, iii) the proton--lepton convolution and iv) the
higher--order {\it outer} radiative corrections, for the analysis of
experimental data on the search of contributions of interactions
beyond the SM has been pointed out in \cite{Ivanov2017b} (see also
\cite{Ivanov2017a,Ivanov2018a}).

\section{Acknowledgements}

We thank Hartmut Abele for discussions stimulating the work under this
paper as a first step towards the analysis of the SM corrections of
order $10^{-5}$, Torleif Ericson for fruitful discussions and useful
comments, and Alexander Andrianov for discussions of renormalizability
of the linear $\sigma$--model. The work of A. N. Ivanov was supported
by the Austrian ``Fonds zur F\"orderung der Wissenschaftlichen
Forschung'' (FWF) under contracts P26781-N20 and P26636-N20 and
``Deutsche F\"orderungsgemeinschaft'' (DFG) AB 128/5-2. The work of
R. H\"ollwieser was supported by the Deutsche Forschungsgemeinschaft
in the SFB/TR 55. The work of M. Wellenzohn was supported by the MA 23
(FH-Call 16) under the project ``Photonik - Stiftungsprofessur f\"ur
Lehre''.

\end{document}